\pdfoutput=1
\documentclass[twocolumn]{WileyMSP-template}
\usepackage[utf8]{inputenc}
\usepackage[T1]{fontenc}
\usepackage{lmodern}

\usepackage{float}
\usepackage{stfloats}

\usepackage{amsmath}
\usepackage[noadjust]{cite}
\usepackage{multirow}
\usepackage{array}
\usepackage{booktabs}
\usepackage{microtype}

\makeatletter
\renewcommand{\section}{\@startsection{section}{1}{\z@}{-1.3ex \@plus -0.2ex \@minus -0.2ex}{0.5ex \@plus 0.1ex}{\normalfont\Large\bfseries}}
\renewcommand{\subsection}{\@startsection{subsection}{2}{\z@}{-1.0ex \@plus -0.2ex \@minus -0.1ex}{0.35ex \@plus 0.1ex}{\normalfont\large\bfseries}}
\makeatother
\AtBeginDocument{%
  \setlength{\abovedisplayskip}{2.5pt plus 1pt minus 1pt}%
  \setlength{\belowdisplayskip}{2.5pt plus 1pt minus 1pt}%
  \setlength{\abovedisplayshortskip}{1.5pt plus 1pt minus 1pt}%
  \setlength{\belowdisplayshortskip}{1.5pt plus 1pt minus 1pt}%
}
\let\origtexttt\texttt
\makeatletter
\renewcommand{\texttt}[1]{%
  \begingroup
  \ttfamily
  \@tfor\@tempa:=#1\do{\@tempa\discretionary{}{}{}}%
  \endgroup
}
\makeatother

\usepackage{amssymb,amsfonts}
\usepackage{tabularx}
\usepackage{xcolor}
\usepackage{enumitem}
\usepackage[colorlinks=true,linkcolor=blue!70!black,citecolor=blue!70!black,urlcolor=blue!70!black]{hyperref}
\newcommand{\calgreen}[1]{#1}
\newcommand{\calyellow}[1]{#1$^{\dagger}$}
\newcommand{\calred}[1]{#1$^{\ddagger}$}
\setlist[itemize]{topsep=3pt,itemsep=2pt,leftmargin=1.5em}
\setlist[enumerate]{topsep=3pt,itemsep=2pt,leftmargin=1.8em}
\begin{document}

\pagestyle{fancy}
\rhead{\includegraphics[width=2.5cm]{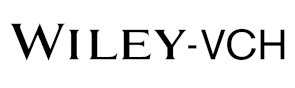}}

\title{Continuous Kinetic Model of Counterion-Mediated Redox Switching in 14-Bit Molecular Memristors for Analog In-Memory Computing}

\twocolumn[
  \begin{@twocolumnfalse}
    \setlength{\RaggedRightParindent}{0pt}
    \setlength{\parindent}{0pt}
    \maketitle

    \author{Logan Larsh$^{1}$}
    \author{Dhiman Biswas$^{2}$}
    \author{Nazmus Saadat As-Saquib$^{3}$}
    \author{Su-in Yi$^{3}$}
    \author{Thirumalai Venkatesan$^{2}$}
    \author{Mike Banad$^{1,*}$}

    \begin{affiliations}
    $^{1}$The University of Oklahoma, Department of Electrical and Computer Engineering, Norman, OK 73019\\
    $^{2}$The University of Oklahoma, Department of Physics and Astronomy, Norman, OK 73019\\
    $^{3}$Texas A\&M University, Department of Electrical and Computer Engineering, College Station, TX 77843\\
    $^*$bana@ou.edu
    \end{affiliations}


    \begin{abstract}

    Kinetic molecular memristors offer a compelling path toward high-density analog in-memory computing by achieving 14-bit precision, wherein discrete redox transitions and counterion displacements dictate non-volatile conductance across a deterministic energy landscape. However, integrating these nanoscale kinetics into macroscale circuit simulators has remained bottlenecked by derivative discontinuities and solver convergence failures. Here, we present a physics-inspired continuous kinetic model that bridges microscopic molecular transport with scalable circuit architectures. By casting discrete intramolecular states, counterion relaxation dynamics, and Marcus-Hush-Chidsey electron transfer rates into continuously differentiable state variables, we derive a robust, Spectre-compatible compact model and a matching vectorized Python solver. This work culminates in a unified computational platform to reliably predict the behavior of complex circuits built on these molecular memristors.

    \end{abstract}
    \vspace{0.4em}
  \end{@twocolumnfalse}
]

\setlength{\parindent}{10pt}
\setlength{\RaggedRightParindent}{10pt}

\section{Introduction}

\begin{figure*}[!t]
    \centering
    \includegraphics[width=0.95\textwidth]{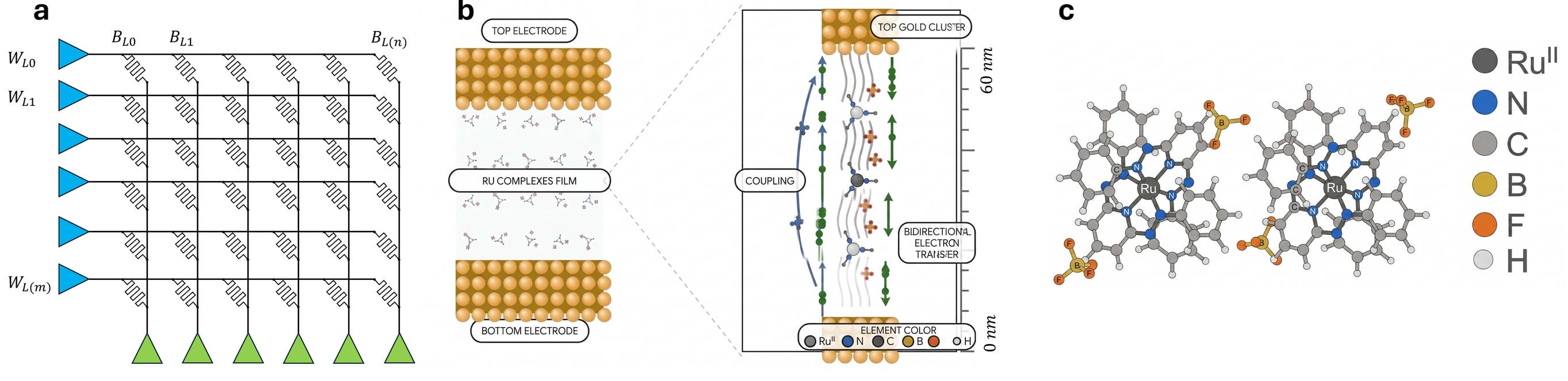}
    \caption{Structural and conceptual overview of the molecular memristor crossbar. \textbf{a}, Circuit schematic of a dense crossbar array with molecular memristors at each crosspoint. \textbf{b}, Device cross section of the 60~nm $[\text{Ru}^{\text{II}}\text{L}_2](\text{BF}_4)_2$ thin film between Au electrodes, with an inset illustrating bidirectional electron transfer. \textbf{c}, Molecular structure of the bis-tridentate ruthenium complex.}
    \label{fig:intro_abstract}
\end{figure*}

Vector-matrix multiplication (VMM) serves as the fundamental mathematical backbone of modern computing algorithms, ranging from deep neural networks to advanced computational theory \cite{krizhevsky2012, jouppi2017, gupta2020}. Conventional digital architectures execute these operations with high latency and energy overhead due to the von Neumann bottleneck, where repeated data transfers between memory hierarchies and processing cores substantially throttle throughput \cite{xia2019, ham2021}. Analog in-memory computing offers a transformative alternative by mapping matrix weights directly to conductances in dense non-volatile memory (NVM) crossbars, enabling input voltage vectors to be multiplied and accumulated natively in $O(1)$ time using Ohm's law for local multiplication and Kirchhoff's current law for parallel summation \cite{burr2017neuromorphic, jeong2016, yi2021a, adam2016, kvatinsky2013, balatti2015, nguyen2017}.

Despite the architectural promise of such crossbar arrays, widespread adoption remains bottlenecked by a ``Hardware Wall'' rooted in the physical limitations of conventional NVM devices. Phase-change memory (PCM) relies on high-energy melt-quench processes that inherently limit endurance and induce abrupt, highly asymmetric RESET transitions \cite{burr2017neuromorphic}. Conversely, filamentary resistive random access memory (RRAM) is plagued by stochastic ion migration and pronounced cycle-to-cycle variability \cite{xia2019, wang2020}. These physical non-idealities restrict traditional NVMs to low analog resolution, typically below 6-bit precision \cite{jiang2016, kim2017}, often forcing reliance on power-hungry mixed-precision architectures that retain a digital von Neumann coprocessor to iteratively correct analog computational errors \cite{legallo2018mixed}.

Re-examining classical continuous state assumptions of memristive systems \cite{strukov2008, chua1971, chua1976}, Goswami and coworkers \cite{goswami2017, goswami2020, sharma2024, williams2024} demonstrated that charge disproportionation in baseline $[\text{Ru}^{\text{II}}\text{L}_2](\text{BF}_4)_2$ bis-azo-aromatic complexes (with mobile $\text{BF}_4^-$ counterions, $r = 2.18$~\AA) resolves this bottleneck alongside related molecular computing architectures \cite{yi2022}. Outperforming roadmap projections, these $\text{BF}_4^-$ devices combine projected sub-100-attojoule write energies for scaled 10~nm $\times$ 10~nm nanocells (alongside 0.19~nJ per nominal write pulse evaluated for macroscopic $1.05~\mu\text{m} \times 1.1~\mu\text{m}$ crossbar cells), $10^{12}$ write cycles enabled by charge-conserving intermolecular redox without filament rupture, and sub-5-nanosecond switching speeds \cite{sharma2024, williams2024}. While empirical pulse characterization indicates no conductance perturbation below 830~mV \cite{sharma2024}, our continuous compact formulation anchors the effective $\text{BF}_4^-$ activation threshold at $V_{\text{th}} = 0.4664$~V ($\approx 0.47$~V), supporting non-destructive sub-threshold readout at $V_{\text{read}} \le 0.20$~V across millions of inference dot products. In a representative junction with a 100~nm $\times$ 100~nm area and a 60~nm film thickness accommodating approximately $10^4$ molecules and $2 \times 10^4$ counterions (scaling to $\sim 10^6$ molecules in $1.05~\mu\text{m} \times 1.1~\mu\text{m}$ cells; fabrication protocol, crossbar metallization, and coordination film properties detailed in Supplementary Note S3), the operational states reflect the collective ligand-centered electronic configurations of interacting $[\text{Ru}^{\text{II}}\text{L}_2]^{2+}$ molecule pairs (where numbers denote additional electrons occupying the azo-aromatic $\pi^*$ antibonding orbitals while the metal center remains low-spin $\text{Ru}^{\text{II}}$): the symmetry-broken, low-conducting (31) ground state, the valence-symmetric (22) state, the intermediate (11) state, and the unreduced-ligand (00) configuration \cite{goswami2020, kundu2024, gaur2024}. While quasi-static voltage sweeps ($>50$~ms per step) produce abrupt digital transitions from (31) to (11) at $\approx 2.0$~V and to (00) at $\approx 2.76$~V, sub-100-nanosecond pulses unlock a kinetically limited analog trajectory \cite{kundu2024}. Because transitioning a molecular pair from (31) to (22) requires a modest $\text{BF}_4^-$ counterion activation barrier of only 130~meV (with an unrelaxed electronic Gibbs free energy difference $\Delta G^0 = 0.10$~eV), which is merely 8\% of the barrier separating the (31) and (11) states \cite{kundu2024}, short positive pulses ($+900$~mV, $80$~ns) kinetically confine excitation to this lower-energy pathway. This pulse duration supplies the energy to surmount the 130~meV barrier, driving linear charge-conserving intermolecular electron transfer into the valence-symmetric (22) state while counterion relaxation into local electrostatic pockets stabilizes the configuration at a $+25$~meV metastable elevation and confers non-volatility to each level \cite{goswami2020, sharma2024, kundu2024}.
\par
This establishes zeroth-order kinetics driven by progressive volumetric state conversion, where deterministic increments in the displaced $\text{BF}_4^-$ counterion population advance carrier hopping probability wavefronts to scale conductance linearly across 16,520 discrete analog levels. Negative pulses ($-750$~mV, $65$~ns) reverse this $\text{BF}_4^-$ counterion motion, symmetrically depressing the device back to the (31) state to enable open-loop weight updates for dot product arrays without costly verify-and-correct feedback cycles \cite{sharma2024, kundu2024}. However, scaling from isolated junctions to macroscopic crossbars couples millions of nonlinear boundary conditions across shared interconnects, where empirical curve fitting fails to capture parasitic IR drops, localized thermal drifts, or subtle screening shifts. Directly incorporating Thomas-Fermi potential screening, Marcus-Hush-Chidsey electron transfer rates, and planar nucleation barriers within a differentiable mathematical formulation is thus essential for the predictive fidelity required to design and verify large-scale dot product engines in CMOS+X architectures.
\par
Even with these physical properties established, transitioning from isolated junctions to system deployment confronts a corresponding ``Software Wall'' at the circuit-to-algorithm interface. Evaluating large crossbars requires translating nanoscale molecular kinetics into macroscopic circuit equations, where standard Newton-Raphson solvers frequently struggle to converge against the sharp mathematical derivatives of raw molecular kinetics \cite{kim2019, biolek2009}. Existing simulation paradigms are forced to compromise: circuit-level macro models (such as NeuroSim) estimate peripheral area and energy but abstract away the transient physical dynamics necessary for hardware-in-the-loop training \cite{chen2018neurosim}, while machine learning libraries (such as MemTorch) offer parallelized network integration but lack the native analog characteristics required for circuit design \cite{lammie2020memtorch}. Although detailed physical network models, such as the thermodynamic network model proposed by Goswami et al. \cite{goswami2021}, capture nanoscale kinetics, they are too computationally intensive for simulating large-scale VMM arrays. Unlocking the full potential of molecular electronics therefore requires a continuous physical model mapping discrete molecular redox kinetics and interfacial screening profiles into differentiable state variables.
\par
Building upon this legacy of physical device innovation, here we present a physics-inspired continuous kinetic model and multi-scale validation workflow optimized for transition-metal complex molecular memristors (Figure 1; architectural overview and 11-section structural glossary in Supplementary Note S1). We incorporate the kinetic hopping transport model formulated by Kundu and Goswami \cite{kundu2024}, in which the 60~nm $[\text{Ru}^{\text{II}}\text{L}_2](\text{BF}_4)_2$ molecular film is discretized into a 30-site hopping framework governed by stationary rate equations combining Marcus electron transfer theory, Thomas-Fermi electrostatic potential screening, and planar nucleation kinetics (microscopic transport derivations in Supplementary Note S4). To translate these transport equations into a scalable simulation platform, we developed two mathematically equivalent engines tailored for complementary design scales: a transistor-level Verilog-A compact model in Cadence Spectre/Virtuoso for interactive CMOS circuit co-simulation, and a vectorized Python solver utilizing $C^1$-continuous monotonic cubic Hermite state transitions for license-free, constant-time $O(1)$ array evaluation (computational solver architecture, memory profiling, and MNA acceleration detailed in Supplementary Note S8). While conventional memristor models in Newton-Raphson circuit simulators frequently suffer from convergence failures caused by discontinuous derivatives and abrupt boundary cutoffs \cite{kim2019, biolek2009}, our continuous formulation promotes robust numerical convergence via direction-aware Hermite smoothstep boundary clamping, softplus voltage thresholding, and closed-form analytical screening profiles. For the baseline $\text{BF}_4^-$ coordination system, this formulation demonstrates high quantitative agreement with empirical hardware measurements across both nominal linear switching ($\text{RMSE} = 36.2~\mu\text{S}$, relative error $0.84\%$ across $33,040$ pulses) and high-overdrive saturation and snapback ($\text{RMSE} = 150.2~\mu\text{S}$, relative error $2.90\%$; Figure 3b). Using this platform, we executed hardware-in-the-loop neural training and spatial frequency processing on high-resolution James Webb Space Telescope imagery, demonstrating how molecular transport physics translates into scalable neuromorphic performance.

\section{Physics Modeling}

\begin{figure*}[!t]
    \centering
    \includegraphics[width=0.95\textwidth]{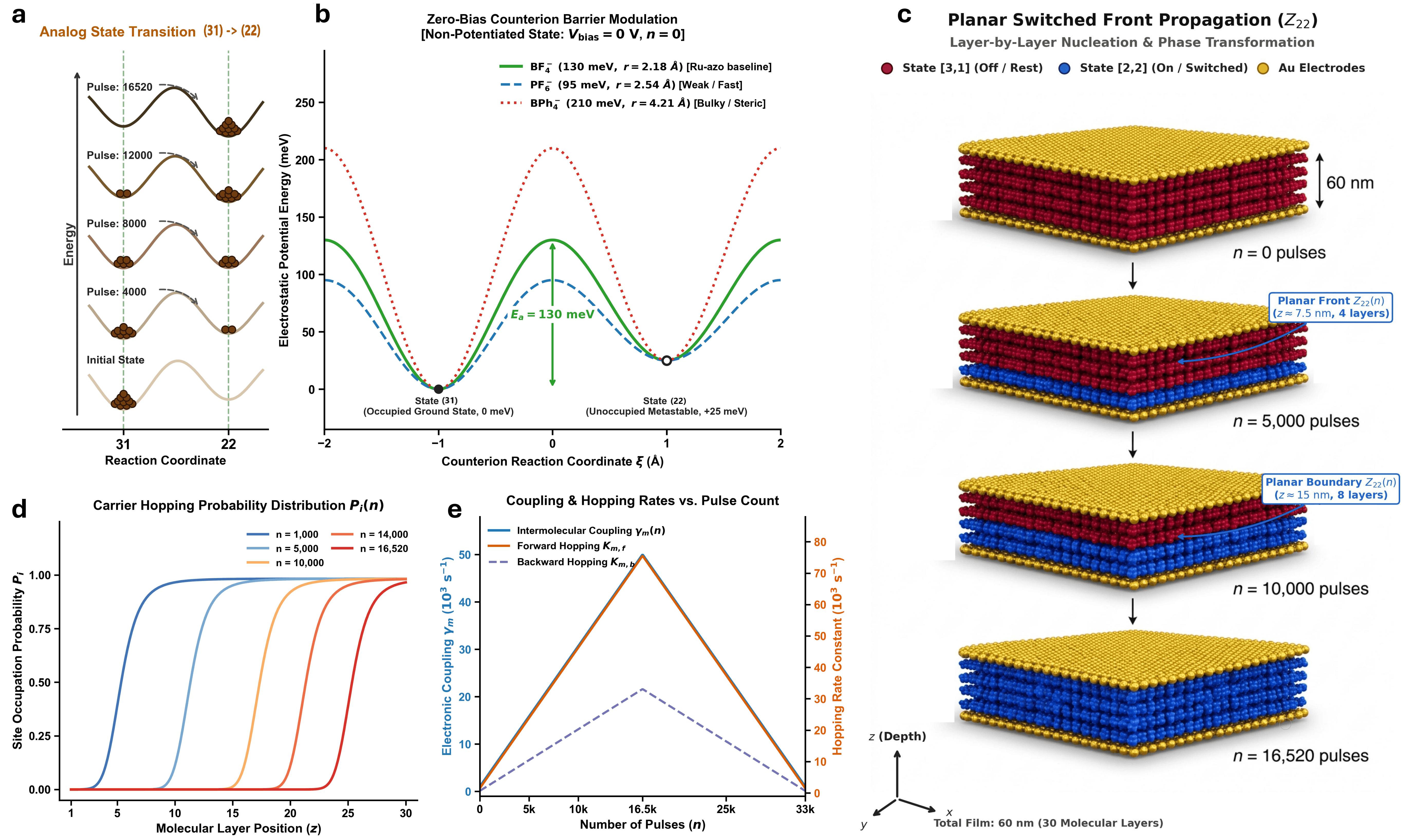}
    \caption{Microscopic physical transport mechanisms and energy landscape. \textbf{a}, Double-well potential landscape tilting and carrier transfer during (31) $\to$ (22) potentiation across 16,520 pulses in baseline $[\text{Ru}^{\text{II}}\text{L}_2](\text{BF}_4)_2$. \textbf{b}, Zero-bias counterion electrostatic potential energy landscape contrasting activation barriers for baseline $\text{BF}_4^-$ ($130$~meV, $r = 2.18$~\AA), lower-barrier $\text{PF}_6^-$ ($95$~meV, $r = 2.54$~\AA), and unmodeled bulky literature reference $\text{BPh}_4^-$ ($210$~meV, $r = 4.21$~\AA) between the ground state ($\text{31}_{31}$, $0$~meV) and counterion-relaxed metastable state ($\text{22}_{22}$, $+25$~meV net elevation for $\text{BF}_4^-$). \textbf{c}, 3D schematic of planar switched front ($Z_{22}$) propagation through the 30-layer $\text{BF}_4^-$ molecular stack at $n = 0$, $5,000$, $10,000$, and $16,520$ pulses (resting state 31 in crimson, switched state 22 in blue). \textbf{d}, Stationary carrier hopping occupation probability distribution $P_i(n)$ across the 30 molecular layers along depth $z$. \textbf{e}, Normalized per-channel intermolecular electronic coupling $\gamma_m(n)$ and directional forward ($K_{m,f}$) and backward ($K_{m,b}$) hopping rate constants across the 33,040-pulse $\text{BF}_4^-$ cycle.}
    \label{fig:physics_modeling}
\end{figure*}

\subsection{Initial Physical Mechanisms}

The benchmarked molecular memristor system investigated here employs a 60~nm coordination film of $[\text{Ru}^{\text{II}}\text{L}_2](\text{BF}_4)_2$ discretized into $L = 30$ bimolecular site layers along depth $z$ ($\approx 2$~nm per layer) within a dense crossbar architecture \cite{sharma2024, kundu2024, gaur2024} (detailed device fabrication protocols, crossbar contact geometries, and electroforming-free switching physics are elaborated in Note S3, Supporting Information), while the normalized spatial coordinate $t = z/L$ and the continuous kinetic formulations derived below generalize across arbitrary film thicknesses, layer counts, and counterion chemistries (Figure 2; complete architectural roadmap in Note S1 and master parameter catalogs in Note S2, Supporting Information). Crucially, the Cadence Spectre/Virtuoso and vectorized Python implementations derived from this framework are mathematically equivalent formulations of the same underlying physical kinetics, Marcus electron transfer rates, and continuous state variables. As detailed in Table \ref{tab:dual_engine_comparison} (with a direct testbench-by-testbench comparison validating $< 0.02\%$ transient parity in Note S7 of the Supporting Information), the Spectre/Virtuoso compact model targets interactive, transistor-level circuit simulation, schematic-driven exploration, and SPICE co-simulation with peripheral CMOS circuitry (such as sense amplifiers, write drivers, and decoders), whereas the vectorized Python solver makes high-throughput crossbar scalability, large-scale array computing, and hardware-in-the-loop AI projects accessible and computationally tractable without commercial EDA licensing barriers.

\begin{table}[H]
\centering
\caption{Comparison of Dual Simulation Framework Engines}
\label{tab:dual_engine_comparison}
\renewcommand{\arraystretch}{1.22}
\resizebox{\columnwidth}{!}{%
\begin{tabular}{lcc}
\toprule
\textbf{Model Dimension} & \textbf{Spectre / Virtuoso} & \textbf{Vectorized Python Solver} \\
\midrule
Kinetic Formulation & $C^1$-Continuous ODE & $C^1$-Continuous ODE \\
Transport Physics & Marcus-Hush-Chidsey & Marcus-Hush-Chidsey \\
Boundary Clamping & Hermite Smoothstep & Hermite Smoothstep \\
Engine Architecture & Verilog-A Compact Model & Vectorized MNA \& PCHIP \\
Simulation Scope & CMOS Co-Simulation & License-Free Array Scaling \\
Target Workloads & Transient, STDP, \& IR-Drop & MNIST \& JWST IDFT \\
Transient Parity & SPICE Reference Baseline & $< 0.02\%$ Error (Note S7) \\
\bottomrule
\end{tabular}%
}
\end{table}

Grounded in the shared physical core summarized in Table \ref{tab:dual_engine_comparison}, each bis-tridentate ruthenium complex in this coordination network coordinates with two mobile tetrafluoroborate ($\text{BF}_4^-$) counterions to maintain local electroneutrality. Unlike filamentary memristors that rely on stochastic ion migration and suffer from conductive filament rupture \cite{xia2019, wang2020, jiang2016, huang2016}, this molecular junction achieves 14-bit analog precision and $10^{12}$-cycle endurance through uniform field-driven intermolecular electron transfer coupled to deterministic counterion displacement that minimizes electrostatic energy \cite{goswami2021}. As illustrated in Figure 2a, applying sequential write pulses indexed by cumulative pulse count $n$ tilts the double-well electrostatic landscape, driving a deterministic carrier ensemble redistribution from the charge-disproportionate ground state (31) into the valence-symmetric conductive state (22) across 16,520 discrete analog levels. At zero bias ($V_{\text{bias}} = 0$~V, $n = 0$), thermodynamic equilibrium is anchored by local counterion electrostatic binding pockets along reaction coordinate $\xi$ (Figure 2b) \cite{rath2022, kundu2024}. Prior to ionic relaxation, fast vertical electron transfer from the occupied ground state ($\text{31}_{31}$, $0$~meV) to the unrelaxed valence-symmetric intermediate ($\text{22}_{31}$) entails an unrelaxed electronic Gibbs free energy difference of $\Delta G^0 = 0.10$~eV ($+100$~meV) and reorganization energy $E_\lambda = 0.04$~eV ($40$~meV) \cite{sharma2024, kundu2024}. Subsequent $\text{BF}_4^-$ counterion relaxation into reconfigured coordination pockets ($\text{22}_{31} \to \text{22}_{22}$) releases $\approx 75$~meV of lattice stabilization energy to anchor the metastable state ($\text{22}_{22}$) at a $+25$~meV net elevation along $\xi$ (Figure 2b) \cite{goswami2020, rath2022, sharma2024, kundu2024}, conferring non-volatility to each programmed level. Spontaneous thermal relaxation is suppressed by the baseline $\text{BF}_4^-$ counterion ($r = 2.18$~\AA), whose intervening activation barrier $E_a = 130$~meV ($\approx 134$~meV experimentally \cite{sharma2024, kundu2024}) far exceeds room-temperature thermal energy ($k_B T \approx 25.7$~meV at $T_{\text{amb}} = 298.15$~K), shielding programmed states against thermal drift. Substituting alternative counterions demonstrates how ionic radius and Coulombic binding modulate this barrier and reshape switching kinetics \cite{kundu2024, gaur2024}: a lower barrier ($95$~meV for modeled alternative $\text{PF}_6^-$, $r = 2.54$~\AA) accelerates domain nucleation into a steeper non-linear regime, whereas the unmodeled bulky tetraphenylborate literature reference ($\text{BPh}_4^-$, $r = 4.21$~\AA, $210$~meV barrier) introduces steric crowding and excluded-volume limits that restrict counterion displacement \cite{williams2024, gaur2024} (crystallographic coordination geometries, steric excluded-volume limits, and Coulombic barrier tuning across alternative counterions are detailed in Note S3, Supporting Information).

As detailed in Note S4 of the Supporting Information, advancing cumulative pulse count $n$ under calibrated $\text{BF}_4^-$ excitation ($+900$~mV, $80$~ns) drives spatial counterion drift across the $L = 30$ layers, dynamically shifting the internal electrostatic potential profile $\Phi_{\text{mol}}(z,n)$ via Thomas-Fermi screening \cite{liang2004electrostatic, nitzan2002electrostatic, kundu2024} and accelerating the local $(31) \to (22)$ electron transfer rate $k_{\text{ET}}(z,n)$ governed by the Marcus-Hush-Chidsey integral over the Fermi-Dirac acceptor distribution $\psi(E)$ \cite{marcus1965theory, migliore2012evaluation, kundu2024}. Because $E_\lambda = 0.04$~eV is comparable to the thermal scale, local potential tilting $e\Phi_{\text{mol}}(z,n)$ enables rapid sub-volt redox switching without destructive electroforming. For layers where the normalized transfer probability $P(z,n) = k_{\text{ET}}(z,n) \gamma_{\text{DA}}^{-1}$ surmounts the critical switching threshold ($\theta(z,n)=1$), nucleation-limited domain switching captures in-plane spatial inhomogeneity \cite{jo2007domain, tagantsev2002non, kundu2024}, evaluating the \mbox{$x$--$y$} fraction of molecular pairs converted from resting (31) to conductive (22) at depth $z$ as a function of cumulative write pulse count $n$:
\begin{equation}
\label{eq:nucleation_domains}
Nr_{xy}(z,n) = \frac{1}{\pi} \arctan\left(\frac{n - \chi(z)}{\kappa_{\text{nuc}}}\right) + \frac{1}{2}
\end{equation}
where $\chi(z)$ is the layer-dependent nucleation center and $\kappa_{\text{nuc}}$ is the nucleation half-width parameter (in pulses) governing the inverse nucleation rate across the \mbox{$x$--$y$} plane \cite{kundu2024}. Because ionic displacement is kinetically slower than electronic hopping, short $+900$~mV, $80$~ns pulses confine switching to this low-energy counterion pathway within a small-perturbation regime, lowering the $130$~meV barrier to displace a deterministic fraction of $\text{BF}_4^-$ anions per pulse without triggering filament formation \cite{kundu2024}. As visualized in the subsampled 12-slab 3D schematic of Figure 2c (where each rendered slab represents $2.5$ physical molecular layers), this domain nucleation manifests macroscopically as a planar switched front $Z_{22}(n)$ advancing from early-stage nucleation ($4$ illustrated slabs in Figure 2c, corresponding to $Z_{22} \approx 9.3$ physical layers or $18.7$~nm at $n = 5,000$ in Equation \eqref{eq:front_regression}) and intermediate conversion ($8$ illustrated slabs in Figure 2c, corresponding to $Z_{22} \approx 18.7$ physical layers or $37.4$~nm at $n = 10,000$) to complete volumetric conversion across all 30 physical layers ($60$~nm, $12$ illustrated slabs) at $n = 16,520$ pulses, suppressing stochastic filament jitter to support uniform weight updates. Averaging $Nr_{xy}(z,n)$ across all 30 layers yields the volumetric switched state fraction as a function of the pulse state coordinate $n$:
\begin{equation}
\label{eq:volume_fraction}
f_{22}(n) = \frac{1}{30}\sum_{z=1}^{30} Nr_{xy}(z,n)
\end{equation}
with $f_{31}(n) = 1 - f_{22}(n)$, which linearly scales the effective intermolecular electronic coupling rate constant:
\begin{equation}
\label{eq:gamma_coupling}
\gamma_m(n) = f_{31}(n)\gamma_{m,31} + f_{22}(n)\gamma_{m,22}
\end{equation}
determining the directional forward and backward intermolecular hopping rate constants during readout:
\begin{equation}
\label{eq:hopping_rate}
K_{m,f/b}(n) = \gamma_m(n) \exp\left[\pm \frac{e a_m V_r}{\eta k_B T}\right]
\end{equation}
where $e$ is the elementary charge, $a_m = 1/30$ is the dimensionless site voltage fraction across the 30-layer chain, $V_r$ is the applied read voltage, and $\eta$ is the conduction ideality factor. In the resting (31) ground state of the baseline $\text{BF}_4^-$ complex, charge disproportionation restricts intermolecular $\pi$-orbital overlap; normalized per channel in Figure 2e (scaling to macroscopic junction current via Equation \eqref{eq:compact_current_params}, corresponding to microscopic ensemble rates of $10^8$ to $2 \times 10^{13}\text{ s}^{-1}$ in Kundu and Goswami \cite{kundu2024}), conversion into the valence-symmetric (22) state expands orbital delocalization to scale $\gamma_m(n)$ nearly 50-fold from $\gamma_{m,31} = 1.0 \times 10^3\text{ s}^{-1}$ to $\gamma_{m,22} = 48 \times 10^3\text{ s}^{-1}$ at $n = 16,520$ pulses before symmetrically reversing over the 33,040-pulse cycle. During sub-threshold readout ($V_{\text{read}} \le 0.20$~V), the $130$~meV $\text{BF}_4^-$ barrier pins the counterions while the read bias directionally tilts carrier hopping ($K_{m,f} = 78 \times 10^3\text{ s}^{-1}$ vs. $K_{m,b} = 33 \times 10^3\text{ s}^{-1}$, $K_{m,f}/K_{m,b} \approx 2.36$; Figure 2e). Solving the 30-site stationary master equation ($\dot{P}_i = 0$, Note S4 in the Supporting Information) \cite{livshits2014long, lehmann2002incoherent, kundu2024} yields the sigmoidal site occupation probability profiles $P_i(n)$ in Figure 2d across pulse counts $n = 1,000$ to $16,520$ (saturating at $P_i \approx 0.98$), illustrating the layer-by-layer percolation of conductive (22) states across the 60~nm film between electrodes and governing the non-destructive stationary read current $I(n, V_r)$ across all 16,520 states.

\subsection{Analytical Potential Screening and Parameter Extraction}

Translating these microscopic mechanisms into a scalable circuit model requires replacing the infinite Thomas-Fermi Fourier sine series for $\Phi_{\text{mol}}(z,n)$ (Notes S4 and S5, Supporting Information), whose spatial progression reflects the finite radius ($r = 2.18$~\AA) and finite-volume ion packing limits of $\text{BF}_4^-$ counterions but is computationally prohibitive to evaluate at every circuit simulator timestep \cite{sharma2024, kundu2024}. We therefore formulated a closed-form analytical screening representation across the normalized film coordinate $t = z/L$ ($L = 30$) that preserves electrostatic conservation laws and decouples bulk Thomas-Fermi screening from contact boundary saturation for constant-time $O(1)$ circuit evaluation: a generalized sigmoidal function $s(t)$ models the field-driven potential drop into the bulk film and maps into an interface-constrained rational potential $u(t)$ enforcing normalized boundary conditions ($u(0) = 0$ and $u(1) = 1$) at the metal contacts:
\begin{equation}
s(t) = \left(1.0 - (1.0 - t)^q\right)^p
\end{equation}
\begin{equation}
\label{eq:rational_potential}
u(t) = \frac{(1.0 + k) s(t)}{1.0 + k s(t)^2}
\end{equation}
Here, shaping exponents $p$ and $q$ capture the non-linear interfacial screening factors $F_m$ governed by local $\text{BF}_4^-$ coordination, while parameter $k$ compresses the gradient near the counter electrode to physically capture finite-volume counterion packing saturation under strong bias without unphysical potential overshoot. To eliminate discrete 30-layer discretization steps in the film bulk while strictly preserving exact contact boundary conditions, a three-point spatial moving-average continuum kernel is blended via a boundary-tapered envelope:
\begin{equation}
u_{\text{ma}}(t) = 0.25 u(t - w) + 0.5 u(t) + 0.25 u(t + w)
\end{equation}
\begin{equation}
\Phi_{\text{mol}}(z,n) = \Phi_{\text{max}} \cdot \left[ (1.0 - a_{\text{eff}}) u(t) + a_{\text{eff}} u_{\text{ma}}(t) \right]
\end{equation}
\begin{figure*}[!t]
    \centering
    \includegraphics[width=0.92\textwidth]{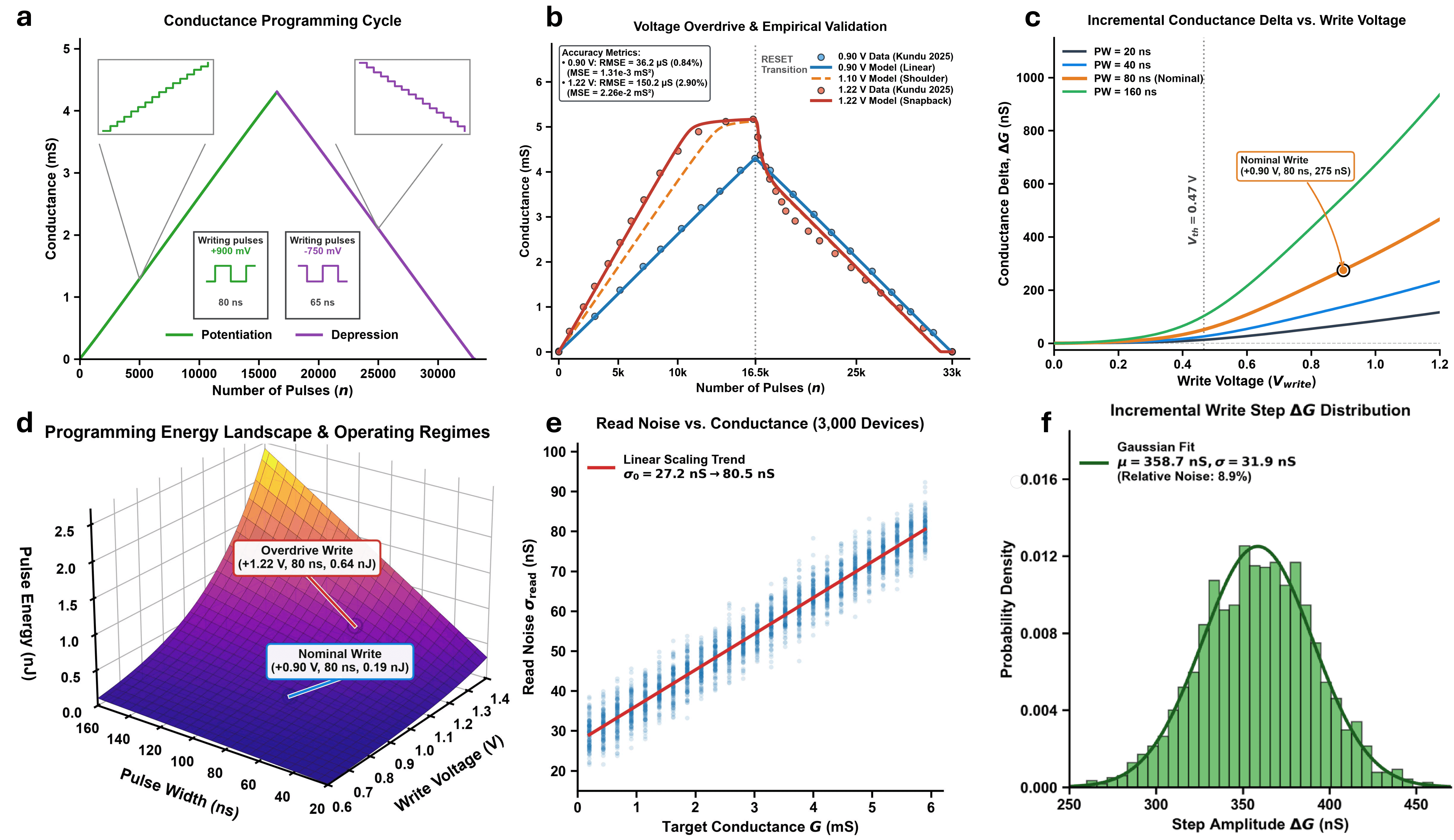}
    \caption{Continuous compact modeling, electrical characteristics, and hardware noise parameterization for baseline $[\text{Ru}^{\text{II}}\text{L}_2](\text{BF}_4)_2$ devices. \textbf{a}, Full 14-bit conductance programming cycle across 33,040 pulses ($16,520$ potentiation pulses at $+900$~mV, $80$~ns; $16,520$ depression pulses at $-750$~mV, $65$~ns) reaching $4.33$~mS under nominal calibration ($\Delta G_{\text{nom}} = 275$~nS/pulse), with insets highlighting discrete staircase write resolution. \textbf{b}, Measured vs. modeled voltage overdrive dynamics under $0.90$~V (RMSE $= 36.2~\mu$S, $0.84\%$ error), $1.10$~V, and $1.22$~V (peak $5.18$~mS, RMSE $= 150.2~\mu$S, $2.90\%$ error) pulses (experimental data from Kundu and Goswami \cite{kundu2024}). \textbf{c}, Incremental conductance update $\Delta G$ as a function of write voltage ($0.0$~V to $1.2$~V) across pulse widths ($20$~ns to $160$~ns), illustrating turn-on at $V_{\text{th}} = 0.47$~V. \textbf{d}, 3D programming energy landscape mapping per-pulse energy across pulse width and voltage ($0.19$~nJ at $+0.90$~V, $80$~ns; $0.64$~nJ at $+1.22$~V, $80$~ns). \textbf{e}, Conductance-dependent read noise $\sigma_{\text{read}}$ parameterized from 3,000 device evaluations \cite{sharma2024} at $V_{\text{read}} = 0.20$~V ($27.2$~nS to $80.5$~nS). \textbf{f}, Incremental write step amplitude $\Delta G$ distribution parameterized across 2,000 cycles \cite{sharma2024}, fit by a Gaussian ($\mu = 358.7$~nS, $\sigma = 31.9$~nS, $8.9\%$ relative write noise).}
    \label{fig:compact_modeling}
\end{figure*}%
where $w$ is the spatial kernel half-width, $\Phi_{\text{max}} = 0.70$~V is the peak internal potential drop under nominal $+0.90$~V $\text{BF}_4^-$ programming, and $a_{\text{eff}} = \alpha_{\text{taper}} \cdot a_L(t) \cdot a_R(t)$ incorporates smoothstep edge tapers $a_L(t) = (t/w)^2 (3.0 - 2.0 t/w)$ and $a_R(t) = ((1.0 - t)/w)^2 (3.0 - 2.0(1.0 - t)/w)$. Physically, these edge tapers extinguish spatial filtering ($a_{\text{eff}} \to 0$) at the contacts ($t = 0$ and $t = 1$), enforcing exact electrode boundary conditions ($u(0) = 0$ and $u(1) = 1$) while preserving a continuously differentiable interior potential profile. Across the 14-bit state space ($n = 0$ to $16,520$), the evolution from the unswitched profile ($p_i, q_i, k_i, \alpha_{\text{taper},i}, w_i$) to the switched profile ($p_f, q_f, k_f, \alpha_{\text{taper},f}, w_f$), coupled by boundary midpoint $z_{\text{mid}}$ and front width $z_{\text{width}}$, is parameterized across all 12 state variables via $C^1$-continuous monotonic piecewise cubic Hermite polynomials (PCHIP) for $O(1)$ lookup without Runge oscillations (analytical surrogate derivations and spline parameterization in Supplementary Note S5).

Similarly, because evaluating the Marcus-Hush-Chidsey integral over $\psi(E)$ ($E_\lambda = 0.04$~eV, $\Delta G^0 = 0.10$~eV; Notes S4 and S5, Supporting Information) requires numerical quadrature that bottlenecks circuit simulation, we represent the physical transfer rate with a closed-form analytical exponential decay:
\begin{equation}
\label{eq:marcus_surrogate}
k_{\text{ET}}(\Phi_{\text{mol}}) = A_{\text{ET}} \exp\left(-B_{\text{ET}} \Phi_{\text{mol}}\right)
\end{equation}
where baseline $\text{BF}_4^-$ pre-factor $A_{\text{ET}} = 1.135 \times 10^3\text{ s}^{-1}$ and effective potential sensitivity $B_{\text{ET}} = 7.12\text{ V}^{-1}$ within the active switching window around the $k_{\text{c}} = 0.30\text{ s}^{-1}$ threshold (corresponding to a full-range logarithmic slope $B_{\text{ET,full}} = 35.97\text{ V}^{-1}$ across the complete $\Phi_{\text{mol}} \in [0, \Phi_{\text{max}}]$ span; Note S5, Supporting Information) match the microscopic integral \cite{kundu2024}, with the negative sign reflecting that positive bias displaces counterions to lower $\Phi_{\text{mol}}$ and accelerate forward electron transfer. Bridging Equation \eqref{eq:marcus_surrogate} to Equation \eqref{eq:nucleation_domains}, redox conversion in layer $z$ becomes active when $k_{\text{ET}}(z,n) > 0.30\text{ s}^{-1}$ ($\theta(z,n) \in \{0, 1\}$), defining the switched front index $Z_{22}(n) = \sum_{z=1}^{30} \theta(z,n)$ ($\Delta z = 2$~nm). Linear regression across the 30-layer profile yields a linear relation between the cumulative write pulse count $n$ and switched layer index $Z_{22}$:
\begin{equation}
\label{eq:front_regression}
n(Z_{22}) = p_1 Z_{22} + p_2
\end{equation}
with planar front slope $p_1 = 534.2\text{ pulses/layer}$ ($267.1\text{ pulses/nm}$ across the 60~nm $[\text{Ru}^{\text{II}}\text{L}_2](\text{BF}_4)_2$ film) and nucleation offset $p_2 = 15.1\text{ pulses}$. Inverting Equation \eqref{eq:front_regression} gives $Z_{22}(n) = (n - p_2)/p_1$, advancing through $Z_{22} \approx 9.3$~layers ($18.7$~nm) at $n = 5,000$, $18.7$~layers ($37.4$~nm) at $n = 10,000$, and complete conversion near $16,520$ pulses across all 30 layers ($60$~nm; Figure 2c,d). This linear spatial progression matches the experimental observations of Sharma et al. \cite{sharma2024} and Kundu and Goswami \cite{kundu2024}, where each voltage pulse drives a uniform ensemble counterion displacement that advances the conductive (22) phase front in planar increments. Parameterizing the nucleation centers in Equation \eqref{eq:nucleation_domains} by $\chi(z) = a_1 z + a_2$ ($a_1 = p_1 = 534.2\text{ pulses/layer}$, $a_2 = p_2 = 15.1\text{ pulses}$) directly links the analytical Marcus rate in Equation \eqref{eq:marcus_surrogate} to the volumetric fraction $f_{22}(n)$ in Equation \eqref{eq:volume_fraction}, underpinning the symmetric 14-bit conductance trajectory in Figure 3a.

\subsection{Continuous Physics-Based Modeling of Crossbar Electrical Response}

Bridging the microscopic switched fraction $f_{22}(n)$ from Equation \eqref{eq:volume_fraction} to macroscale crossbar simulation requires a continuous dynamical compact model that sequentially couples dynamic state switching, boundary regularization, self-heating feedback, macroscopic carrier transport, and empirical hardware noise across the six baseline $[\text{Ru}^{\text{II}}\text{L}_2](\text{BF}_4)_2$ validation benchmarks in Figure 3: the 14-bit bidirectional staircase cycle (Figure 3a), measured voltage-overdrive saturation and snapback dynamics (Figure 3b), voltage and pulse-width $\Delta G$ turn-on kinetics (Figure 3c), the 3D per-pulse programming energy landscape (Figure 3d), conductance-dependent read noise scaling (Figure 3e), and the cycle-to-cycle write noise distribution (Figure 3f). Whereas the stationary hopping rate $\gamma_m(n)$ ($\sim 10^3$ to $4.8 \times 10^4\text{ s}^{-1}$ per normalized channel) governs steady-state carrier flux during sub-threshold readout, dynamic state switching during nanosecond write pulses is driven by field-accelerated $\text{BF}_4^-$ counterion drift over the $130$~meV barrier. To eliminate hard threshold derivative discontinuities that stall Newton-Raphson circuit solvers \cite{kim2019, biolek2009}, we first formulate the continuous switching rate governing the time derivative of the pulse state coordinate $dn/dt$ (in $\text{s}^{-1}$):
\begin{equation}
\label{eq:rate_equation}
\begin{split}
\frac{dn}{dt} = v_{\tanh} &\cdot \left[ \frac{V_{\text{smooth}} \cdot \ln\left(1.0 + \exp\left(\frac{|V_{\text{int}}| - V_{\text{th}}}{V_{\text{smooth}}}\right)\right)}{\left||V_{\text{ref}}| - V_{\text{th}}\right| + V_{\varepsilon}} \right]^{\alpha_{\text{eff}}} \\
&\cdot \kappa_{\text{clamp}}(n) \cdot k_{\text{thermal}}(T)
\end{split}
\end{equation}
where $n(t) \in [0, n_{\max}]$ ($N_{\text{pot}} = 16,520$, $n_{\max} = 33,040 \ge N_{\text{pot}}$) is the continuous cumulative write pulse state coordinate tracking the switched molecular ensemble across the junction ($N_{\text{mol}} \sim 10^6$ in $1.05~\mu\text{m} \times 1.1~\mu\text{m}$ crossbar cells and $\sim 10^4$ in $100\text{ nm} \times 100\text{ nm}$ nanocells), $v_{\tanh} = \tanh(V_{\text{int}} / V_{\text{smooth}})$ smoothly activates switching polarity across zero bias, and $V_{\varepsilon} = 10^{-6}$~V guards denominator normalization. Here, for the baseline $\text{BF}_4^-$ compact model, we distinguish the empirical pulse-update cutoff ($\approx 830$~mV reported in Sharma et al. \cite{sharma2024}, below which discrete nanosecond pulses produce no discernible conductance step) from our extracted mathematical activation threshold $V_{\text{th}} = 0.4664$~V ($\approx 0.47$~V), which pairs with smoothing parameter $V_{\text{smooth}} = 0.1038$~V to anchor softplus differentiability across turn-on (Figure 3c) and facilitate non-destructive readout ($dn/dt \approx 0$) at $V_{\text{read}} \le 0.20$~V (with the Verilog-A compact model additionally applying a conditional sub-threshold fast path setting $dn/dt = 0$ below $V_{\text{bypass}} = V_{\text{th}} - 3V_{\text{smooth}} = 0.155$~V when no overdrive discharge is pending). Reference potentials $V_{\text{ref,pos}} = +0.90$~V and $V_{\text{ref,neg}} = -0.75$~V (evaluated via $|V_{\text{ref}}| - V_{\text{th}} = 0.4336$~V for potentiation and $|-0.75\text{ V}| - V_{\text{th}} = 0.2836$~V for depression) serve as behavioral normalization anchors calibrated to the baseline $\text{BF}_4^-$ device (master physical constants, kinetic rate parameters, and testbench catalogs consolidated in Supplementary Note S2, Tables S1--S5) so that the over-threshold voltage bracket evaluates to unity ($1.0$) under nominal excitation, enabling straightforward re-anchoring for alternative counterion chemistries (multi-ion coordination scaling and barrier tuning in Supplementary Note S3). At $V_{\text{applied}} = +0.90$~V, the normalized bracket equals $1.0$ and $dn/dt$ equals the fundamental $\text{BF}_4^-$ transition velocity $\kappa_{\text{switch}} = 1.173 \times 10^7\text{ s}^{-1}$ ($\tau_{\text{switch}} = \kappa_{\text{switch}}^{-1} \approx 85.3$~ns), yielding a continuous state increment $\Delta n = (dn/dt)\Delta t \approx 0.94$ per $80$~ns nominal pulse (accumulating $\approx 15,590$ continuous state units across the $N_{\text{pot}} = 16,520$ potentiation pulses, with local nominal step $\Delta G_{\text{nom}} = 275$~nS/pulse in Figure 3c and full-cycle chord average $(4.33\text{ mS} - 0.20\text{ mS})/16,520 \approx 262.1$~nS/pulse in Figure 3a). Under depression ($v_{\tanh} < 0$), pairing $V_{\text{ref,neg}} = -0.75$~V ($65$~ns pulse) with kinetic symmetry ratio $r_{\text{asym}} = 0.986$ ($\kappa_{\text{eff}} = r_{\text{asym}} \kappa_{\text{switch}}$) yields a reverse state increment $\Delta n \approx -0.77$ per $65$~ns pulse, with the calibrated depression branch of the $f_{22}(n)$ lookup trajectory restoring a symmetric conductance return across the $N_{\text{cycle}} = 33,040$ pulse cycle (Figure 3a,b) \cite{sharma2024, kundu2024}, while expending only $0.19$~nJ per nominal potentiation pulse, $0.13$~nJ per depression pulse, and $0.64$~nJ at $+1.22$~V overdrive (Figure 3d). Baseline $\text{BF}_4^-$ exponent $\alpha = 0.932$ and overdrive coupling $\beta_\alpha = 0.125$ govern $\alpha_{\text{eff}} = \alpha + \beta_\alpha \tanh(\delta_v)$ with $\delta_v = (|V_{\text{int}}| - V_{\text{ref,pos}}) / (V_{\text{ref,pos}} - V_{\text{th}} + V_{\varepsilon})$, capturing the upward curvature of $\Delta G$ above $+0.90$~V in Figure 3c as strong fields overcome internal Thomas-Fermi screening.
\par
To prevent unbounded growth of $n(t)$ when displaced $\text{BF}_4^-$ counterions ($r = 2.18$~\AA) reach interfacial close-packing without step cutoffs causing Newton-Raphson Jacobian singularities \cite{kim2019, biolek2009, williams2024, jeong2016}, $\kappa_{\text{clamp}}(n) = \kappa_{\text{switch}} \cdot C_{\text{clamp}}(n)$ applies a $C^1$-continuous direction-aware Hermite smoothstep:
\begin{equation}
C_{\text{clamp}}(n) = \begin{cases} 1.0 & n_{\max} - n \ge w_{\text{sat}} \\ 0.0 & n_{\max} - n \le 0.0 \\ z_{\text{hi}}^2 (3.0 - 2.0 z_{\text{hi}}) & \text{with } z_{\text{hi}} = \frac{n_{\max} - n}{w_{\text{sat}}} \end{cases}
\label{eq:hermite_clamping}
\end{equation}
where $N_{\text{smooth}} = 17.06$ and the effective upper saturation width $w_{\text{sat}} = \max(50.0, 1.5 N_{\text{smooth,eff}})$ against the hard upper capacity bound $n_{\max} = 33,040$ govern the boundary transition over which space-charge screening decelerates (22)-state conversion under overpotentiation \cite{kundu2024}, and lower guard $w_{\text{zero}} = 30$ stabilizes return to the (31) ground state. Together with overdrive memory register $c_{\text{discharge}} = \max(c_{\text{discharge}}, (V_{\text{applied}} - V_{\text{ref,pos}})/(V_{\text{ref,pos}} - V_{\text{th}}))$ and clamped factor $\text{od}_{\text{clamp}} = \min(1.0, \max(0.0, c_{\text{discharge}} k_{\text{overdrive}}))$, Equation \eqref{eq:hermite_clamping} captures the four-stage physical overdrive lifecycle in Figure 3b (multi-regime electrical stressing schedules and transient snapback verification in Supplementary Note S6) \cite{sharma2024, kundu2024, gaur2024}: (1) \textit{overdrive saturation plateau}, where high fields ($V_{\text{applied}} > 0.90$~V) pack counterions against the electrode and space-charge screening forms an elevated shoulder reaching $5.18$~mS at $+1.22$~V; (2) \textit{rapid initial RESET snapback}, where nominal cycles ($\text{od}_{\text{clamp}} = 0$) exhibit return shelf latency $r_{\text{latency}} = 0.879$, whereas overdrive ($\text{od}_{\text{clamp}} \to 1$) collapses $r_{\text{lat,eff}} = r_{\text{latency}}(1.0 - \text{od}_{\text{clamp}}) \to 0$ so bias reversal immediately discharges interfacial counterion pileup ($k_{\text{discharge}} = 362.88\text{ s}^{-1}$); (3) \textit{the knee transition}, where relaxation toward shifted landing state $n_{\text{base}} = 0.5 n_{\max} [1.0 - \text{od}_{\text{clamp}}(1.0 - r_{\text{bottom\_blend}})]$ ($r_{\text{bottom\_blend}} = 0.714$) follows power-law excursion $u^{p_{\text{decay}}}$ ($u = (n - n_{\text{base}})/(n_{\max} - n_{\text{base}})$) governed by knee fraction $f_{\text{discharge}} = 0.375$ and exponent $p_{\text{decay}} = 1/f_{\text{discharge}} = 2.67$; and (4) \textit{baseline linear return}, where planar front migration resumes along the linear depression slope ($\text{RMSE} = 150.2~\mu\text{S}$, $2.90\%$ error relative to the $5.18$~mS peak at $+1.22$~V, evaluated across the $1.10$~V and $1.22$~V overdrive trajectories; Figure 3b).
\par
Concurrently, power dissipation $P_{\text{diss}} = |V_{\text{int}} I|$ drives localized self-heating governed by Newton cooling $dT/dt = (T_{\text{amb}} - T)/\tau_{\text{th}} + R_{\text{th}} P_{\text{diss}} / \tau_{\text{th}}$, where baseline $\text{BF}_4^-$ parameters $R_{\text{th}} = 52.14$~K/W and $\tau_{\text{th}} = 127.8$~ns reflect rapid heat sinking into the gold contacts ($\Delta T < 0.02$~K across low-to-intermediate conductance states in Figure 4a, and $< 0.058$~K at peak nominal conductance). Junction temperature $T(t)$ feeds back into $dn/dt$ in Equation \eqref{eq:rate_equation} via the Arrhenius factor $k_{\text{thermal}}(T) = \exp\left[\frac{E_a}{k_B}\left(\frac{1}{T_{\text{amb}}} - \frac{1}{T}\right)\right]$ ($T_{\text{amb}} = 298.15$~K), whose extracted macroscopic activation energy $E_a = 0.156$~eV ($155.9$~meV) independently aligns with the microscopic $130$~meV ($134$~meV experimental) $\text{BF}_4^-$ barrier in Figure 2b \cite{sharma2024, kundu2024}. Macroscopic carrier transport at pulse state $n(t)$ is evaluated via field-assisted hyperbolic sine hopping:
\begin{equation}
\label{eq:compact_current_params}
I(n, V_{\text{int}}) = I_{\text{prod}}(n) \sinh\left(\frac{V_{\text{int}}}{V_{\text{scale,eff}}(n)}\right)
\end{equation}
\begin{figure*}[!b]
    \centering
    \includegraphics[width=0.94\textwidth]{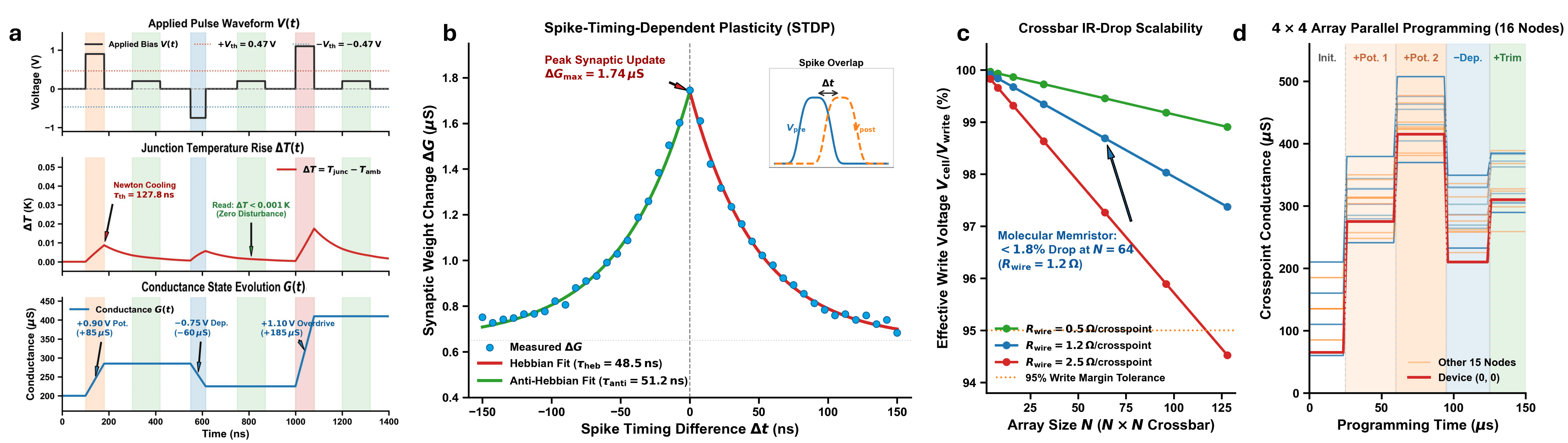}
    \caption{Dynamic switching kinetics, synaptic plasticity, and crossbar network scalability for baseline $[\text{Ru}^{\text{II}}\text{L}_2](\text{BF}_4)_2$ devices. \textbf{a}, Transient pulse response ($+0.90$~V potentiation, $-0.75$~V depression, $+1.10$~V overdrive) interleaved with 200~mV sub-threshold reads, illustrating conductance transitions ($+85~\mu\text{S}$, $-60~\mu\text{S}$, $+185~\mu\text{S}$), zero read disturbance, bounded self-heating ($\Delta T < 0.02$~K), and thermal relaxation ($\tau_{\text{th}} = 127.8$~ns). \textbf{b}, Spike-timing-dependent plasticity (STDP) illustrating synaptic weight change $\Delta G$ versus relative spike timing $\Delta t$ with Hebbian ($\tau_{\text{heb}} = 48.5$~ns) and anti-Hebbian ($\tau_{\text{anti}} = 51.2$~ns) fits (experimental data from Sharma et al. \cite{sharma2024} and Gaur et al. \cite{gaur2024}; inset illustrates pulse overlap). \textbf{c}, Interconnect IR-drop write margin degradation at diagonal crosspoint $(N,N)$ across array sizes $N = 4$ to $128$ for $R_{\text{wire}} = 0.5, 1.2, 2.5~\Omega/\text{crosspoint}$ ($1.3\%$ drop, $< 1.8\%$ bound, at $N = 64$ for $1.2~\Omega$). \textbf{d}, Synchronous 16-node parallel conductance tracking in a $4\times 4$ array over $150~\mu$s, highlighting device (0,0) (red curve).}
    \label{fig:circuit_simulation}
\end{figure*}%
where $I_{\text{prod}}(n) = I_{\text{scale}}(n) \cdot g_{\text{scale}} \cdot f_{22}(n)$ and $V_{\text{scale,eff}}(n) = V_{\text{scale}}(n) / (1.0 + \gamma_{\text{comp}} f_{22}(n))$, with baseline $\text{BF}_4^-$ barrier compression $\gamma_{\text{comp}} = -0.012$ and contact resistance $R_c = 8.03~\Omega$ setting $V_{\text{int}} = V_{\text{applied}} - 2 I R_c$ across $-0.1$~V to $+0.6$~V via local Newton-Raphson root-finding. In the sub-threshold limit ($V_{\text{int}} \ll V_{\text{scale,eff}}$, $\sinh(x) \approx x$), Equation \eqref{eq:compact_current_params} reduces to linear Ohm's law $I(n, V_{\text{int}}) \approx (I_{\text{prod}}(n) / V_{\text{scale,eff}}(n)) V_{\text{int}}$, recovering the stationary read current $I(n, V_r)$ (Note S4, Supporting Information). Through $I_{\text{scale}}(n)$ and baseline $\text{BF}_4^-$ technology scaling factor $g_{\text{scale}} = 6.23$, Equation \eqref{eq:compact_current_params} maps the normalized rates $\gamma_{m,31}$ and $\gamma_{m,22}$ from Equation \eqref{eq:gamma_coupling} across three progressive segments along the same continuous $\text{BF}_4^-$ potentiation sweep and its extended overdrive/saturation continuum: (i) the \textit{early-to-mid linear regime}, the initial non-saturating linear segment ($200$~nS to $2.05$~mS across the initial $8,000$ pulses of the potentiation sweep with chord slope $\approx 256$~nS/pulse at $80$~ns pulse width, alongside $\Delta G = 112$~nS/pulse for $20$~ns pulse widths in Figure 3c); (ii) the \textit{nominal full-range potentiation trajectory}, the complete calibrated $N_{\text{pot}} = 16,520$-pulse programming cycle ($200$~nS to $4.33$~mS under nominal $+0.90$~V, $80$~ns pulses, $\Delta G_{\text{nom}} = 275$~nS/pulse, $\text{RMSE} = 36.2~\mu\text{S}$, $0.84\%$ relative error across the $33,040$-pulse cycle; Figure 3a--c), extending into the elevated overdrive plateau reaching $5.18$~mS at $+1.22$~V overdrive ($\text{RMSE} = 150.2~\mu\text{S}$, $2.90\%$ error in Figure 3b); and (iii) the \textit{deep-saturation continuum}, the extended asymptotic continuation of the same physical sweep reaching $5.04$~mS at $25,000$ pulses in extended transient verification and approaching the physical $5.90$~mS close-packing capacity ceiling out to $52,000$ pulses \cite{sharma2024, kundu2024}.
\par
Collectively, Equations \eqref{eq:nucleation_domains}--\eqref{eq:compact_current_params} form a closed-loop physical cycle: planar nucleation in Equation \eqref{eq:nucleation_domains} updates $f_{22}(n)$ in Equation \eqref{eq:volume_fraction} to scale coupling in Equations \eqref{eq:gamma_coupling}--\eqref{eq:hopping_rate}, while write pulses drive $dn/dt$ in Equation \eqref{eq:rate_equation} with clamping in Equation \eqref{eq:hermite_clamping} to modulate current $I(n, V_{\text{int}})$ in Equation \eqref{eq:compact_current_params}, driving Joule dissipation $P_{\text{diss}} = |V_{\text{int}} I|$ and Arrhenius feedback $k_{\text{thermal}}(T)$. Finally, the model parameterizes three empirical $[\text{Ru}^{\text{II}}\text{L}_2](\text{BF}_4)_2$ hardware noise sources \cite{sharma2024, gaur2024} (statistical noise extraction and inverse-CDF quantile sampling detailed in Supplementary Note S9): (1) conductance-dependent read noise across 3,000 devices at $V_{\text{read}} = 0.20$~V scaling linearly from $\sigma_{\text{read},0} = 27.2$~nS to $80.5$~nS at $5.90$~mS (Figure 3e); (2) cycle-to-cycle write noise across 2,000 full-range hardware cycles spanning the $5.90$~mS saturation window ($5.90\text{ mS} / 16,520 \approx 357.1$~nS/pulse) in Figure 3f \cite{sharma2024}, where Gaussian mean $\mu_{\Delta G} = 358.7$~nS and spread $\sigma_{\text{write}} = 31.9$~nS ($8.9\%$ relative noise) define statistical write-noise distribution parameters rather than a separate operating mode; and (3) device-to-device spatial variation across 75,000 measurements ($\sigma = 107.2$~nS extracted across 3,000 devices in Sharma et al. \cite{sharma2024}, alongside the $73.6$~nS spread across 4,096 crosspoints in Gaur et al. \cite{gaur2024}, sampled via an inverse-CDF table) for Sections 3 and 4.

\section{Circuit Dynamics and Crossbar Scalability}

When integrated into dynamic circuits and scalable crossbars, the baseline $[\text{Ru}^{\text{II}}\text{L}_2](\text{BF}_4)_2$ molecular memristor exhibits transient switching, thermal self-regulation, synaptic plasticity, and interconnect-dependent response characteristics governed by the continuous transport formulations in Equations \eqref{eq:rate_equation}--\eqref{eq:compact_current_params} (Figure 4). Because softplus thresholding, polarity-weighted $\tanh(V_{\text{int}} / V_{\text{smooth}})$ zero-bias regularization, and Hermite clamping preserve continuous $C^1$ differentiability above the Verilog-A sub-threshold fast-path cutoff ($V_{\text{bypass}} = 0.155$~V) inside the Newton-Raphson Jacobian loop without artificial step cutoffs \cite{kim2019, biolek2009}, these physical behaviors translate directly into numerically stable circuit dynamics across isolated junctions and multi-node crossbars, maintaining $< 0.02\%$ testbench-by-testbench transient discrepancy between our vectorized Python solver and Spectre/Virtuoso Verilog-A model (Table \ref{tab:dual_engine_comparison}; Note S7, Supporting Information).

Under high-speed write and read sequences, integrating the continuous rate equation $dn/dt$ in Equation \eqref{eq:rate_equation} alongside hyperbolic sine transport in Equation \eqref{eq:compact_current_params} yields deterministic state transitions coupled to localized thermal dissipation for the baseline $\text{BF}_4^-$ junction (Figure 4a; single-device pulse dynamics, staircase microstepping, and multi-phase array reprogramming are detailed in Note S6, Supporting Information). Stimulated by sequential macro-step write bursts ($\sim 300$-pulse burst envelopes at $+0.90$~V potentiation, $-0.75$~V depression, and $+1.10$~V overdrive, corresponding to the cumulative effect of $275$~nS single-pulse microsteps in Figure 3a,c), the device conductance advances predictably from its $200~\mu$S baseline: stepping by $+85~\mu$S to $285~\mu$S under nominal potentiation, decreasing by $-60~\mu$S to $225~\mu$S upon depression, and jumping by $+185~\mu$S to $410~\mu$S under overdrive as $\alpha_{\text{eff}}$ accelerates counterion drift. During intervening interrogation windows, 200~mV read pulses remain well below the softplus threshold ($V_{\text{th}} = 0.47$~V, $dn/dt \approx 0$), leaving programmed conductances undisturbed ($\Delta G = 0$) across millions of reads with negligible heating ($\Delta T < 0.001$~K). Concurrently, active switching power $P_{\text{diss}} = |V_{\text{int}} I|$ drives transient Joule self-heating governed by Newton cooling ($\tau_{\text{th}} = 127.8$~ns, $R_{\text{th}} = 52.14$~K/W), raising junction temperature by only $\approx 0.009$~K during potentiation, $\approx 0.006$~K during depression, and $\approx 0.018$~K under overdrive ($\Delta T < 0.02$~K). Rapid heat sinking into the gold contacts relaxes $T(t)$ between pulses, keeping the Arrhenius feedback factor $k_{\text{thermal}}(T)$ ($E_a = 0.156$~eV) near unity and preventing thermal runaway at megahertz rates.

When subjected to overlapping analog waveforms rather than rectangular pulses, the baseline $[\text{Ru}^{\text{II}}\text{L}_2](\text{BF}_4)_2$ junction naturally exhibits local synaptic adaptation through spike-timing-dependent plasticity (STDP, Figure 4b). In spiking neuromorphic circuits, STDP serves as an essential mechanism for local, unsupervised learning, where the polarity and magnitude of synaptic weight modification $\Delta G$ are governed by the relative arrival time $\Delta t = t_{\text{post}} - t_{\text{pre}}$ between pre-synaptic and post-synaptic action potentials \cite{wang2017}. When continuous action potentials arrive across the two terminals (Figure 4b, inset), their temporal superposition $V_{\text{int}}(t) = V_{\text{pre}}(t) - V_{\text{post}}(t)$ continuously tilts the internal potential $\Phi_{\text{mol}}(z,n)$, exponentially modulating the analytical Marcus-Hush-Chidsey transfer rate $k_{\text{ET}}(\Phi_{\text{mol}})$ in Equation \eqref{eq:marcus_surrogate} and integrating $dn/dt$ in Equation \eqref{eq:rate_equation} across a $-150$~ns to $+150$~ns coincidence window. Experimental $\text{BF}_4^-$ measurements (cyan markers, reported by Sharma et al. \cite{sharma2024} and Gaur et al. \cite{gaur2024} using $80$~ns spikes with $50$~ns exponential tails) demonstrate quantitative agreement with our continuous kinetic model, producing a symmetric temporal coincidence adaptation profile: causal spike ordering ($\Delta t > 0$) drives Hebbian potentiation with exponential decay constant $\tau_{\text{heb}} = 48.5$~ns, whereas acausal timing ($\Delta t < 0$) drives anti-Hebbian depression with $\tau_{\text{anti}} = 51.2$~ns, peaking at $\Delta G_{\max} = 1.74~\mu$S above a $0.65~\mu$S baseline floor when spikes coincide ($\Delta t \to 0$). This native analog interaction illustrates how continuous redox kinetics (\(\tau_{\text{switch}} \approx 85.3\)~ns) translate voltage overlaps into biorealistic synaptic plasticity, executing adaptation nearly six orders of magnitude faster than biological synapses ($10$~ms to $50$~ms) without CMOS pulse-shaping circuitry.
\par
In scaled $[\text{Ru}^{\text{II}}\text{L}_2](\text{BF}_4)_2$ crossbar networks, interconnect wire resistance ($R_{\text{wire}}$) introduces distributed ohmic line loading that progressively attenuates the effective write bias delivered to interior nodes, reaching maximum degradation at the opposite diagonal crosspoint $(N,N)$ (Figure 4c). Across representative wire resistances of $0.5$, $1.2$, and $2.5~\Omega/\text{crosspoint}$ in arrays up to $N = 128$, margin degradation is tempered by the $\text{BF}_4^-$ device's operational conductance window: because molecular switching draws moderate programming currents ($I_{\text{write}} \approx 0.35$~mA), parasitic line loading remains substantially lower than in low-resistance filamentary arrays. Under nominal metallization ($1.2~\Omega/\text{crosspoint}$), worst-case voltage attenuation at $N = 64$ is $1.3\%$ (strictly within the $< 1.8\%$ design bound, $V_{\text{cell}} / V_{\text{write}} = 98.7\%$), preserving write margins well above the 95\% tolerance boundary. In a $128\times 128$ crossbar, the available write margin remains above 97\% for nominal wiring and above 99\% for low-resistance metallization ($0.5~\Omega/\text{crosspoint}$), crossing below the 95\% threshold only when wire resistance reaches $2.5~\Omega/\text{crosspoint}$ in arrays exceeding $100\times 100$. This benign loading characteristic scales passive crossbars across hundreds of nodes without active write-boosting drivers or line-drop compensation.
\par
\begin{figure*}[!b]
    \centering
    \includegraphics[width=0.76\textwidth]{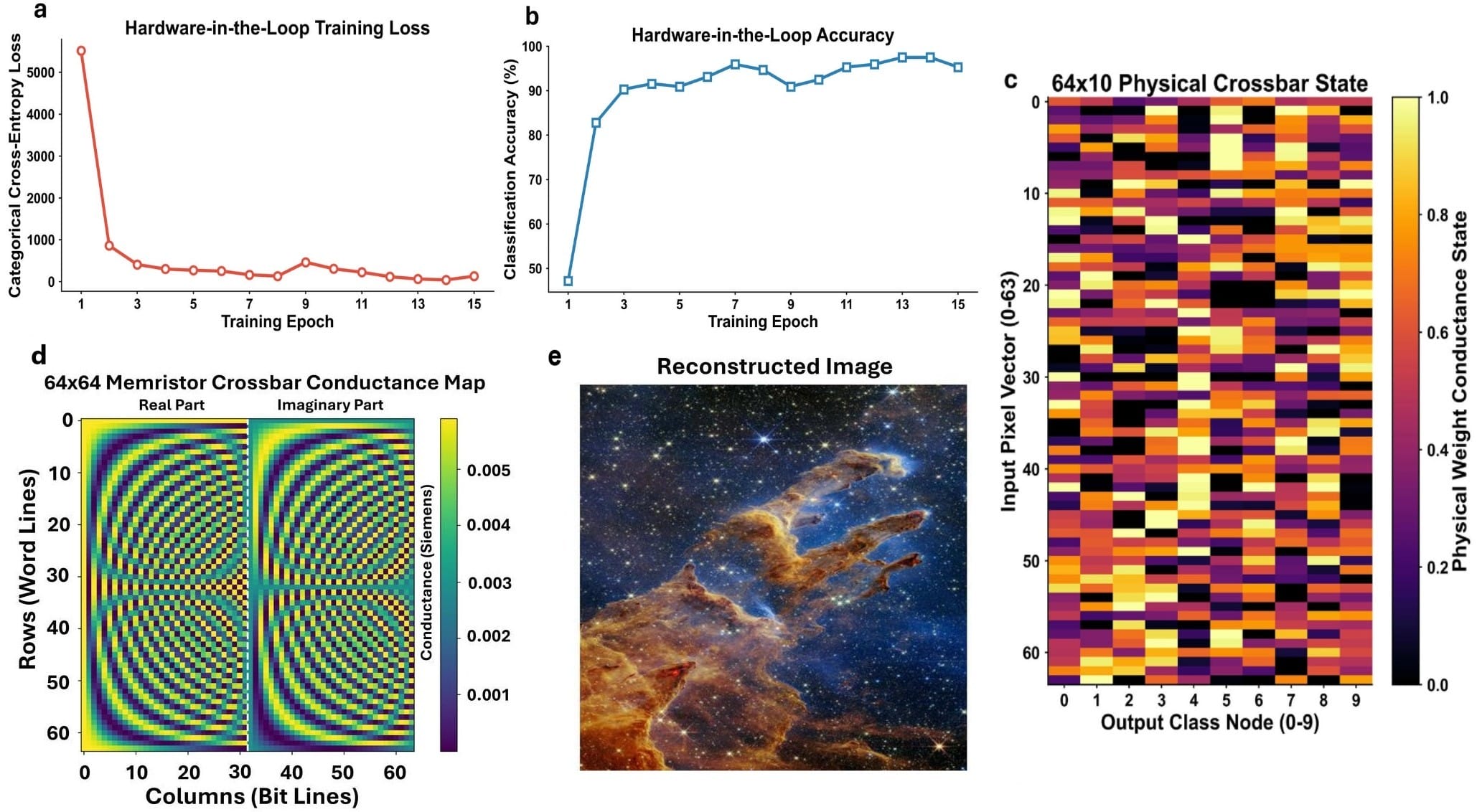}
    \caption{Hardware-in-the-loop neural network training and high-dynamic-range spatial image processing on $[\text{Ru}^{\text{II}}\text{L}_2](\text{BF}_4)_2$ molecular crossbars. \textbf{a}, Categorical cross-entropy training loss progression across 15 epochs on downsampled $8\times 8$ MNIST digits. \textbf{b}, Validation classification accuracy trajectory rising to a peak of $97.50\%$ (epoch 13) and stabilizing at $95.31\%$. \textbf{c}, Final programmed physical conductance state of the $64\times 10$ crossbar layer after training convergence. \textbf{d}, Programmed $64\times 64$ physical memristor crossbar conductance map for the 64-point inverse discrete Fourier transform (IDFT) basis partitioned into real (columns 0--31) and imaginary (columns 32--63) bitline components. \textbf{e}, Reconstructed full-color analog image ($1000\times 1000$ Pillars of Creation; source image credit: NASA, ESA, CSA, STScI) following physical crossbar dot-product matrix multiplication, achieving $99.30\%$ structural similarity (SSIM) and $43.0$~dB SNR. (The uncompressed digital reference image and complete spatial color-plane and phase-plane decompositions are provided in Figure S6 of the Supporting Information.)}
    \label{fig:system_applications}
\end{figure*}
When integrated into parallel multi-device circuits, the individual $[\text{Ru}^{\text{II}}\text{L}_2](\text{BF}_4)_2$ memristors interact collectively while maintaining distinct, non-interfering state trajectories (Figure 4d). In a representative $4\times 4$ array operated over a 150~$\mu$s schedule across initialization (`Init.', $0$ to $25~\mu$s), two potentiation stages (`+Pot. 1', $25$ to $60~\mu$s; `+Pot. 2', $60$ to $95~\mu$s), depression (`$-$Dep.', $95$ to $125~\mu$s), and fine-tuning (`+Trim', $125$ to $150~\mu$s), all 16 crosspoints modulate their conductances concurrently across the $60~\mu$S to $520~\mu$S operational range. Device (0,0) (red curve), for example, steps through $65~\mu$S, $275~\mu$S, $415~\mu$S, $210~\mu$S, and trims to $310~\mu$S, reflecting local pulse stimuli without distortion from sneak paths or parasitic current feedback from neighboring nodes. Because the underlying continuous rate formulation eliminates Jacobian singularities across concurrent multi-node transitions, these collective dynamics exhibit smooth, non-oscillatory convergence (computational solver architecture, memory profiling, and MNA acceleration detailed in Supplementary Note S8), providing the electrical stability that underpins the neural network and spatial image processing benchmarks in Section 4.
\par
\section{Applications}

Deploying the calibrated $[\text{Ru}^{\text{II}}\text{L}_2](\text{BF}_4)_2$ compact model across system-level machine learning and scientific computing benchmarks validates the scalability and predictive utility of our dual-engine framework. To facilitate seamless adoption across electronic design automation (EDA) and academic research workflows, the unified simulation platform is architected for cross-platform portability across operating systems, processor architectures, and software ecosystems (Table \ref{tab:platform_compatibility}; Note S8, Supporting Information). While the behavioral Verilog-A compact model integrates directly into Cadence Spectre/Virtuoso workflows on enterprise Linux computing clusters for tapeout-grade CMOS co-design, the vectorized Python and TensorFlow modules execute with zero proprietary licensing prerequisites across Linux, Windows, and macOS environments, enabling interactive exploration on consumer laptops, desktop workstations, and distributed GPU acceleration nodes (supporting both NVIDIA CUDA and AMD ROCm) alike.

\begin{table}[!t]
\centering
\caption{Cross-Platform Simulation and Computing Compatibility}
\label{tab:platform_compatibility}
\renewcommand{\arraystretch}{1.02}
\resizebox{\columnwidth}{!}{%
\begin{tabular}{ll}
\toprule
\textbf{Platform Dimension} & \textbf{Supported Environments and Toolchains} \\
\midrule
Operating Systems & Linux (RHEL, Ubuntu), Windows (10/11), macOS \\
Compute Architectures & x86\_64 (Intel, AMD), ARM64 (Apple Silicon), GPUs (NVIDIA CUDA, AMD ROCm) \\
Circuit EDA Tools & Cadence Virtuoso (IC6.1.8+ / ICADVM20.1+), Spectre / APS (19.x--23.x) \\
Compact Model HDL & Verilog-A (Accellera Verilog-AMS 2.4, $C^1$-continuous) \\
Algorithmic Stack & Python (3.9--3.12), NumPy, SciPy (Monotonic PCHIP) \\
Machine Learning Stack & TensorFlow (2.x) with Custom Gradient STE \\
Deployment Model & Publicly available repository, license-free Python runtime \\
\bottomrule
\end{tabular}%
}
\end{table}

\subsection{Artificial Intelligence Training} 

Using the calibrated baseline $[\text{Ru}^{\text{II}}\text{L}_2](\text{BF}_4)_2$ compact model, algorithmic validation was performed via a custom hardware-in-the-loop dense crossbar layer integrated natively within TensorFlow (version 2.x) to classify the Modified National Institute of Standards and Technology (MNIST) handwritten digit database ($60,000$ training and $10,000$ testing $28\times 28$ grayscale images across 10 classes, 0 through 9) \cite{lecun1998, li2018, hu2018} (differential synaptic crossbar mapping, custom straight-through estimator backpropagation, and per-epoch energy accounting are detailed in Note S10, Supporting Information). To match the physical $64\times 10$ crossbar array without auxiliary hidden layers, each $28\times 28$ image is downsampled via spatial average pooling into an $8\times 8$ feature grid, flattening into a 64-element analog vector driving the 64 word lines. During the forward pass, the intrinsic linearity and bidirectional symmetry ($r_{\text{asym}} = 0.986$) of planar front propagation ($p_1 = 534.2\text{ pulses/layer}$ in Equation \eqref{eq:front_regression}) map bipolar weights onto physical target conductances via fast open-loop pulse programming ($+0.90$~V, $80$~ns potentiation; $-0.75$~V, $65$~ns depression), eliminating the costly ADC/DAC write-and-verify cycles required by filamentary RRAM and PCM \cite{legallo2018mixed}. Pooled activations are scaled into sub-threshold read voltages ($V_{\text{read}} \le 0.20$~V, guarded by $V_{\text{th}} = 0.47$~V) to execute non-destructive vector-matrix multiplication via Equation \eqref{eq:compact_current_params}, while a custom \texttt{@tf.custom\_gradient} straight-through estimator (STE) routes analytical gradients to the Adam optimizer during backpropagation \cite{bengio2013, li2018}.

Crucially, hardware-in-the-loop training incorporates all three empirical $\text{BF}_4^-$ hardware noise channels parameterized in Section 2: conductance-dependent read noise ($\sigma_{\text{read}} = 27.2$~nS to $80.5$~nS, Figure 3e), cycle-to-cycle write step variation ($\sigma_{\text{write}} = 31.9$~nS, $8.9\%$ relative noise, Figure 3f), and device-to-device spatial mismatch ($\sigma = 107.2$~nS). Whereas conventional RRAM and PCM accelerators frequently suffer from weight divergence below 90\% accuracy due to update asymmetry and coarse bit quantization \cite{chen2018neurosim}, our 14-bit symmetric molecular architecture absorbed these hardware fluctuations without divergence. Over 15 training epochs, categorical cross-entropy loss dropped from $5516.39$ to $132.15$ (Figure 5a), with validation accuracy peaking at $97.50\%$ (epoch 13) and stabilizing at $95.31\%$ (epoch 15, Figure 5b), closely tracking the 32-bit floating-point baseline ($\approx 96.1\%$). The final programmed $64\times 10$ conductance matrix (Figure 5c) and 15-epoch workflow consume only $0.0028$~J ($0.0020$~J for weight updates and $0.0008$~J for inference reads; complete per-epoch loss trajectories and energy accounting in Supplementary Note S10).

\subsection{James Webb Image Processing}

Evaluating the baseline $[\text{Ru}^{\text{II}}\text{L}_2](\text{BF}_4)_2$ framework on high-dynamic-range scientific image processing expands verification beyond classification \cite{wang2017, kumar2017, kumar2020, canli2018}. Operating the crossbar as a physical Fourier dot-product engine reduces $O(n^2)$ matrix-vector multiplication complexity to $O(1)$ single-step Kirchhoff current summations \cite{hu2018, canli2018}. The 1D 64-point inverse discrete Fourier transform (IDFT) basis matrix $W_{p,q} = \exp(j 2\pi p q / 64)$ ($p, q \in \{0, \dots, 63\}$) is mapped onto a $64\times 64$ 1T1R molecular crossbar and programmed via two-pass closed-loop 14-bit tuning (\texttt{fineTune=True}, \texttt{weightBits=14}, \texttt{inputBits=14}, \texttt{outputBits=14}; Figure 5d). Because passive molecular junctions maintain strictly non-negative conductances, the architecture drives the 64 word lines with frequency indices $p \in \{0, \dots, 63\}$ and partitions the bitline columns into real ($\text{Re}(W_{:,0:32})$, columns 0--31) and imaginary ($\text{Im}(W_{:,0:32})$, columns 32--63) sub-arrays to evaluate spatial output indices $q \in \{0, \dots, 31\}$ directly in hardware, while conjugate indices $q \in \{33, \dots, 63\}$ are recovered via Hermitian symmetry and the Nyquist index $q = 32$ is evaluated digitally. The signed basis entries $W \in [-1, +1]$ are shifted via $W_{\text{target}} = (W + 1)/2 \in [0, 1]$ across the $200$~nS to $5.90$~mS $\text{BF}_4^-$ conductance window around the common-mode midpoint $G_{\text{mid}} = 2.95$~mS, with the midpoint offset removed digitally after readout ($Y = 2 Y_{\text{hw}} - \sum_p V_p$ in normalized units; Note S11, Supporting Information). The $1000\times 1000$ James Webb Space Telescope Pillars of Creation image (source image credit: NASA, ESA, CSA, STScI) is zero-padded to $1024\times 1024$ and partitioned into $16\times 16 = 256$ blocks of $64\times 64$ pixels per RGB color channel ($768$ blocks total; global RGB color-plane and 2D phase-spectrum visualizations are provided in Figure S6, Supporting Information).

Within each $64\times 64$ block, each row undergoes a 1D 64-point fast Fourier transform, and the resulting complex frequency vectors are split into four non-negative unipolar rails ($\text{Re}_+$, $\text{Re}_-$, $\text{Im}_+$, and $\text{Im}_-$), each normalized per vector by its peak magnitude and applied to the 64 word lines via 14-bit pulse-width-modulated readout ($V_{\text{read}} \le 0.20$~V) across four batched 64-vector evaluations per block. Because the $64\times 64$ IDFT basis is programmed once and reused across all $768$ blocks without reprogramming, the vectorized Python solver executes the complete three-channel reconstruction in $3,072$ batched crossbar evaluations ($196,608$ vector-matrix multiplications; complete 768-block batched evaluation schedule and SSIM/SNR formulations detailed in Supplementary Note S11). Following complex magnitude assembly and global min-max dynamic range rescaling cropped to the original $1000\times 1000$ dimensions, the reconstructed analog image (Figure 5e) achieves $99.30\%$ structural similarity (SSIM) and $43.0$~dB signal-to-noise ratio ($\text{SNR} = 10\log_{10}(\langle I_{\text{orig}}^2 \rangle / \text{MSE})$ relative to the original $1000\times 1000$ image; Figure S6, Supporting Information), expending only $0.0246$~J ($0.0004$~J for initial array programming and $0.0242$~J for inference). Because this benchmark evaluates deterministic array fidelity with stochastic variability channels and interconnect resistance disabled, the $43.0$~dB full-array SNR relative to isolated single-device demonstrations ($74.0$~dB SNR in Sharma et al. \cite{sharma2024}) reflects finite 14-bit input, weight, and output quantization, residual two-pass programming tolerance across the $4,096$ crosspoints, and nonlinear hyperbolic sine ($\sinh$) transport across the readout window, whereas enabling the empirical device-to-device and write dispersion channels of Note S9 would introduce additional array-level SNR degradation.

\section{Conclusion}

\begin{table*}[!b]
\centering
\caption{Comparative Benchmarking of Device Physics and Neuromorphic Simulation Platforms}
\label{tab:hardware_comparison}
\renewcommand{\arraystretch}{0.97}
\resizebox{\textwidth}{!}{
\begin{tabular}{lcccc}
\toprule
\multicolumn{5}{l}{\textbf{a, Non-Volatile Memory Device Physics and Switching Characteristics}} \\
\midrule
\textbf{Technology Type} & \textbf{Switching Mechanism} & \textbf{Conductance States} & \textbf{I-V Linearity} & \textbf{Update Symmetry} \\
\midrule
Phase-Change Memory (PCM) \cite{burr2017neuromorphic} & Melt-Quench / Crystallization & $< 100$ & Highly Nonlinear & Asymmetric (Abrupt RESET) \\
Filamentary RRAM \cite{xia2019, wang2020} & Stochastic Ion Migration & $< 64$ ($< 6$-bit) & Nonlinear & Asymmetric \\
Molecular Memristor ($[\text{Ru}^{\text{II}}\text{L}_2](\text{BF}_4)_2$, Modeled Here) \cite{sharma2024, kundu2024} & Kinetic Counterion-Mediated Redox & 16,520 (14-bit) & Linear & Symmetric \\
\midrule[\heavyrulewidth]
\multicolumn{5}{l}{\textbf{b, Neuromorphic Modeling Platforms and Simulation Scope}} \\
\midrule
\textbf{Framework} & \textbf{Primary Focus} & \textbf{Deep Learning Integration} & \textbf{Native Circuit Verification} & \textbf{Transient Physics Modeling} \\
\midrule
NeuroSim \cite{chen2018neurosim} & Circuit-Level Macro Modeling & Yes (Trace-Based) & No (Macro Estimations) & No (Abstracted) \\
MemTorch \cite{lammie2020memtorch} & ML Algorithm Mapping & Yes (PyTorch-Native) & No & Abstracted / Stochastic \\
Unified Molecular Model (This Work) & Device-to-Algorithm Bridging & Yes (TensorFlow via STE) & Yes (Spectre-Native) & Yes (Continuous Kinetic Integral) \\
\bottomrule
\end{tabular}
}
\end{table*}

In this work, we established a continuous kinetic modeling framework for counterion-mediated molecular memristors across device, circuit, and algorithmic scales, validated against published experimental characterization datasets \cite{sharma2024, kundu2024, gaur2024}. By casting discrete 30-layer $(31) \leftrightarrow (22)$ ligand redox disproportionation, Thomas-Fermi electrostatic screening series, Marcus-Hush-Chidsey electron transfer integrals, and Cauchy-Lorentz planar domain nucleation into $C^1$-continuous analytical surrogates with softplus thresholding ($V_{\text{th}} = 0.4664$~V, $V_{\text{smooth}} = 0.1038$~V) and direction-aware cubic Hermite clamping, our formulation eliminates the Jacobian singularities that stall Newton-Raphson solvers in SPICE and Cadence Spectre. Across five circuit verification testbenches, our Verilog-A compact model and vectorized Python solver maintain $< 0.02\%$ transient waveform discrepancy (Note S7, Supporting Information), reproducing nominal 14-bit switching ($\text{RMSE} = 36.2~\mu\text{S}$, $0.84\%$ error across $33,040$ pulses), overdrive saturation and RESET snapback ($\text{RMSE} = 150.2~\mu\text{S}$, $2.90\%$ error), sub-threshold read immunity ($V_{\text{read}} \le 0.20$~V), bounded Newton self-heating ($\Delta T < 0.02$~K), nanosecond STDP ($\tau_{\text{heb}} = 48.5$~ns, $\tau_{\text{anti}} = 51.2$~ns), and $> 98\%$ write margins across $64\times 64$ passive crossbars under asymmetric sneak-path loading.

At the system scale, embedding empirical hardware noise distributions extracted from $75,000$ measurements across $3,000$ devices and $2,000$ write cycles (Note S9, Supporting Information) confirms that true 14-bit precision persists under wafer-scale non-idealities. Incorporating conductance-dependent read noise ($27.2$~nS to $80.5$~nS), cycle-to-cycle write dispersion ($\mu_{\Delta G} = 358.7$~nS, $\sigma_{\text{write}} = 31.9$~nS, $8.9\%$ relative noise), and inverse-CDF spatial mismatch ($\sigma = 107.2$~nS), our hardware-in-the-loop MNIST classifier achieved $95.31\%$ validation accuracy ($97.50\%$ peak) at $0.0028$~J. Furthermore, the $64\times 64$ Fourier dot-product engine executed full-resolution processing of $1000\times 1000$ RGB James Webb Space Telescope imagery across $3,072$ batched crossbar evaluations ($768$ padded $64\times 64$ spatial blocks across three color channels, comprising $196,608$ vector-matrix multiplications) with $99.30\%$ SSIM and $43.0$~dB SNR at $0.0246$~J total active energy. As summarized in Table \ref{tab:hardware_comparison}a, whereas PCM and filamentary RRAM remain limited by melt-quench asymmetry and filament chatter ($< 100$ states), the $[\text{Ru}^{\text{II}}\text{L}_2](\text{BF}_4)_2$ junction harnesses deterministic counterion-stabilized redox conversion across $16,520$ linear, symmetric states to support open-loop weight updates without digital error-correction co-processors. Moreover, because counterion transport barriers scale predictably across chemical species ($\text{PF}_6^-$ at $95$~meV, $\text{BF}_4^-$ at $130$~meV, and $\text{BPh}_4^-$ at $210$~meV), this architecture provides a chemically tunable retention-versus-speed trade-off spanning cryogenic sensing to high-retention edge storage.

Finally, translating these nanoscale advantages into CMOS+X accelerators requires simulation tools uniting transistor-level accuracy with array throughput. As shown in Table \ref{tab:hardware_comparison}b, prior platforms remain split between circuit macro-estimators such as NeuroSim \cite{chen2018neurosim} that abstract away transient state kinetics, and deep learning wrappers such as MemTorch \cite{lammie2020memtorch} that omit Kirchhoff circuit verification. Pairing our Accellera Verilog-AMS 2.4-compliant Spectre Verilog-A compact model for CMOS co-design with a vectorized Python engine combining 12-parameter monotonic PCHIP splines, $61.8\%$ event-driven point skipping, Level-3 BLAS column-partitioned nodal solvers ($> 50,158\text{ steps/s}$ on CPU), and batched GPU acceleration supporting NVIDIA CUDA and AMD ROCm ($64,500\text{ steps/s}$ at $64\times 64$ with linear $O(N_{\text{devices}})$ scaling; Note S8, Supporting Information) provides an open-source, physically verified foundation to design, simulate, and tape out hybrid CMOS-molecular neuromorphic computing systems.

\vspace{0.55em}
\noindent{\normalsize\bfseries Conflict of Interest}\par\vspace{0.15em}\noindent
The authors declare no conflict of interest.\par

\vspace{0.55em}
\noindent{\normalsize\bfseries Acknowledgements}\par\vspace{0.15em}\noindent
The authors acknowledge valuable discussions with members of the INQUIRE Lab and express sincere gratitude to Sreetosh Goswami and the Centre for Nano Science and Engineering (CeNSE) at the Indian Institute of Science (IISc), Bangalore, for insightful discussions on molecular memristor device physics and experimental mechanisms.\par

\vspace{0.55em}
\noindent{\normalsize\bfseries Funding}\par\vspace{0.15em}\noindent
This work was supported in part by the U.S. National Science Foundation under Award No. 2432082. Any opinions, findings, and conclusions or recommendations are those of the authors and do not necessarily reflect the views of the National Science Foundation.\par

\vspace{0.55em}
\noindent{\normalsize\bfseries Author Contributions}\par\vspace{0.15em}\noindent
L.L. developed the dual simulation framework including the Python and Spectre engines and derived the continuous compact model. D.B. implemented the James Webb Space Telescope image processing. N.S.A. developed the TensorFlow neural network integration. S.Y. provided algorithmic and physical validation of the simulated crossbars. T.V. provided expert insight into the physical transport mechanisms. M.B. supervised the project and reviewed and edited the manuscript.\par

\vspace{0.55em}
\noindent{\normalsize\bfseries Code and Data Availability}\par\vspace{0.15em}\noindent
The simulation framework, compact models, parameter extraction pipelines, and validation datasets are available on GitHub at \texttt{https://github.com/INQUIRELAB/\allowbreak MolecularMemristor\allowbreak SpectreandPythonModel}.\par

\clearpage
\newgeometry{margin=0.75in}
\onecolumn
\pagestyle{plain}
\let\texttt\origtexttt
\setcounter{page}{1}
\setcounter{section}{0}
\setcounter{subsection}{0}
\setcounter{equation}{0}
\setcounter{figure}{0}
\setcounter{table}{0}
\renewcommand{\theequation}{S\arabic{equation}}
\renewcommand{\thefigure}{S\arabic{figure}}
\renewcommand{\thetable}{S\arabic{table}}
\renewcommand{\thesection}{Note S\arabic{section}}
\renewcommand{\thesubsection}{S\arabic{section}.\arabic{subsection}}
\makeatletter
\renewcommand{\section}{\@startsection{section}{1}{0pt}{1.4ex plus 0.4ex minus 0.2ex}{0.7ex plus 0.2ex}{\normalfont\large\bfseries}}
\renewcommand{\subsection}{\@startsection{subsection}{2}{0pt}{1.0ex plus 0.3ex minus 0.1ex}{0.4ex plus 0.1ex}{\normalfont\normalsize\bfseries}}
\renewcommand{\subsubsection}{\@startsection{subsubsection}{3}{0pt}{0.8ex plus 0.2ex minus 0.1ex}{0.3ex plus 0.1ex}{\normalfont\normalsize\bfseries}}
\makeatother
\captionsetup{font=small,labelfont=bf,labelsep=colon,skip=4pt}
\setlength{\abovedisplayskip}{6pt plus 2pt minus 2pt}
\setlength{\belowdisplayskip}{6pt plus 2pt minus 2pt}
\setlength{\abovedisplayshortskip}{3pt plus 1pt}
\setlength{\belowdisplayshortskip}{3pt plus 1pt}
\justifying
\setlength{\parindent}{0pt}
\setlength{\RaggedRightParindent}{0pt}
\setlength{\parskip}{0.45em}
\tolerance=1
\emergencystretch=\maxdimen
\sloppy

\begin{center}
    {\Large\bfseries Supporting Information}\\[0.5em]
    {\large\bfseries Continuous Kinetic Model of Counterion-Mediated Redox Switching in 14-Bit Molecular Memristors for Analog In-Memory Computing}\\[0.8em]
    {\normalsize Logan Larsh$^{1}$, Dhiman Biswas$^{2}$, Nazmus Saadat As-Saquib$^{3}$, Su-in Yi$^{3}$, Thirumalai Venkatesan$^{2}$, and Mike Banad$^{1,*}$}\\[0.5em]
    {\footnotesize
    $^{1}$~Department of Electrical and Computer Engineering, The University of Oklahoma, Norman, OK, 73019, United States\\
    $^{2}$~Department of Physics and Astronomy, The University of Oklahoma, Norman, OK, 73019, United States\\
    $^{3}$~Department of Electrical and Computer Engineering, Texas A\&M University, College Station, TX, 77843, United States\\
    $^{*}$~Corresponding Author: \texttt{bana@ou.edu}
    }
\end{center}
\hrule
\medskip

\section{Executive Overview, Framework Architecture, and Section Glossary}
\label{sec:noteS1}

\subsection{The Non-Volatile Memory Hardware Wall and Molecular Redox Motivation}
Modern artificial intelligence and scientific computing workloads are dominated by dense vector-matrix multiplication (VMM) operations~\cite{si:krizhevsky2012,si:jouppi2017,si:gupta2020}. In conventional von Neumann digital architectures, executing VMM requires repeatedly shuttling weight tensors and intermediate activations between off-chip dynamic random access memory (DRAM), static random access memory (SRAM) caches, and arithmetic logic units, creating an energy and latency bottleneck that scales poorly with model size~\cite{si:xia2019,si:ham2021}. Analog in-memory computing (AIMC) circumvents this bottleneck by storing synaptic weights directly as physical conductance states within non-volatile memory (NVM) crossbar arrays, executing parallel multiplication via Ohm's law ($I_{ij} = V_i G_{ij}$) and current summation via Kirchhoff's current law ($I_j = \sum_i I_{ij}$) within a single physical clock cycle~\cite{si:burr2017,si:jeong2016,si:hu2018,si:yi2021}.

Despite extensive architectural investigation, deploying analog crossbar accelerators at commercial scale has been impeded by a fundamental ``Hardware Wall'' rooted in the stochastic solid-state physics of conventional inorganic memristors~\cite{si:strukov2008,si:chua1971,si:chua1976,si:wang2020resistive}. Phase-change memory (PCM) cells operate through Joule-heated amorphization (melt-quench RESET) and progressive crystallization (SET), which require high programming currents, limit cycling endurance to $10^6$ to $10^9$ operations, and introduce pronounced update asymmetry between gradual potentiation and abrupt RESET collapse~\cite{si:burr2017}. Filamentary resistive random access memory (RRAM) relies on stochastic electroforming and atomic migration of oxygen vacancies or metal cations to bridge electrodes through sub-10~nm conductive filaments~\cite{si:huang2016,si:jiang2016,si:kim2017}. Because filament formation is governed by localized electric field crowding and positive thermal feedback, filamentary devices exhibit pronounced cycle-to-cycle (C2C) write noise, device-to-device (D2D) spatial dispersion, non-linear current-voltage ($I$--$V$) characteristics, and low effective bit resolution (typically fewer than 64 distinguishable states, or $< 6$~bits)~\cite{si:xia2019,si:li2018}. While both inorganic and molecular memristors can be integrated with access transistors in one-transistor-one-resistor (1T1R) hybrid CMOS arrays for row selection, sneak-path cutoff, and peripheral control, conventional RRAM and PCM accelerators are heavily constrained by mandatory closed-loop write-and-verify feedback algorithms, multi-cycle pulse-tuning protocols, and auxiliary digital co-processors required to combat filament stochasticity and asymmetric melt-quench dynamics across stateful logic and in-memory computing arrays~\cite{si:legallo2018,si:chen2018neurosim,si:adam2016,si:kvatinsky2014,si:balatti2015,si:nguyen2017}.

A transformative solution to these stochastic switching limitations was established when Goswami and coworkers demonstrated 14-bit linear, symmetric kinetic switching through field-driven counterion displacement coupled to ligand-centered redox charge disproportionation in transition-metal bis-azo-aromatic coordination complexes~\cite{si:goswami2017,si:goswami2020,si:goswami2021,si:sharma2024,si:williams2024,si:kundu2024,si:gaur2024,si:yi2022}. In thin films of the bis-tridentate ruthenium complex $[\mathrm{Ru^{II}L_2}](\mathrm{BF_4})_2$ (where $\mathrm{L} = 2\text{-(phenylazo)pyridine}$), conductance modulation does not involve metallic filament growth or dielectric breakdown. Instead, nanosecond voltage pulses drive uniform, charge-conserving intermolecular electron transfer between neighboring ligand-centered $\pi^*$ (pi-antibonding) molecular orbitals coupled to the deterministic sub-angstrom displacement of mobile tetrafluoroborate ($\mathrm{BF_4^-}$) counterions across a $60\ \mathrm{nm}$ coordination lattice~\cite{si:goswami2020,si:sharma2024,si:kundu2024}. Whether deployed in selector-less passive arrays or 1T1R hybrid CMOS crossbars, this intrinsic linearity and bidirectional symmetry enable fast open-loop single-pulse weight updates across $16,520$ discrete analog conductance levels ($33,040$-pulse full cycle). Eliminating iterative analog-to-digital and digital-to-analog (ADC/DAC) write-and-verify feedback loops bypasses the verification penalty that typically consumes over $70\%$ of the total programming energy budget in traditional NVM arrays~\cite{si:legallo2018,si:sharma2024}. Consequently, these coordination junctions combine projected sub-100-attojoule switching energies in scaled $10\ \mathrm{nm} \times 10\ \mathrm{nm}$ nanocells, $0.19\ \mathrm{nJ}$ nominal write pulses in macroscopic $1.05\ \mathrm{\mu m} \times 1.1\ \mathrm{\mu m}$ crossbars, sub-5-nanosecond intrinsic switching speeds, and $10^{12}$-cycle endurance without electroforming or filament degradation~\cite{si:sharma2024,si:williams2024,si:kundu2024}. Furthermore, the $E_a = 130\ \mathrm{meV}$ counterion activation barrier establishes an intrinsic physical voltage threshold ($V_{\mathrm{th}} \approx 0.47\ \mathrm{V}$ in our continuous compact model) that protects non-destructive read integrity at $V_{\mathrm{read}} \le 0.20\ \mathrm{V}$ while providing natural half-select sneak-path attenuation in passive crossbars~\cite{si:sharma2024,si:kundu2024}.

\enlargethispage{\baselineskip}
To systematically evaluate how counterion steric volume and Coulombic binding govern these switching kinetics, this work establishes a three-tier counterion taxonomy across our physical derivations and simulation codebase (detailed in Notes~S2 and S3):
\begin{enumerate}[topsep=2pt,itemsep=1.5pt,parsep=0pt]
    \item \textit{$\mathrm{BF_4^-}$ (Tetrafluoroborate, Primary Baseline):} Exhibiting an effective crystallographic radius of $r = 2.18\ \text{\AA}$ and a zero-bias activation barrier of $E_a = 130\ \mathrm{meV}$ ($\approx 134\ \mathrm{meV}$ experimentally), $\mathrm{BF_4^-}$ represents the optimal baseline coordination system~\cite{si:sharma2024,si:kundu2024}. Its barrier shields programmed states against room-temperature thermal relaxation ($k_B T \approx 25.7\ \mathrm{meV}$ at $T_{\mathrm{amb}} = 298.15\ \mathrm{K}$) while enabling $16,520$ linear, symmetric conductance states under sub-100~ns open-loop pulses. This system is fully modeled, parameterized, and empirically validated across all circuit and algorithmic benchmarks in this study.
    \item \textit{$\mathrm{PF_6^-}$ (Hexafluorophosphate, Modeled Alternative):} Exhibiting a larger ionic radius of $r = 2.54\ \text{\AA}$ and a reduced activation barrier of $E_a = 95\ \mathrm{meV}$, $\mathrm{PF_6^-}$ weakens local electrostatic pinning~\cite{si:kundu2024,si:gaur2024}. Consequently, switching initiates at lower thresholds and saturates earlier, producing steeper non-linear initial conductance increments that deviate by $\approx 25\%$ from linear calibration at intermediate pulse counts. This system is fully modeled and parameterized in our simulation codebase (Tables~\ref{tab:S1} and \ref{tab:S3}) to demonstrate comparative ionic tuning.
    \item \textit{$\mathrm{BPh_4^-}$ (Tetraphenylborate, Steric Literature Reference):} Exhibiting a bulky radius of $r = 4.21\ \text{\AA}$ and an elevated activation barrier of $E_{a,\mathrm{micro}} = 210\ \mathrm{meV}$, this organic counterion introduces substantial steric crowding and excluded-volume constraints that strongly impede interstitial ionic translation~\cite{si:williams2024,si:kundu2024,si:gaur2024}. Because this steric lockout suppresses sub-$100\ \mathrm{ns}$ incremental analog front propagation in favor of high-voltage ($>2.0\ \mathrm{V}$) quasi-static digital transitions, dynamic pulse-programming datasets are physically absent in the literature. Nevertheless, $\mathrm{BPh_4^-}$ serves as an essential crystallographic and thermodynamic literature benchmark (Figure~2b of the main manuscript and Table~\ref{tab:S3}), delineating the upper physical boundary where counterion size precludes continuous analog operation.
\end{enumerate}

\subsection{The Multiscale Simulation Challenge and Modular Organization}
While the microscopic transport theory of $[\mathrm{Ru^{II}L_2}](\mathrm{BF_4})_2$ junctions has been established at the single-device level~\cite{si:kundu2024}, translating these quantum-chemical and electrostatic equations into large-scale electronic design automation (EDA) and machine learning frameworks presents three major computational obstacles:
\begin{enumerate}[topsep=2pt,itemsep=1.5pt,parsep=0pt]
    \item \textit{Infinite Series and Numerical Quadrature Overhead:} Evaluating the spatial electrostatic potential $\Phi_{\mathrm{mol}}(z,n)$ via an infinite Thomas-Fermi Fourier sine series~\cite{si:liang2004,si:nitzan2002} and computing the Marcus-Hush-Chidsey (MHC) electron transfer rate $k_{\mathrm{ET}}(z,n)$ via numerical quadrature over the Fermi-Dirac distribution~\cite{si:marcus1965,si:migliore2012} at each of the $L = 30$ molecular layers requires thousands of floating-point operations per device per timestep, rendering transient crossbar simulations intractable.
    \item \textit{Jacobian Singularities in Newton-Raphson Solvers:} Raw physical switching thresholds ($\theta(z,n) \in \{0,1\}$), piecewise voltage cutoffs, and hard state-space clipping at $n = 0$ and $n = n_{\max}$ introduce discontinuous first derivatives ($\partial I/\partial V$ and $\partial(dn/dt)/\partial V$) that cause Newton-Raphson matrix solvers in commercial SPICE and Cadence Spectre simulators to oscillate or fail to converge~\cite{si:kim2019,si:biolek2009}.
    \item \textit{Lack of Unified Circuit-to-Algorithm Verification:} Existing neuromorphic simulators either abstract away transient circuit physics into macro-level lookup estimators (such as NeuroSim~\cite{si:chen2018neurosim}) or wrap idealized PyTorch weight tensors without transistor-level Kirchhoff current summation or electro-thermal feedback (such as MemTorch~\cite{si:lammie2020memtorch}).
\end{enumerate}

\newpage

To resolve these challenges, this Supporting Information manual is organized into four progressive phases spanning eleven self-contained technical notes (Figure~\ref{fig:S0} and Table~\ref{tab:S0}). Note S2 consolidates all master reference parameter tables and the publicly available Python and Cadence Spectre simulation testbench catalogs (Tables~\ref{tab:S1}--\ref{tab:S5}) in a single dedicated section so subsequent derivations and benchmarks can reference them cleanly. Notes S3--S11 then progress systematically from experimental coordination chemistry and microscopic quantum transport through analytical surrogate compilation, single-device and array transient simulations, dual-engine Cadence Spectre versus Python parity verification, solver memory and throughput profiling, empirical hardware noise quantile sampling, hardware-in-the-loop MNIST neural network training, and row-wise 64-point Fourier astronomical image reconstruction.

\begin{figure}[!h]
    \setcounter{figure}{-1}
    \centering
    \vspace{0.4em}
    \includegraphics[width=\textwidth,height=0.62\textheight,keepaspectratio]{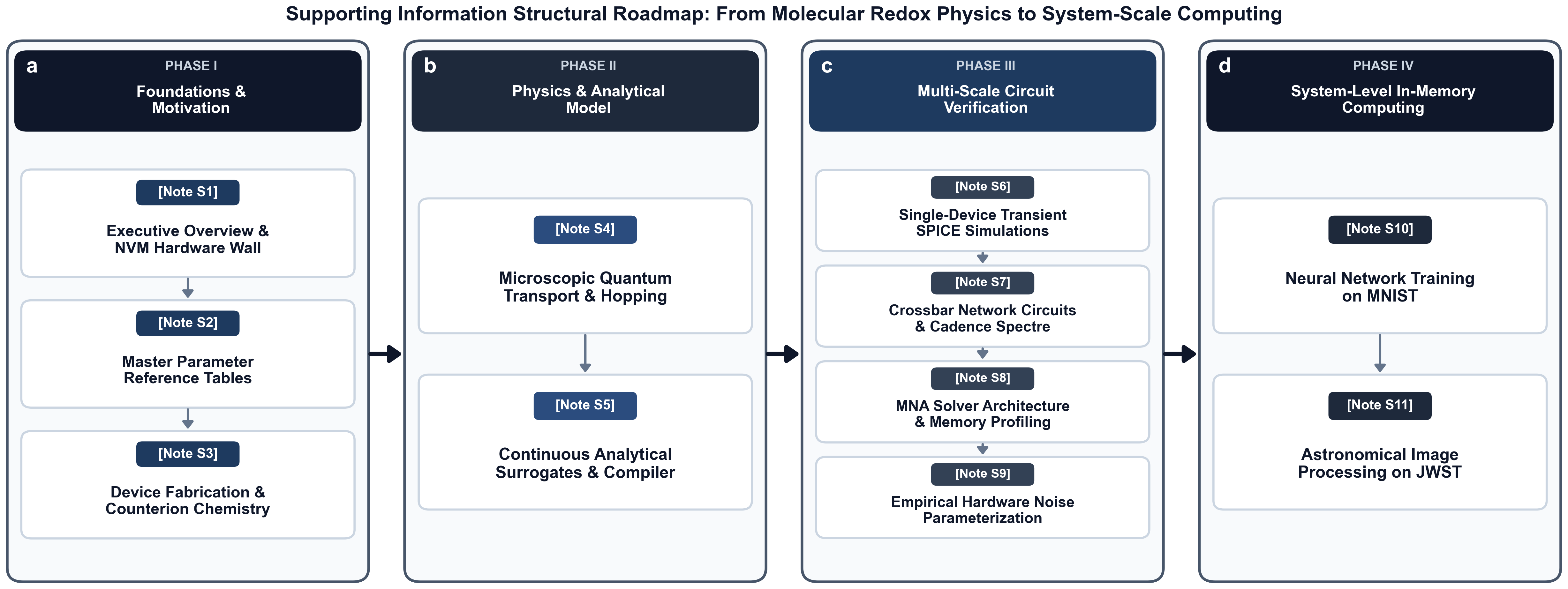}
    \vspace{0.4em}
    \caption{\textbf{Graphical table of contents and four-phase structural roadmap of the Supporting Information.} \textbf{a}, Phase~I: Foundations and Motivation (Notes~S1--S3), encompassing the executive overview of the NVM Hardware Wall, the consolidated Master Parameter Vault (Tables~\ref{tab:S1}--\ref{tab:S5}), and the three-tier coordination counterion taxonomy ($\mathrm{BF_4^-}$, $\mathrm{PF_6^-}$, and $\mathrm{BPh_4^-}$). \textbf{b}, Phase~II: Physics to Analytical Surrogates (Notes~S4 and S5), bridging 30-layer microscopic Thomas-Fermi screening, Marcus-Hush-Chidsey electron transfer, and Cauchy-Lorentz planar nucleation into $C^1$-continuous boundary-regularized rational potentials, smoothstep edge tapers, and 12-parameter PCHIP splines. \textbf{c}, Phase~III: Multi-Scale Circuit Verification and Algorithmic Solvers (Notes~S6--S9), covering baseline $[\mathrm{Ru^{II}L_2}](\mathrm{BF_4})_2$ ($\mathrm{BF_4^-}$) single-device and $4\times 4$ array transient simulations with Newton thermal dissipation, dual-engine Cadence Spectre vs.\ vectorized Python parity verification ($< 0.02\%$ discrepancy), vectorized Level-3 BLAS MNA solver profiling ($> 50,158\ \mathrm{steps/s}$), and empirical $\mathrm{BF_4^-}$ hardware noise quantile extraction across $75,000$ crosspoints. \textbf{d}, Phase~IV: Large-Scale System Benchmarks (Notes~S10 and S11) on baseline $[\mathrm{Ru^{II}L_2}](\mathrm{BF_4})_2$ ($\mathrm{BF_4^-}$) crossbars, demonstrating hardware-in-the-loop TensorFlow MNIST training via a custom Straight-Through Estimator ($95.31\%$ accuracy, $2.8\ \mathrm{mJ}$) and 64-point Inverse Discrete Fourier Transform (IDFT) reconstruction of James Webb Space Telescope (JWST) imagery ($99.30\%$ SSIM, $43.0\ \mathrm{dB}$ SNR, $24.6\ \mathrm{mJ}$). (Note: Graphical module badges in panels \textbf{c} and \textbf{d} denote thematic domain groupings mapped to the 11-Note sequence in Table~\ref{tab:S0}, where Note~S8 details the MNA solver, Note~S9 details hardware noise, Note~S10 details MNIST training, and Note~S11 details JWST IDFT reconstruction.)}
    \label{fig:S0}
\end{figure}

\newpage

\begin{table}[!p]
    \setcounter{table}{-1}
    \centering
    \caption{\textbf{Structural Glossary and Roadmap of the 11-Note Supporting Information Manual.} Each Note isolates a specific physical, mathematical, circuit, or system-level domain, cross-referenced to its associated figures, tables, equations, and main manuscript sections.}
    \label{tab:S0}
    \scriptsize
    \setlength{\tabcolsep}{3.5pt}
    \renewcommand{\arraystretch}{0.82}
    \begin{tabularx}{\textwidth}{@{}l >{\raggedright\arraybackslash}p{3.4cm} >{\raggedright\arraybackslash}X >{\raggedright\arraybackslash}p{3.5cm} l@{}}
        \toprule
        \textbf{Section} & \textbf{Technical Topic} & \textbf{Physical / Computational Purpose and Scope} & \textbf{Figures, Tables \& Equations} & \textbf{Main Text} \\
        \midrule
        Note S1 & Executive Overview, Framework Architecture \& Section Glossary & Defines the NVM Hardware Wall, open-loop molecular redox motivation, three-tier counterion taxonomy, and 4-phase roadmap. & Figure~\ref{fig:S0}, Table~\ref{tab:S0} & Sec.~1, Fig.~1 \\
        Note S2 & Master Reference Tables (Dedicated Tables Section) & Consolidates physical constants, compact model parameters ($\mathrm{BF_4^-}$ and $\mathrm{PF_6^-}$), $\mathrm{BF_4^-}$ noise quantiles, counterion metrics, and Python/Spectre testbench catalogs. & Tables~\ref{tab:S1}--\ref{tab:S5} & Sec.~2--4, Tables~1--2 \\
        Note S3 & Device Fabrication, Coordination Chemistry \& Counterion Physics & Documents the $\mathrm{Au} / 60\ \mathrm{nm}$ $[\mathrm{Ru^{II}L_2}](\mathrm{BF_4})_2 / \mathrm{Au}$ stack, electroforming-free operation, and coordination barrier physics. & Table~\ref{tab:S3} & Sec.~1, Sec.~2.1, Fig.~2a--b \\
        Note S4 & Microscopic Quantum Chemical \& Electrostatic Formulations & Derives the baseline $\mathrm{BF_4^-}$ Thomas-Fermi Fourier series $\Phi_{\mathrm{mol}}(z,n)$, MHC integral $k_{\mathrm{ET}}$, Cauchy-Lorentz $\mathit{Nr}_{xy}(z,n)$, and 30-site master equations. & Eqs.~(S1)--(S19), Table~\ref{tab:S1} & Sec.~2.1, Eqs.~(1)--(4), Fig.~2c--e \\
        Note S5 & Closed-Form Analytical Surrogates \& Boundary Regularization & Derives the baseline $\mathrm{BF_4^-}$ rational potential $u(t)$, smoothstep edge tapers $a_L, a_R$, 12-parameter physical PCHIP splines, and exponential MHC surrogate. & Figure~\ref{fig:S1}, Eqs.~(S20)--(S30) & Sec.~2.2, Eqs.~(5)--(10), Fig.~3a \\
        Note S6 & Circuit-Level Transient Simulations \& Device Verification & Validates baseline $\mathrm{BF_4^-}$ symbolic compact equations, $4\times 4$ crossbar control, single-pulse kinetics, multi-phase reprogramming, staircase microsteps, write-read immunity, Newton cooling, and overdrive stress. & Figures~\ref{fig:S2}, \ref{fig:S3}, Eqs.~(S31)--(S41), Tables~\ref{tab:S4}--\ref{tab:S5} & Sec.~2.3, Sec.~3, Figs.~3--4 \\
        Note S7 & Dual-Engine Verification: Cadence Spectre vs.\ Python Parity & Establishes theoretical equivalence and validates $<0.02\%$ transient waveform parity between Cadence Spectre Verilog-A and the Python engine across 5 testbenches. & Figure~\ref{fig:S4}, Eq.~(S42), Tables~\ref{tab:S4}, \ref{tab:S5}, \ref{tab:parity_summary} & Sec.~2.1, Sec.~3, Table~2 \\
        Note S8 & Computational Solver Architecture, Memory \& Scalability & Details vectorized BLAS-3 MNA solver, adaptive point skipping ($>50,158\ \mathrm{steps/s}$), $O(N_{\mathrm{devices}})$ RAM/VRAM scaling, and platform specs. & Figure~\ref{fig:S5}, Eqs.~(S43)--(S47), Tables~\ref{tab:S4}, \ref{tab:solver_specifications} & Sec.~3, Table~2 \\
        Note S9 & Empirical Hardware Noise Parameterization \& Quantiles & Details extraction of baseline $\mathrm{BF_4^-}$ read noise, C2C write noise, and D2D mismatch inverse-CDF lookup tables across 75,000 measurements. & Table~\ref{tab:S2}, Eqs.~(S48)--(S51), Tables~\ref{tab:S4}--\ref{tab:S5} & Sec.~2.3, Fig.~3e--f \\
        Note S10 & Machine Learning \& Neural Network Training (MNIST) & Details baseline $\mathrm{BF_4^-}$ $64\times 10$ differential crossbar mapping, custom STE backpropagation, and per-epoch energy accounting. & Eqs.~(S52)--(S55), Table~\ref{tab:S4} & Sec.~4.1, Fig.~5a--c, Table~1 \\
        Note S11 & Astronomical Spatial Frequency Image Processing (JWST) & Derives baseline $\mathrm{BF_4^-}$ split real/imaginary 64-point IDFT crossbar mapping, 768-block (3,072 batched evaluation) RGB tiling schedule, and SSIM/SNR fidelity metrics. & Figure~\ref{fig:S6}, Eqs.~(S56)--(S60), Table~\ref{tab:S4} & Sec.~4.2, Fig.~5d,e \\
        \bottomrule
    \end{tabularx}
\end{table}

\clearpage

\section{Master Reference Tables}
\label{sec:noteS2}

To provide a unified quantitative reference across all subsequent derivations and circuit simulations, this section consolidates the five master reference tables of our framework:
\begin{itemize}[topsep=1pt,itemsep=0pt,parsep=0pt]
    \item \textit{Table~\ref{tab:S1}:} Master physical, microscopic electrostatic, continuous compact conduction, dynamic switching, overdrive memory, and electro-thermal parameters for baseline $[\mathrm{Ru^{II}L_2}](\mathrm{BF_4})_2$ ($\mathrm{BF_4^-}$) and alternative $[\mathrm{Ru^{II}L_2}](\mathrm{PF_6})_2$ ($\mathrm{PF_6^-}$) devices.
    \item \textit{Table~\ref{tab:S2}:} Empirical hardware noise parameters and inverse-CDF quantile checkpoints for the baseline $[\mathrm{Ru^{II}L_2}](\mathrm{BF_4})_2$ ($\mathrm{BF_4^-}$) system extracted from $3,000$ physical devices, $2,000$ full potentiation cycles, and $75,000$ crosspoint measurements.
    \item \textit{Table~\ref{tab:S3}:} Counterion chemical, crystallographic, thermodynamic, and kinetic comparison across baseline tetrafluoroborate ($\mathrm{BF_4^-}$), modeled alternative hexafluorophosphate ($\mathrm{PF_6^-}$), and steric literature reference tetraphenylborate ($\mathrm{BPh_4^-}$) coordination complexes~\cite{si:sharma2024,si:williams2024,si:kundu2024,si:gaur2024}.
    \item \textit{Table~\ref{tab:S4}:} Exhaustive catalog of Python verification testbenches, array scaling benchmarks, AI workloads, and API-accessible device models in the publicly available simulation repository (\texttt{PythonSimulator/}, \texttt{HardwareTestTB/}, and \texttt{JamesWebb/}).
    \item \textit{Table~\ref{tab:S5}:} Complete catalog of Cadence Spectre / Virtuoso circuit netlists, transient testbenches, automated Ocean waveform scripts, and core Verilog-A model definitions (\texttt{MolmemVirtuosoSpectreV1.0/}).
\end{itemize}

\subsection{Counterion Empirical Calibration Hierarchy and Fidelity Legend}
Before consulting Tables~\ref{tab:S1} and \ref{tab:S3}, it is essential to distinguish the empirical calibration depth and simulation status of the three coordination counterions evaluated in this work:
\begin{enumerate}[topsep=1pt,itemsep=1pt,parsep=0pt]
    \item \textit{Baseline $\mathrm{BF_4^-}$ (Tetrafluoroborate) --- Production-Grade Baseline:}
    \begin{itemize}[topsep=0pt,itemsep=0pt,parsep=0pt]
        \item \textit{Empirical Status:} Tightly fitted and empirically cross-validated across complete $33,040$-pulse potentiation and depression cycles, multi-voltage overdrive saturation sweeps ($0.90\ \mathrm{V}$, $1.10\ \mathrm{V}$, and $1.22\ \mathrm{V}$), $3,000$-device read noise distributions ($75,000$ crosspoint measurements), and $2,000$-cycle write variability datasets~\cite{si:sharma2024,si:kundu2024}.
        \item \textit{Simulation Role:} Fully calibrated production compact model implemented identically in both the vectorized Python MNA engine and Cadence Spectre Verilog-A circuit netlists.
    \end{itemize}
    \item \textit{Alternative $\mathrm{PF_6^-}$ (Hexafluorophosphate) --- Limited-Subset Model$^{\dagger}$:}
    \begin{itemize}[topsep=0pt,itemsep=0pt,parsep=0pt]
        \item \textit{Empirical Status:} Calibrated against a limited experimental characterization subset demonstrating that a reduced activation barrier ($E_a = 95\ \mathrm{meV}$ vs.\ $130\ \mathrm{meV}$) and larger ionic radius ($r = 2.54\ \text{\AA}$ vs.\ $2.18\ \text{\AA}$) cause earlier turn-on and steeper non-linear switching (deviating by $\approx 25\%$ from linear calibration at intermediate states)~\cite{si:kundu2024,si:gaur2024}.
        \item \textit{Simulation Role:} Implemented in our simulation codebase to validate chemical tunability across counterion substitutions (Note~S3 and Tables~\ref{tab:S1} and \ref{tab:S3}), with higher parameter sensitivity in deep-saturation overdrive and electro-thermal relaxation regimes.
    \end{itemize}
    \item \textit{Literature Benchmark $\mathrm{BPh_4^-}$ (Tetraphenylborate) --- Steric Reference$^{\ddagger}$:}
    \begin{itemize}[topsep=0pt,itemsep=0pt,parsep=0pt]
        \item \textit{Empirical Status:} Dynamic pulse-programming datasets are absent in the literature because steric crowding ($r = 4.21\ \text{\AA}$) and an elevated barrier ($E_{a,\mathrm{micro}} = 210\ \mathrm{meV}$) preclude sub-$100\ \mathrm{ns}$ analog switching, exhibiting only high-voltage ($> 2.0\ \mathrm{V}$) quasi-static digital transitions~\cite{si:williams2024,si:kundu2024,si:gaur2024}.
        \item \textit{Theoretical Role:} Serves as a qualitative crystallographic benchmark in Note~S3 and Table~\ref{tab:S3} to establish the upper physical boundary where steric hindrance prevents counterion-mediated analog switching.
    \end{itemize}
\end{enumerate}

\subsubsection*{\texorpdfstring{Calibration Fidelity Legend (Tables~\ref{tab:S1} and \ref{tab:S3})}{Calibration Fidelity Legend (Tables tab:S1 and tab:S3)}}
\begin{itemize}[topsep=1pt,itemsep=0pt,parsep=0pt]
    \item \textit{Unmarked / Standard Text:} Fully verified parameters tightly calibrated against multi-voltage empirical hardware datasets ($\mathrm{BF_4^-}$ production baseline).
    \item \textit{Superscript Dagger ($^{\dagger}$):} Parameters calibrated against a limited experimental characterization subset or exhibiting higher sensitivity in deep overdrive and thermal regimes ($\mathrm{PF_6^-}$ alternative).
    \item \textit{Superscript Double Dagger ($^{\ddagger}$):} No available empirical pulse dataset in the literature; included as a qualitative crystallographic benchmark ($\mathrm{BPh_4^-}$ steric reference).
\end{itemize}

\begin{table}[H]
    \centering
    \caption{\textbf{Master Physical, Kinetic, Electro-Thermal, and Overdrive Parameters for Baseline $[\mathrm{Ru^{II}L_2}](\mathrm{BF_4})_2$ and Alternative $[\mathrm{Ru^{II}L_2}](\mathrm{PF_6})_2$ Molecular Memristors (Groups 1 and 2: Microscopic Film and Compact Conduction).} Fidelity legend: Unmarked = Fully verified baseline ($\mathrm{BF_4^-}$ multi-voltage hardware fit); $^{\dagger}$ = Limited-subset fit ($\mathrm{PF_6^-}$ characterization subset with potential underfitting in deep overdrive or thermal regimes).}
    \label{tab:S1}
    \scriptsize
    \setlength{\tabcolsep}{3.4pt}
    \renewcommand{\arraystretch}{0.95}
    \begin{tabularx}{\textwidth}{@{}l >{\raggedright\arraybackslash}X l l l >{\raggedright\arraybackslash}p{4.25cm}@{}}
        \toprule
        \textbf{Parameter} & \textbf{Physical Meaning and Compact Model Role} & \textbf{$\mathrm{BF_4^-}$} & \textbf{$\mathrm{PF_6^-}$} & \textbf{Units} & \textbf{Literature Comparison / Source} \\
        \midrule
        Calibration Status & Empirical model calibration and verification tier & \parbox[t]{2.2cm}{\raggedright Production Baseline (full multi-voltage hardware fit)} & \parbox[t]{2.2cm}{\raggedright Limited-Subset Fit$^{\dagger}$ (characterization subset)} & Tier & Defined across all Python and Spectre testbenches (Note S2) \\
        \midrule
        \multicolumn{6}{@{}l}{\textit{\textbf{Group 1: Microscopic Coordination Film and Electrostatic Parameters (Notes S3--S5)}}} \\
        $d_{\mathrm{film}}, L, \Delta z$ & Film thickness, molecular layers, lattice spacing & $60.0, 30, 2.0$ & $60.0, 30, 2.0$ & $\mathrm{nm}, 1, \mathrm{nm}$ & Lit.: $d_{\mathrm{film}} = 60\ \mathrm{nm}$~\cite{si:sharma2024}, $L = 30$ layers ($\Delta z = 2.0\ \mathrm{nm}$)~\cite{si:kundu2024} \\
        $r_{\mathrm{ion}}$ & Effective crystallographic counterion radius & \calgreen{$2.18$} & $2.54$ & \AA & Lit.: $2.18\ \text{\AA}$ ($\mathrm{BF_4^-}$), $2.54\ \text{\AA}$ ($\mathrm{PF_6^-}$) via DFT/X-ray~\cite{si:kundu2024,si:gaur2024} \\
        $E_{a,\mathrm{micro}}$ & Zero-bias counterion barrier ($134\ \mathrm{meV}$ exp.) & \calgreen{$130.0$} & $95.0$ & $\mathrm{meV}$ & Lit.: $134\ \mathrm{meV}$ exp.~\cite{si:sharma2024} vs.\ $130\ \mathrm{meV}$ modeled~\cite{si:kundu2024}; $95\ \mathrm{meV}$ ($\mathrm{PF_6^-}$)~\cite{si:kundu2024,si:gaur2024} \\
        $\Delta G^0, E_\lambda$ & Electronic free energy and reorganization energy & $0.10, 0.04$ & $0.10, 0.04$ & $\mathrm{eV}$ & Lit.: $\Delta G^0 = 0.10\ \mathrm{eV}$ ($100\ \mathrm{meV}$), $E_\lambda = 0.04\ \mathrm{eV}$ ($40\ \mathrm{meV}$)~\cite{si:kundu2024} \\
        $\Delta E_{\mathrm{meta}}$ & Metastable elevation of relaxed $22_{22}$ state & $+25.0$ & $+18.0$ & $\mathrm{meV}$ & Lit.: $+25\ \mathrm{meV}$ ($\mathrm{BF_4^-}$, $\approx 75\ \mathrm{meV}$ lattice stabilization)~\cite{si:sharma2024,si:kundu2024} \\
        $A_{\mathrm{ET}}, B_{\mathrm{ET}}$ & MHC surrogate pre-factor and active-window slope ($35.97\ \mathrm{V^{-1}}$ full span) & $1.135\times 10^3, 7.12$ & $1.420\times 10^3, 6.85$ & $\mathrm{s^{-1}}, \mathrm{V^{-1}}$ & No Literature Comparison / Unique to This Compact Model (Note S5) \\
        $k, \mathbf{\Theta}_{\mathrm{PCHIP}}$ & Rational screening exponent $k$ \& PCHIP splines & $54\ \mathrm{nodes}$ & $54\ \mathrm{nodes}$ & Unitless & No Literature Comparison / Unique to This Compact Model (Note S5) \\
        $p_1, p_2$ & Planar switched front slope and offset & $534.2, 15.1$ & $412.5, 11.4$ & $\mathrm{pulses}$ & Lit.: 30-layer linear front ($p_1 = 534.2\ \mathrm{pulses/layer}$)~\cite{si:kundu2024} \\
        $\kappa_{\mathrm{nuc,pot}}, \kappa_{\mathrm{nuc,dep}}$ & In-plane Cauchy-Lorentz nucleation half-width & $800.0, 500.0$ & $620.0, 440.0$ & $\mathrm{pulses}$ & Lit.: $\kappa_{\mathrm{nuc}} = 800$ (SET), $500$ (RESET) pulses~\cite{si:kundu2024} \\
        $\gamma_{m,31}, \gamma_{m,22}$ & Hopping rates in $(31)$ and $(22)$ states & $1.0\times 10^3, 4.8\times 10^4$ & $1.2\times 10^3, 5.4\times 10^4$ & $\mathrm{s^{-1}}$ & Lit.: 30-site master eq.\ ($48\times$ ratio $\gamma_{m,22}/\gamma_{m,31}$)~\cite{si:kundu2024} \\
        \midrule
        \multicolumn{6}{@{}l}{\textit{\textbf{Group 2: Continuous Compact Conduction and State-Space Parameters}}} \\
        \texttt{nMax} ($n_{\max}$) & Upper Hermite clamping capacity bound ($2N_{\mathrm{pot}}$, $N_{\mathrm{pot}} = 16,520$) & $33,040$ & $33,040$ & $\mathrm{pulses}$ & Lit.: $N_{\mathrm{pot}} = 16,520$ states ($33,040$-pulse full cycle)~\cite{si:sharma2024} \\
        \texttt{f22StartVal} & Initial ground-state switched fraction $f_{22}(0)$ & $0.12037256$ & $0.00051660$ & Unitless & Lit.: Cauchy-Lorentz domain boundary~\cite{si:kundu2024} (PCHIP spline fit) \\
        \texttt{f22PeakVal} & Peak switched volumetric fraction $f_{22}(N_{\mathrm{pot}})$ & $0.87791974$ & $0.99999985$ & Unitless & Lit.: Cauchy-Lorentz domain apex~\cite{si:kundu2024} (PCHIP spline fit) \\
        \texttt{f22EndVal} & Final full-cycle switched fraction $f_{22}(n_{\max})$ & $0.13955489$ & $0.00210450$ & Unitless & Lit.: Full-cycle return boundary~\cite{si:sharma2024,si:kundu2024} (PCHIP spline fit) \\
        \texttt{gScale} ($g_{\mathrm{scale}}$) & Parallel-channel transconductance gain & $6.227303$ & $6.582740$ & Unitless & No Literature Comparison / Unique to This Compact Model \\
        \texttt{rC} ($R_c$) & Single-contact series resistance ($2R_c$ total) & \calgreen{$8.033$} & \calyellow{$37.281$} & $\Omega$ & No Literature Comparison / Unique to This Compact Model \\
        \texttt{gamma} ($\gamma_{\mathrm{comp}}$) & High-conductance barrier compression factor & $-0.012230$ & $-0.000117$ & Unitless & No Literature Comparison / Unique to This Compact Model \\
        $G_{\min}, G_{\mathrm{nom}}, G_{\max}$ & Ground, nominal $+0.90\ \mathrm{V}$, and max conductance & \calgreen{$0.0002, 4.33, 5.90$} & \calyellow{$0.0002, 6.52, 6.58$} & $\mathrm{mS}$ & Lit.: $[200\ \mathrm{nS}, 5.90\ \mathrm{mS}]$ span ($4.33\ \mathrm{mS}$ nom.)~\cite{si:sharma2024} \\
        \bottomrule
    \end{tabularx}
\end{table}

\begin{table}[H]
    \setcounter{table}{0}
    \centering
    \caption{\textbf{(Continued) Master Physical, Kinetic, Electro-Thermal, and Overdrive Parameters for Baseline $[\mathrm{Ru^{II}L_2}](\mathrm{BF_4})_2$ and Alternative $[\mathrm{Ru^{II}L_2}](\mathrm{PF_6})_2$ Molecular Memristors (Groups 3 and 4: Dynamic Switching, Overdrive Memory, and Electro-Thermal Relaxation).} Superscript daggers ($^{\dagger}$) denote $\mathrm{PF_6^-}$ parameters calibrated against a limited experimental characterization subset with potential underfitting in deep overdrive or thermal regimes.}
    \scriptsize
    \setlength{\tabcolsep}{3.4pt}
    \renewcommand{\arraystretch}{0.95}
    \begin{tabularx}{\textwidth}{@{}l >{\raggedright\arraybackslash}X l l l >{\raggedright\arraybackslash}p{4.25cm}@{}}
        \toprule
        \textbf{Parameter} & \textbf{Physical Meaning and Compact Model Role} & \textbf{$\mathrm{BF_4^-}$} & \textbf{$\mathrm{PF_6^-}$} & \textbf{Units} & \textbf{Literature Comparison / Source} \\
        \midrule
        \multicolumn{6}{@{}l}{\textit{\textbf{Group 3: Dynamic Switching Kinetics, Softplus Thresholds, and Overdrive Memory (Note S6)}}} \\
        \texttt{kappa} ($\kappa_{\mathrm{switch}}$) & Base transition velocity ($\tau_{\mathrm{switch}} \approx 85.3\ \mathrm{ns}$) & \calgreen{$1.172865 \times 10^7$} & $1.755291 \times 10^7$ & $\mathrm{s^{-1}}$ & Lit.: $80\ \mathrm{ns}$ pulse switching velocity~\cite{si:sharma2024,si:kundu2024} \\
        \texttt{vTh} ($V_{\mathrm{th}}$) & Effective compact switching threshold voltage & \calgreen{$0.466409$} & $0.510364$ & $\mathrm{V}$ & Lit.: Protects $V_{\mathrm{read}} \le 0.20\ \mathrm{V}$ read immunity~\cite{si:sharma2024} \\
        \texttt{vSmooth} ($V_{\mathrm{smooth}}$) & Softplus threshold and polarity smoothing scale & $0.103776$ & $0.041015$ & $\mathrm{V}$ & No Literature Comparison / Unique to This Compact Model \\
        \texttt{nSmooth} ($N_{\mathrm{smooth}}$) & Hermite boundary clamping transition scale ($w_{\mathrm{sat}} = \max(50, 1.5 N_{\mathrm{smooth,eff}})$) & $17.06$ & \calyellow{$177.61$} & $\mathrm{pulses}$ & No Literature Comparison / Unique to This Compact Model \\
        \texttt{vRefPos, vRefNeg} & Reference calibration voltages (SET / RESET) & $0.9000, 0.7500$ & $0.9000, 0.7500$ & $\mathrm{V}$ & Lit.: $+0.90\ \mathrm{V}$ ($80\ \mathrm{ns}$ SET), $-0.75\ \mathrm{V}$ ($65\ \mathrm{ns}$ RESET)~\cite{si:sharma2024} \\
        \texttt{alpha} ($\alpha$) & Baseline over-threshold voltage exponent & \calgreen{$0.931756$} & \calyellow{$4.862785$} & Unitless & Lit.: Linear drift ($\mathrm{BF_4^-}$) vs.\ $\approx 25\%$ nonlinear ($\mathrm{PF_6^-}$)~\cite{si:sharma2024,si:kundu2024} \\
        \texttt{beta\_alpha} ($\beta_\alpha$) & Overdrive voltage exponent coupling coefficient & \calgreen{$0.125186$} & \calyellow{$1.588296$} & Unitless & No Literature Comparison / Unique to This Compact Model \\
        \texttt{beta\_s} ($\beta_s$) & Upper saturation transition overdrive slope & $-4.382853$ & \calyellow{$-0.787981$} & Unitless & No Literature Comparison / Unique to This Compact Model \\
        \texttt{rate\_asymmetry} ($r_{\mathrm{asym}}$) & Reverse-to-forward kinetic symmetry ratio & \calgreen{$0.985694$} & $0.915890$ & Unitless & Lit.: Symmetric SET/RESET kinetics ($80\ \mathrm{ns} / 65\ \mathrm{ns}$)~\cite{si:sharma2024,si:kundu2024} \\
        \texttt{rTopBlend} & Upper saturation transition overdrive blend factor & $0.050000$ & $0.050000$ & Unitless & No Literature Comparison / Unique to This Compact Model \\
        \texttt{kDischarge} ($k_{\mathrm{discharge}}$) & Overdrive RESET snapback discharge rate & $362.88$ & \calyellow{$60.62$} & $\mathrm{s^{-1}}$ & Lit.: Overdrive interfacial ion discharge~\cite{si:kundu2024} (compact rate fit) \\
        \texttt{fDischarge} ($f_{\mathrm{discharge}}$) & Overdrive snapback knee ($p_{\mathrm{decay}} = 1/f_{\mathrm{discharge}}$) & $0.374848$ & \calyellow{$0.699784$} & Unitless & No Literature Comparison / Unique to This Compact Model \\
        \texttt{rBottomBlend} & Overdrive depression floor blend factor & $0.714488$ & $0.647027$ & Unitless & No Literature Comparison / Unique to This Compact Model \\
        \texttt{rLatency} ($r_{\mathrm{latency}}$) & Nominal return shelf latency before depression & $0.878506$ & \calyellow{$0.001523$} & Unitless & No Literature Comparison / Unique to This Compact Model \\
        \texttt{k\_overdrive, vBypass} & Overdrive register gain and bypass cutoff & $7.296635, 0.1551$ & $7.296635, 0.0500$ & $1, \mathrm{V}$ & No Literature Comparison / Unique to This Compact Model \\
        \midrule
        \multicolumn{6}{@{}l}{\textit{\textbf{Group 4: Coupled Electro-Thermal Relaxation Parameters (Note S6)}}} \\
        \texttt{E\_a} ($E_{a,\mathrm{macro}}$) & Macroscopic Arrhenius barrier in $k_{\mathrm{thermal}}(T)$ & \calgreen{$155.9$ [$0.1559$]} & \calyellow{$73.3$ [$0.0733$]} & $\mathrm{meV}$ [$\mathrm{eV}$] & Lit.: $134\ \mathrm{meV}$ exp.~\cite{si:sharma2024} / $130\ \mathrm{meV}$ DFT ($\mathrm{BF_4^-}$); $95\ \mathrm{meV}$ ($\mathrm{PF_6^-}$)~\cite{si:kundu2024} \\
        \texttt{R\_th} ($R_{\mathrm{th}}$) & Effective junction-to-electrode thermal resistance & \calgreen{$52.14$} & \calyellow{$9994.72$} & $\mathrm{K/W}$ & No Literature Comparison / Unique to This Compact Model \\
        \texttt{tau\_th} ($\tau_{\mathrm{th}}$) & Newton thermal time constant ($R_{\mathrm{th}}C_{\mathrm{th}}$) & \calgreen{$127.8$} & \calyellow{$2.955$} & $\mathrm{ns}$ & No Literature Comparison / Unique to This Compact Model \\
        $T_{\mathrm{amb}}$ & Ambient reference operating temperature & $298.15$ & $298.15$ & $\mathrm{K}$ & Lit.: Standard room-temperature operation ($298.15\ \mathrm{K}$)~\cite{si:sharma2024} \\
        \bottomrule
    \end{tabularx}
\end{table}

\begin{table}[H]
    \centering
    \caption{\textbf{Empirical Hardware Noise Parameters and Inverse-CDF Quantile Checkpoints for Baseline $[\mathrm{Ru^{II}L_2}](\mathrm{BF_4})_2$ ($\mathrm{BF_4^-}$) Molecular Memristors Extracted Across $75,000$ Physical Crosspoint Measurements.}}
    \label{tab:S2}
    \scriptsize
    \setlength{\tabcolsep}{5pt}
    \renewcommand{\arraystretch}{1.10}
    \begin{tabularx}{\textwidth}{@{}l >{\raggedright\arraybackslash}X l l@{}}
        \toprule
        \textbf{Noise Metric / Quantile} & \textbf{Statistical Definition and Empirical Lookup Table Role} & \textbf{Calibrated Value} & \textbf{Units} \\
        \midrule
        \multicolumn{4}{@{}l}{\textit{\textbf{1. Global Characterization Dataset Dimensions}}} \\
        $V_{\mathrm{read}}$ & Sub-threshold non-destructive read interrogation bias & $0.200$ & $\mathrm{V}$ \\
        $N_{\mathrm{tested\_devices}}$ & Number of physical devices evaluated for read and D2D statistics~\cite{si:sharma2024} & $3,000$ & $\mathrm{devices}$ \\
        $N_{\mathrm{benchmark\_levels}}$ & Target conductance levels per device ($3,000 \times 25 = 75,000\ \mathrm{pts}$) & $25$ & $\mathrm{levels}$ \\
        $N_{\mathrm{write\_cycles}}$ & Full potentiation cycles evaluated for C2C write variability~\cite{si:sharma2024} & $2,000$ & $\mathrm{cycles}$ \\
        $N_{\mathrm{pulses\_per\_cycle}}$ & Programming pulses per potentiation sweep~\cite{si:sharma2024} & $16,520$ & $\mathrm{pulses}$ \\
        \midrule
        \multicolumn{4}{@{}l}{\textit{\textbf{2. Conductance-Dependent Read Noise Profile (100-Point Empirical Lookup Table)}}} \\
        $\sigma_{\mathrm{read}}(g_{\mathrm{norm}} = 0.00)$ & Baseline read noise standard deviation floor at $G_{\min} = 200\ \mathrm{nS}$ & $27.23$ ($2.723 \times 10^{-8}$) & $\mathrm{nS}$ ($\mathrm{S}$) \\
        $\sigma_{\mathrm{read}}(g_{\mathrm{norm}} = 0.25)$ & Read noise standard deviation at $25\%$ normalized conductance & $40.66$ ($4.066 \times 10^{-8}$) & $\mathrm{nS}$ ($\mathrm{S}$) \\
        $\sigma_{\mathrm{read}}(g_{\mathrm{norm}} = 0.50)$ & Read noise standard deviation at $50\%$ normalized conductance & $54.10$ ($5.410 \times 10^{-8}$) & $\mathrm{nS}$ ($\mathrm{S}$) \\
        $\sigma_{\mathrm{read}}(g_{\mathrm{norm}} = 0.75)$ & Read noise standard deviation at $75\%$ normalized conductance & $67.32$ ($6.732 \times 10^{-8}$) & $\mathrm{nS}$ ($\mathrm{S}$) \\
        $\sigma_{\mathrm{read}}(g_{\mathrm{norm}} = 1.00)$ & Maximum read noise ceiling at $G_{\max} = 5.90\ \mathrm{mS}$ ($>97\ \mathrm{dB}$ SNR) & $80.54$ ($8.054 \times 10^{-8}$) & $\mathrm{nS}$ ($\mathrm{S}$) \\
        \midrule
        \multicolumn{4}{@{}l}{\textit{\textbf{3. Cycle-to-Cycle (C2C) Write Step Variability (100-Point Empirical Lookup Table)}}} \\
        $\mu_{\Delta G}$ & Ensemble mean single-pulse conductance increment (full hardware range)~\cite{si:sharma2024} & $358.73$ ($3.587 \times 10^{-7}$) & $\mathrm{nS}$ ($\mathrm{S}$) \\
        $\sigma_{\mathrm{write,mean}}$ & Ensemble mean standard deviation of single-pulse step $\Delta G$~\cite{si:sharma2024} & $31.92$ ($3.192 \times 10^{-8}$) & $\mathrm{nS}$ ($\mathrm{S}$) \\
        $\eta_{\mathrm{write,global}}$ & Global relative write noise ratio ($\sigma_{\mathrm{write,mean}} / \mu_{\Delta G}$, $8.9\%$ hardware fit~\cite{si:sharma2024}) & $8.90\%$ ($0.08899$) & $\%$ ($\mathrm{ratio}$) \\
        $\eta_{\mathrm{write}}(g_{\mathrm{norm}} = 0.00)$ & Relative C2C write noise at ground-state onset ($n \to 0$) & $5.06\%$ ($0.05056$) & $\%$ ($\mathrm{ratio}$) \\
        $\eta_{\mathrm{write}}(g_{\mathrm{norm}} = 0.50)$ & Relative C2C write noise at mid-scale conductance ($n = 8,260$) & $8.82\%$ ($0.08818$) & $\%$ ($\mathrm{ratio}$) \\
        $\eta_{\mathrm{write}}(g_{\mathrm{norm}} = 1.00)$ & Relative C2C write noise at upper saturation ($n = 16,520$) & $12.58\%$ ($0.12576$) & $\%$ ($\mathrm{ratio}$) \\
        \midrule
        \multicolumn{4}{@{}l}{\textit{\textbf{4. Device-to-Device (D2D) Spatial Mismatch Quantiles (102-Point Inverse CDF)}}} \\
        $\sigma_{\mathrm{D2D,mean}}$ & Ensemble D2D spatial conductance standard deviation ($75,000\ \mathrm{pts}$) & $107.17$ ($1.072 \times 10^{-7}$) & $\mathrm{nS}$ ($\mathrm{S}$) \\
        $Q_{\mathrm{D2D}}(u = 0.000)$ & Minimum empirical lower tail offset ($0$-th percentile) & $-481.28$ & $\mathrm{nS}$ \\
        $Q_{\mathrm{D2D}}(u = 0.050)$ & $5$-th percentile D2D spatial conductance offset & $-167.20$ & $\mathrm{nS}$ \\
        $Q_{\mathrm{D2D}}(u = 0.250)$ & $25$-th percentile (first quartile) D2D conductance offset & $-74.02$ & $\mathrm{nS}$ \\
        $Q_{\mathrm{D2D}}(u = 0.500)$ & $50$-th percentile (median) D2D conductance offset & $-0.79$ & $\mathrm{nS}$ \\
        $Q_{\mathrm{D2D}}(u = 0.750)$ & $75$-th percentile (third quartile) D2D conductance offset & $+71.40$ & $\mathrm{nS}$ \\
        $Q_{\mathrm{D2D}}(u = 0.950)$ & $95$-th percentile D2D spatial conductance offset & $+168.79$ & $\mathrm{nS}$ \\
        $Q_{\mathrm{D2D}}(u = 1.000)$ & Maximum empirical upper tail offset ($100$-th percentile) & $+433.90$ & $\mathrm{nS}$ \\
        \bottomrule
    \end{tabularx}
\end{table}

\begin{table}[H]
    \centering
    \caption{\textbf{Counterion Chemical, Crystallographic, Thermodynamic, and Kinetic Comparison Across $[\mathrm{Ru^{II}L_2}](\mathrm{X})_2$ Molecular Memristors ($\mathrm{X} = \mathrm{BF_4^-}, \mathrm{PF_6^-}, \mathrm{BPh_4^-}$)~\cite{si:sharma2024,si:williams2024,si:kundu2024,si:gaur2024}.} Fidelity legend: Unmarked = Fully verified production baseline ($\mathrm{BF_4^-}$); $^{\dagger}$ = Calibrated to limited empirical characterization subset ($\mathrm{PF_6^-}$); $^{\ddagger}$ No available empirical dataset (included as qualitative literature comparison).}
    \label{tab:S3}
    \scriptsize
    \setlength{\tabcolsep}{3.5pt}
    \renewcommand{\arraystretch}{1.15}
    \begin{tabularx}{\textwidth}{@{}l >{\raggedright\arraybackslash}p{2.8cm} >{\raggedright\arraybackslash}p{3.15cm} >{\raggedright\arraybackslash}p{3.2cm} >{\raggedright\arraybackslash}X@{}}
        \toprule
        \textbf{Property / Metric} & \textbf{Tetrafluoroborate ($\mathrm{BF_4^-}$)} & \textbf{Hexafluorophosphate ($\mathrm{PF_6^-}$)} & \textbf{Tetraphenylborate ($\mathrm{BPh_4^-}$)} & \textbf{Literature Comparison \& Architectural Implication} \\
        \midrule
        Chemical Formula & $[\mathrm{Ru^{II}L_2}](\mathrm{BF_4})_2$ & $[\mathrm{Ru^{II}L_2}](\mathrm{PF_6})_2$ & $[\mathrm{Ru^{II}L_2}](\mathrm{BPh_4})_2$ & Lit.: Cationic $[\mathrm{Ru^{II}L_2}]^{2+}$ core invariant~\cite{si:sharma2024,si:kundu2024,si:gaur2024} \\
        Calibration Status & \calgreen{Production Baseline} (full multi-voltage \& noise fit) & \calyellow{Limited-Subset Fit} (characterization subset) & \calred{No Available Dataset} (Literature Benchmark~\cite{si:kundu2024,si:gaur2024}) & Defines exact simulation tier across Python and Spectre engines \\
        Anion Geometry & Compact Tetrahedral & Octahedral & Bulky Tetra-Phenyl Steric Cage & Lit.: Controls interstitial coordination pocket packing~\cite{si:kundu2024,si:gaur2024} \\
        Ionic Radius $r_{\mathrm{ion}}$ (\AA) & \calgreen{$2.18\ \text{\AA}$} & $2.54\ \text{\AA}$ & $4.21\ \text{\AA}$~\cite{si:kundu2024,si:gaur2024} & Lit.: DFT/X-ray radii~\cite{si:kundu2024,si:gaur2024}; nearly $2\times$ expansion to $\mathrm{BPh_4^-}$ \\
        Microscopic Barrier $E_{a,\mathrm{micro}}$ & \calgreen{$130\ \mathrm{meV}$} ($134\ \mathrm{meV}$ exp.) & $95\ \mathrm{meV}$ & $210\ \mathrm{meV}$~\cite{si:kundu2024,si:gaur2024} & Lit.: $134\ \mathrm{meV}$ exp.~\cite{si:sharma2024} vs.\ $130\ \mathrm{meV}$ DFT~\cite{si:kundu2024}; $>5 k_B T$ retention \\
        Macroscopic $E_{a,\mathrm{macro}}$ & \calgreen{$155.9\ \mathrm{meV}$} [$0.1559\ \mathrm{eV}$] & \calyellow{$73.3\ \mathrm{meV}$} [$0.0733\ \mathrm{eV}$] & \calred{\text{N/A}} (Quasi-static digital regime~\cite{si:kundu2024}) & Unique continuum Arrhenius self-heating fit $k_{\mathrm{thermal}}(T)$ \\
        Relaxed $22_{22}$ Elevation & $+25\ \mathrm{meV}$ & $+18\ \mathrm{meV}$ & \calred{\text{N/A}} (Sterically hindered~\cite{si:gaur2024}) & Lit.: $\approx 75\ \mathrm{meV}$ stabilization from $+100\ \mathrm{meV}$ $22_{31}$ ($\mathrm{BF_4^-}$)~\cite{si:sharma2024,si:kundu2024} \\
        Compact Threshold $V_{\mathrm{th}}$ & \calgreen{$0.4664\ \mathrm{V}$} & $0.5104\ \mathrm{V}$ & \calred{\text{N/A}} ($> 2.0\ \mathrm{V}$ quasi-static~\cite{si:kundu2024}) & Lit.: Protects sub-threshold read immunity at $V_{\mathrm{read}} \le 0.20\ \mathrm{V}$~\cite{si:sharma2024} \\
        Switching Velocity $\kappa_{\mathrm{switch}}$ & \calgreen{$1.173 \times 10^7\ \mathrm{s^{-1}}$} & $1.755 \times 10^7\ \mathrm{s^{-1}}$ & \calred{\text{N/A}} (Sub-$100\ \mathrm{ns}$ suppressed) & Lit.: Matches $80\ \mathrm{ns}$ SET pulse velocity~\cite{si:sharma2024,si:kundu2024}; $\mathrm{PF_6^-}$ $1.50\times$ faster \\
        Voltage Exponent $\alpha$ & \calgreen{$0.9318$} (quasi-linear) & \calyellow{$4.8628$} (steep nonlinear) & \calred{\text{N/A}} (Abrupt digital transition) & Lit.: Linear drift ($\mathrm{BF_4^-}$) vs.\ $\approx 25\%$ nonlinear ($\mathrm{PF_6^-}$)~\cite{si:sharma2024,si:kundu2024} \\
        Kinetic Symmetry $r_{\mathrm{asym}}$ & \calgreen{$0.9857$} (symmetric) & $0.9159$ (single-slope compact approx.) & \calred{\text{N/A}} (Non-analog regime) & Lit.: Symmetric open-loop $\Delta n$ updates without write-verify~\cite{si:sharma2024} \\
        Peak Conductance $G_{\max}$ & \calgreen{$4.33\ \mathrm{mS}$} nom.\ ($5.90\ \mathrm{mS}$ sat.) & \calyellow{$6.58\ \mathrm{mS}$} (characterization subset) & \calred{Digital only} (quasi-static transition $\approx 5.8\ \mathrm{mS}$ at $> 2.7\ \mathrm{V}$; sub-$100\ \mathrm{ns}$ analog switching sterically suppressed~\cite{si:gaur2024}) & Lit.: $16,520$ linear states across $[200\ \mathrm{nS}, 5.90\ \mathrm{mS}]$ ($\mathrm{BF_4^-}$)~\cite{si:sharma2024} \\
        \bottomrule
    \end{tabularx}
\end{table}

\clearpage

\subsection{Simulation Framework and Repository Testbench Catalog}
The simulation framework, compact device models, parameter extraction pipelines, and validation datasets are publicly available on GitHub at \url{https://github.com/INQUIRELAB/MolecularMemristorSpectreandPythonModel}. To facilitate reproducibility and transparent cross-referencing with our publicly available codebase, Table~\ref{tab:S4} catalogs the complete Python testbench suite and API-accessible device models (\texttt{PythonSimulator/}, \texttt{HardwareTestTB/}, and \texttt{JamesWebb/}), and Table~\ref{tab:S5} catalogs the corresponding Cadence Spectre / Virtuoso circuit netlists, automated Ocean waveform scripts, and core Verilog-A model definitions (\texttt{MolmemVirtuosoSpectreV1.0/}).

\begin{table}[H]
    \centering
    \caption{\textbf{Exhaustive Catalog of Python Verification Testbenches, Array Scaling Benchmarks, AI Workloads, and API-Accessible Device Models in the Publicly Available Repository (Evaluated on Baseline $[\mathrm{Ru^{II}L_2}](\mathrm{BF_4})_2$ Unless Otherwise Specified).}}
    \label{tab:S4}
    \scriptsize
    \setlength{\tabcolsep}{4.5pt}
    \renewcommand{\arraystretch}{1.04}
    \begin{tabularx}{\textwidth}{@{}l l >{\raggedright\arraybackslash}X@{}}
        \toprule
        \textbf{Script / Component} & \textbf{Directory} & \textbf{Physical Simulation Domain and Execution Scope} \\
        \midrule
        \multicolumn{3}{@{}l}{\textit{\textbf{1. Circuit-Level Testbenches (\texttt{PythonSimulator/tb\_circuit\_level/})}}} \\
        \texttt{tb\_dynamic.py} & \texttt{tb\_circuit\_level} & Nominal 14-bit potentiation and full return-to-zero depression cycle ($39,062$ total pulses: $16,520$ SET + $22,542$ RESET) tracking $V, n, f_{22}, G, I, T$ \\
        \texttt{tb\_dynamic\_extended.py} & \texttt{tb\_circuit\_level} & Extended full-cycle saturation and recovery ($60,000$ total pulses: $25,000$ SET + $35,000$ RESET) evaluating deep-saturation ceiling \\
        \texttt{tb\_standard\_cycle.py} & \texttt{tb\_circuit\_level} & Calibrated $16,520$-pulse potentiation/depression reference cycle ($33,040$ total pulses) and baseline transconductance \\
        \texttt{tb\_write\_read.py} & \texttt{tb\_circuit\_level} & Interleaved $\pm 0.90\ \mathrm{V}$ write pulses and sub-threshold $200\ \mathrm{mV}$ non-destructive reads (zero disturbance) \\
        \texttt{tb\_stress.py} & \texttt{tb\_circuit\_level} & Multi-stage pulse overdrive stress testing ($+1.2\ \mathrm{V}$, $-1.6\ \mathrm{V}$) and zero-bias hold stability \\
        \texttt{tb\_thermal\_relaxation.py} & \texttt{tb\_circuit\_level} & Sustained DC read Joule heating and exponential Newton cooling relaxation ($\tau_{\mathrm{th}} = 127.8\ \mathrm{ns}$) \\
        \texttt{tb\_crossbar\_8x8.py} & \texttt{tb\_circuit\_level} & Parallel $8\times 8$ crossbar vector-matrix multiplication and Kirchhoff current summation \\
        \texttt{tb\_crossbar\_write.py} & \texttt{tb\_circuit\_level} & Interconnect line loading, worst-case diagonal IR drop, and write margins across array scales \\
        \texttt{tb\_noise\_validation.py} & \texttt{tb\_circuit\_level} & Three-channel empirical hardware noise injection (read noise, C2C write steps, D2D spatial quantiles) \\
        \texttt{tb\_memory\_profiling.py} & \texttt{tb\_circuit\_level} & Vectorized BLAS-3 MNA solver memory footprint, runtime scaling, and point-skipping benchmarks \\
        \midrule
        \multicolumn{3}{@{}l}{\textit{\textbf{2. Software-Level and Array Scaling Testbenches (\texttt{PythonSimulator/tb\_software\_level/})}}} \\
        \texttt{tb\_parallel\_crossbars.py} & \texttt{tb\_software\_level} & Multi-array parallel crossbar execution and multi-core dispatch \\
        \texttt{tb\_tiled\_simulator.py} & \texttt{tb\_software\_level} & Large-scale tiled crossbar batch evaluation and partitioned matrix scheduling \\
        \texttt{tb\_tiled\_batch\_input.py} & \texttt{tb\_software\_level} & Pipelined batch vector evaluation across partitioned physical tiles \\
        \texttt{tb\_multi\_program\_4x4.py} & \texttt{tb\_software\_level} & Synchronous multi-device parallel conductance programming across a $4\times 4$ array \\
        \texttt{tb\_batch\_input.py} & \texttt{tb\_software\_level} & Batch vector streaming through single crossbars for inference throughput profiling \\
        \texttt{tb\_quantized\_weights.py} & \texttt{tb\_software\_level} & Fixed-point weight discretization and quantization error analysis on physical crossbars \\
        \texttt{tb\_compare\_backend\_performance.py} & \texttt{tb\_software\_level} & Computational benchmark comparing CPU MNA vs.\ GPU (NVIDIA CUDA / AMD ROCm) and hybrid backends \\
        \texttt{tb\_update\_tile.py} & \texttt{tb\_software\_level} & Dynamic weight refresh and partial sub-array reprogramming algorithms \\
        \texttt{tb\_tf\_training.py} & \texttt{tb\_software\_level} & TensorFlow custom gradient integration and software-level backpropagation verification \\
        \midrule
        \multicolumn{3}{@{}l}{\textit{\textbf{3. Hardware Diagnostics and AI Workloads (\texttt{HardwareTestTB/}, \texttt{JamesWebb/}, \texttt{PythonSimulator/})}}} \\
        \texttt{HardwareTestTB.py} & \texttt{HardwareTestTB} & Comprehensive automated test suite validating multi-voltage hardware non-idealities and stability \\
        \texttt{software\_train\_mnist.py} & \texttt{PythonSimulator} & Hardware-in-the-loop $64\times 10$ MNIST neural network training using Straight-Through Estimator (STE) \\
        \texttt{memristorimageprocess.py} & \texttt{JamesWebb} & Row-wise 64-point Inverse Discrete Fourier Transform (IDFT) spatial image reconstruction on JWST Pillars of Creation \\
        \texttt{Multiplier.py} & \texttt{PythonSimulator} & Standalone analog matrix-vector multiplier utility demonstrating $O(1)$ physical dot products \\
        \midrule
        \multicolumn{3}{@{}l}{\textit{\textbf{4. API-Accessible Device Models (\texttt{molmem\_lib/devices.py})}}} \\
        \texttt{default} / \texttt{Ru\_azo\_BF4} & \texttt{molmem\_lib} & Production baseline $[\mathrm{Ru^{II}L_2}](\mathrm{BF_4})_2$ calibrated to $33,040$ pulses, multi-voltage sweeps, and $75,000$ noise points \\
        \texttt{Ru\_azo\_PF6} & \texttt{molmem\_lib} & Modeled alternative $[\mathrm{Ru^{II}L_2}](\mathrm{PF_6})_2$ with reduced $95\ \mathrm{meV}$ barrier and steeper nonlinear switching \\
        \bottomrule
    \end{tabularx}
\end{table}

\clearpage

\begin{table}[H]
    \centering
    \caption{\textbf{Catalog of Cadence Spectre / Virtuoso Netlists, Testbenches, and Ocean Scripts (\texttt{MolmemVirtuosoSpectreV1.0/}).}}
    \label{tab:S5}
    \scriptsize
    \setlength{\tabcolsep}{4.5pt}
    \renewcommand{\arraystretch}{1.10}
    \begin{tabularx}{\textwidth}{@{}l l >{\raggedright\arraybackslash}X@{}}
        \toprule
        \textbf{Netlist / Script Filename} & \textbf{Directory} & \textbf{Simulation Scope and Circuit Function} \\
        \midrule
        \multicolumn{3}{@{}l}{\textit{\textbf{1. Circuit Netlist Testbenches (\texttt{testbenches/})}}} \\
        \texttt{tb\_dynamic.scs} & \texttt{testbenches} & Transistor-level $39,062$-pulse ($16,520$ SET + $22,542$ RESET) full return-to-zero pulse train transient simulation \\
        \texttt{tb\_dynamic\_extended.scs} & \texttt{testbenches} & Long-duration $60,000$-pulse ($25,000$ SET + $35,000$ RESET) cycling netlist evaluating deep-saturation and thermal equilibrium \\
        \texttt{tb\_standard\_cycle.scs} & \texttt{testbenches} & Calibrated $16,520$ SET + $16,520$ RESET ($33,040$-pulse) cycle matching Python reference \\
        \texttt{tb\_write\_read.scs} & \texttt{testbenches} & Dynamic write/read testbench verifying sub-threshold read disturbance immunity \\
        \texttt{tb\_stress.scs} & \texttt{testbenches} & Multi-stage overvoltage stress netlist ($+1.2\ \mathrm{V}$, $-1.6\ \mathrm{V}$) evaluating boundary clamping stability \\
        \texttt{tb\_thermal\_relaxation.scs} & \texttt{testbenches} & Sustained DC bias heating and transient Newton cooling verification \\
        \texttt{tb\_iv\_sweep.scs} & \texttt{testbenches} & Quasi-static and transient $I$--$V$ DC sweep across positive and negative bias regimes \\
        \texttt{tb\_crossbar\_8x8.scs} & \texttt{testbenches} & Full $8\times 8$ crossbar array transient simulation with TIA virtual-ground column readout \\
        \texttt{tb\_crossbar\_write.scs} & \texttt{testbenches} & Crossbar interconnect wire parasitic loading and worst-case diagonal IR-drop evaluation \\
        \texttt{tb\_noise\_validation.scs} & \texttt{testbenches} & Stochastic Monte Carlo circuit simulation incorporating extracted hardware noise \\
        \midrule
        \multicolumn{3}{@{}l}{\textit{\textbf{2. Cadence Ocean Automated Waveform Scripts (\texttt{ocean/})}}} \\
        \texttt{tb\_dynamic.ocn} & \texttt{ocean} & Automated 6-strip waveform extraction and display in Cadence ViVA ($V, n, f_{22}, G, I, T$) \\
        \texttt{tb\_dynamic\_extended.ocn} & \texttt{ocean} & Multi-strip visualization of extended $60,000$-pulse cycling and saturation dynamics \\
        \texttt{tb\_standard\_cycle.ocn} & \texttt{ocean} & Automated 4-strip plotting of standard potentiation/depression calibration waveforms \\
        \texttt{tb\_stress.ocn} & \texttt{ocean} & Multi-strip waveform extraction under extreme pulse overdrive conditions \\
        \texttt{tb\_thermal\_relaxation.ocn} & \texttt{ocean} & High-resolution thermal transient and exponential cooling extraction \\
        \texttt{tb\_iv\_sweep.ocn} & \texttt{ocean} & $I$--$V$ hysteretic curve tracing and switching threshold extraction in ViVA \\
        \midrule
        \multicolumn{3}{@{}l}{\textit{\textbf{3. Core Model Definitions and Hierarchy (\texttt{models/})}}} \\
        \texttt{molmem\_va.va} & \texttt{models} & Continuous $C^1$ Verilog-A compact model with softplus thresholds and Hermite clamping \\
        \texttt{molmem.scs} & \texttt{models} & Spectre model library definition and diagnostic subcircuit definitions (\texttt{molmem\_device\_mon}) \\
        \texttt{crossbar\_8x8.scs} & \texttt{models} & Hierarchical subcircuit netlist instantiating 64 memristor crosspoints with parasitics \\
        \bottomrule
    \end{tabularx}
\end{table}

\section{Device Fabrication, Coordination Chemistry, and Counterion Physics}
\label{sec:noteS3}

\subsection{Coordination Film Stack and Crossbar Fabrication Protocol}
The physical devices modeled throughout this work correspond to the vertical metal-molecule-metal crossbar junctions reported by Sharma et al.~\cite{si:sharma2024} and analyzed microscopically by Kundu and Goswami~\cite{si:kundu2024} and Gaur et al.~\cite{si:gaur2024}. The vertically stacked architecture comprises three primary physical layers fabricated on thermally oxidized silicon ($\mathrm{Si/SiO_2}$) substrates:
\begin{enumerate}
    \item \textit{Bottom Electrode (Wordline):} A $40\ \mathrm{nm}$ thick gold ($\mathrm{Au}$) bottom contact (with a thin titanium or chromium adhesion underlayer) patterned via electron-beam or optical lithography and thermal evaporation, providing an atomically flat, chemically inert polycrystalline work-function reference ($\Phi_{\mathrm{Au}} \approx 5.1\ \mathrm{eV}$).
    \item \textit{Active Molecular Coordination Film:} A uniform $60\ \mathrm{nm}$ thin film of the cationic bis-tridentate ruthenium complex $[\mathrm{Ru^{II}L_2}](\mathrm{BF_4})_2$ (where $\mathrm{L}$ denotes the redox-active azo-aromatic ligand $2\text{-(phenylazo)pyridine}$ or its tridentate $2,6\text{-bis(phenylazo)pyridine}$ derivative) deposited from filtered acetonitrile solution via controlled spin-coating followed by vacuum drying~\cite{si:goswami2017,si:goswami2020,si:goswami2021,si:sharma2024}. Given a bimolecular coordination spacing of $\Delta z \approx 2.0\ \mathrm{nm}$ along the vertical conduction axis, the $60\ \mathrm{nm}$ film thickness corresponds to a discrete chain of $L = 30$ molecular layers between the electrodes~\cite{si:kundu2024}. Within the coordination lattice, each divalent $[\mathrm{Ru^{II}L_2}]^{2+}$ complex is charge-balanced by two mobile monovalent tetrafluoroborate ($\mathrm{BF_4^-}$) counterions occupying interstitial electrostatic binding pockets~\cite{si:rath2022,si:kundu2024}. For a nanoscale crosspoint of $100\ \mathrm{nm} \times 100\ \mathrm{nm}$ active area, the junction volume ($6 \times 10^{-22}\ \mathrm{m^3}$) encloses approximately $N_{\mathrm{mol}} \approx 10^4$ ruthenium complexes and $2 \times 10^4$ mobile $\mathrm{BF_4^-}$ anions; in micron-scale benchmark crossbars ($1.05\ \mathrm{\mu m} \times 1.1\ \mathrm{\mu m}$ active area~\cite{si:sharma2024}), the active volume encloses $\approx 10^6$ molecular complexes operating in parallel.
    \item \textit{Top Electrode (Bitline):} A perpendicular $40\ \mathrm{nm}$ $\mathrm{Au}$ top electrode deposited via low-energy cooled evaporation techniques to prevent metal atom penetration or thermal degradation of the underlying coordination film~\cite{si:goswami2017,si:goswami2021,si:sharma2024}.
\end{enumerate}

\subsection{Electroforming-Free Operation and Endurance Physics}
A critical distinction between $[\mathrm{Ru^{II}L_2}](\mathrm{BF_4})_2$ molecular memristors and conventional inorganic oxide RRAM ($\mathrm{HfO_2}$, $\mathrm{TaO_x}$, $\mathrm{TiO_2}$) is the complete absence of an electroforming step~\cite{si:sharma2024,si:williams2024}. In filamentary oxides, pristine junctions are stoichiometric insulators requiring an initial high-voltage dielectric breakdown pulse ($3\ \mathrm{V}$ to $8\ \mathrm{V}$) to create a localized percolation path of oxygen vacancies; this stochastic breakdown permanently damages the host lattice and introduces wide device-to-device resistance dispersion~\cite{si:xia2019,si:jiang2016}. In contrast, the as-deposited $[\mathrm{Ru^{II}L_2}](\mathrm{BF_4})_2$ coordination thin film settles natively into its thermodynamic ground state: the symmetry-broken, charge-disproportionate $(31)$ electronic configuration, in which adjacent molecular complexes carry asymmetric ligand electron populations (three versus one electrons in the ligand-centered $\pi^*$ (pi-antibonding) molecular orbitals while the metal center remains low-spin $4d^6\ \mathrm{Ru^{II}}$)~\cite{si:goswami2020,si:sharma2024,si:kundu2024,si:gaur2024}.

Because the unprogrammed junction naturally resides in this stable, charge-disproportionate $(31)$ thermodynamic equilibrium, it completely bypasses the need for an electrical electroforming process. Instead, as-fabricated baseline $[\mathrm{Ru^{II}L_2}](\mathrm{BF_4})_2$ ($\mathrm{BF_4^-}$) crosspoints respond immediately and deterministically to calibrated sub-$100\ \mathrm{ns}$ programming pulses ($+0.90\ \mathrm{V}$, $80\ \mathrm{ns}$ potentiation; $-0.75\ \mathrm{V}$, $65\ \mathrm{ns}$ depression) from the initial write cycle, promoting uniform yield and predictable conductance across multi-terminal crossbars~\cite{si:sharma2024,si:kundu2024}. Because state modulation occurs through localized, charge-conserving intermolecular redox accompanied by sub-angstrom $\mathrm{BF_4^-}$ counterion displacement within fixed coordination pockets (without atomic filament rupture, melt-quench amorphization, or electrode migration), the coordination framework remains structurally intact, supporting cycling endurance exceeding $10^{12}$ write operations without resistance window degradation~\cite{si:sharma2024,si:williams2024}.

\subsection{\texorpdfstring{Counterion Modulation Physics: $\mathrm{BF_4^-}$, $\mathrm{PF_6^-}$, and $\mathrm{BPh_4^-}$}{Counterion Modulation Physics: BF4-, PF6-, and BPh4-}}
While intermolecular electron transfer between adjacent azo-aromatic ligand-centered $\pi^*$ (pi-antibonding) molecular orbitals occurs on picosecond timescales, the non-volatile retention, threshold voltage $V_{\mathrm{th}}$, and pulse-by-pulse linearity of the molecular memristor are governed by the electrostatic binding energy and steric mobility of the counterion species (Table~\ref{tab:S3})~\cite{si:rath2022,si:kundu2024,si:gaur2024}. As illustrated in Figure~2b of the main manuscript, transitioning a molecular pair from the resting ground state ($31_{31}$, referenced at $0\ \mathrm{meV}$) to the unrelaxed valence-symmetric electronic state ($22_{31}$) requires an unrelaxed electronic Gibbs free energy difference of $\Delta G^0 = 0.10\ \mathrm{eV}$ ($+100\ \mathrm{meV}$) with an intramolecular reorganization energy of $E_\lambda = 0.04\ \mathrm{eV}$ ($40\ \mathrm{meV}$)~\cite{si:kundu2024}. In the baseline $\mathrm{BF_4^-}$ system ($r = 2.18\ \text{\AA}$), following electron transfer, the surrounding $\mathrm{BF_4^-}$ counterions surmount a zero-bias microscopic activation barrier of $E_{a,\mathrm{micro}} = 130\ \mathrm{meV}$ ($134\ \mathrm{meV}$ experimental~\cite{si:sharma2024,si:kundu2024}) to displace into reconfigured Coulombic coordination pockets ($22_{31} \to 22_{22}$), releasing $\approx 75\ \mathrm{meV}$ of $\mathrm{BF_4^-}$-specific electrostatic lattice stabilization energy to anchor the switched $22_{22}$ state at a net metastable elevation of $+25\ \mathrm{meV}$ above the ground state~\cite{si:goswami2020,si:rath2022,si:sharma2024,si:kundu2024}.

By substituting the counterion during chemical synthesis while keeping the cationic $[\mathrm{Ru^{II}L_2}]^{2+}$ backbone invariant, the height of the intervening activation barrier $E_a$ along reaction coordinate $\xi$ and the spatial packing density at the electrode interface can be systematically tuned across three distinct physical regimes~\cite{si:williams2024,si:kundu2024,si:gaur2024}:
\begin{enumerate}
    \item \textit{Baseline Tetrafluoroborate ($\mathrm{BF_4^-}$, Ionic Radius $r = 2.18\ \text{\AA}$, $E_{a,\mathrm{micro}} = 130\ \mathrm{meV}$):} The compact tetrahedral $\mathrm{BF_4^-}$ anion establishes an optimal balance between electrostatic coordination binding and interstitial mobility. Its microscopic zero-bias barrier $E_{a,\mathrm{micro}} = 130\ \mathrm{meV}$ ($134\ \mathrm{meV}$ experimental~\cite{si:sharma2024,si:kundu2024}), corresponding to an extracted macroscopic Arrhenius parameter of $E_{a,\mathrm{macro}} = 155.9\ \mathrm{meV}$ [$0.1559\ \mathrm{eV}$] in Table~\ref{tab:S1}, is more than five times larger than room-temperature thermal energy ($k_B T \approx 25.7\ \mathrm{meV}$ at $298.15\ \mathrm{K}$), preventing spontaneous thermal back-relaxation and providing non-destructive read immunity at biases up to $V_{\mathrm{read}} \le 0.20\ \mathrm{V}$ (protected by the compact threshold $V_{\mathrm{th}} = 0.4664\ \mathrm{V}$). Simultaneously, under $+0.90\ \mathrm{V}$ ($80\ \mathrm{ns}$) write pulses, electric-field lowering of the $130\ \mathrm{meV}$ barrier displaces a constant ensemble fraction of $\mathrm{BF_4^-}$ ions per pulse, advancing the conductive $(22)$ phase front at a constant linear rate of $p_1 = 534.2\ \mathrm{pulses/layer}$ across all 30 molecular layers ($16,520$ states, kinetic symmetry ratio $r_{\mathrm{asym}} = 0.9857$).
    \item \textit{Hexafluorophosphate ($\mathrm{PF_6^-}$, Ionic Radius $r = 2.54\ \text{\AA}$, $E_{a,\mathrm{micro}} = 95\ \mathrm{meV}$):} Replacing $\mathrm{BF_4^-}$ with the larger octahedral $\mathrm{PF_6^-}$ anion increases the cation-anion center-to-center equilibrium separation, weakening the local Coulombic binding pocket and lowering the microscopic activation barrier to $E_{a,\mathrm{micro}} = 95\ \mathrm{meV}$ (corresponding to an extracted macroscopic Arrhenius parameter of $E_{a,\mathrm{macro}} = 73.3\ \mathrm{meV}$ [$0.0733\ \mathrm{eV}$] in Table~\ref{tab:S1})~\cite{si:kundu2024,si:gaur2024}. Physically, $E_{a,\mathrm{micro}} = 95\ \mathrm{meV}$ represents the theoretical single-counterion microscopic hopping barrier from DFT and Marcus theory~\cite{si:kundu2024}, whereas $E_{a,\mathrm{macro}} = 73.3\ \mathrm{meV}$ [$0.0733\ \mathrm{eV}$] represents the effective macroscopic Arrhenius parameter extracted across the continuum device under coupled self-heating. Because the reduced $95\ \mathrm{meV}$ barrier is more readily surmounted under applied electric fields, $\mathrm{PF_6^-}$ devices exhibit accelerated domain nucleation ($\kappa_{\mathrm{switch}} = 1.755 \times 10^7\ \mathrm{s^{-1}}$ vs.\ $1.173 \times 10^7\ \mathrm{s^{-1}}$ for $\mathrm{BF_4^-}$), producing higher peak conductance ($\approx 6.58\ \mathrm{mS}$) accompanied by a two-slope potentiation trajectory. Within the compact model, these steeper switching kinetics are parameterized by an increased voltage nonlinearity exponent ($\alpha = 4.863$, $\beta_\alpha = 1.588$) and an elevated effective thermal coupling coefficient ($R_{\mathrm{th}} = 9,994.7\ \mathrm{K/W}$), capturing the enhanced field and self-heating sensitivity characteristic of the lower-barrier $\mathrm{PF_6^-}$ system across the empirical characterization range (detailed in Tables~\ref{tab:S1} and \ref{tab:S3}).
    \item \textit{Tetraphenylborate ($\mathrm{BPh_4^-}$, Ionic Radius $r = 4.21\ \text{\AA}$, $E_{a,\mathrm{micro}} = 210\ \mathrm{meV}$):} Introducing the bulky, sterically hindered $\mathrm{BPh_4^-}$ anion nearly doubles the ionic radius to $4.21\ \text{\AA}$, increasing the effective steric activation barrier to $E_{a,\mathrm{micro}} = 210\ \mathrm{meV}$ (Table~\ref{tab:S3})~\cite{si:williams2024,si:kundu2024,si:gaur2024}. The large excluded volume of the four phenyl rings restricts interstitial counterion translation through the coordination lattice, suppressing sub-$100\ \mathrm{ns}$ incremental analog front propagation and shifting switching toward high-voltage ($> 2.0\ \mathrm{V}$) quasi-static digital transitions~\cite{si:kundu2024,si:gaur2024}. Consequently, dynamic pulse-programming datasets are physically absent in the literature. In our comparative framework, $\mathrm{BPh_4^-}$ serves as an essential crystallographic and thermodynamic literature benchmark (from Gaur et al.~\cite{si:gaur2024} and Kundu and Goswami~\cite{si:kundu2024}; Table~\ref{tab:S3}) establishing the steric boundary where analog operation is suppressed.
\end{enumerate}

\section{Microscopic Quantum Chemical and Electrostatic Transport Formulations}
\label{sec:noteS4}

In this note, we present the complete mathematical derivation of the microscopic 30-layer physical model developed by Kundu and Goswami~\cite{si:kundu2024} for the baseline $[\mathrm{Ru^{II}L_2}](\mathrm{BF_4})_2$ ($\mathrm{BF_4^-}$) coordination system and implemented in our microscopic physics extraction pipeline, which serves as the ground-truth physical foundation for our analytical surrogates (Note~S5) and dual circuit simulation engines (Notes~S6--S8).

\subsection{\texorpdfstring{Thomas-Fermi Electrostatic Potential Profile $\Phi_{\mathrm{mol}}(z,n)$}{Thomas-Fermi Electrostatic Potential Profile (z,n)}}
Consider a baseline $[\mathrm{Ru^{II}L_2}](\mathrm{BF_4})_2$ molecular thin film of total thickness $d = 60\ \mathrm{nm}$ discretized into $L = 30$ bimolecular coordination planes indexed by $z \in \{1, 2, \dots, 30\}$ (with normalized depth coordinate $t = z/L \in [0, 1]$, distinct from physical time $t$ in the rate equations below, and quasi-stationary pulse state index $n \in [0, 16,520]$). When an external programming voltage $V_{\mathrm{applied}}$ is applied across the gold electrodes, mobile tetrafluoroborate counterions ($\mathrm{BF_4^-}$) drift and redistribute across the $L$ layers to screen the applied electric field. Following the electrostatic molecular conductor theory of Liang et al.~\cite{si:liang2004} and Nitzan et al.~\cite{si:nitzan2002} adapted for counterion-mediated coordination films~\cite{si:kundu2024}, the one-dimensional Poisson equation coupled to linearized Thomas-Fermi ionic screening along depth $z$ is governed by:
\begin{equation}
    \frac{\partial^2 \Phi_{\mathrm{mol}}(z,n)}{\partial z^2} - k_{\mathrm{TF}}^2(z,n)\left[\Phi_{\mathrm{mol}}(z,n) - \Phi_{\mathrm{linear}}(z)\right] = 0
    \label{eq:S1_poisson}
\end{equation}
where $\Phi_{\mathrm{linear}}(z) = \Phi_{\max}(z/L)$ is the unscreened Laplace ramp potential ($\Phi_{\max} = 0.70\ \mathrm{V}$, corresponding to the $0.70\text{--}0.72\ \mathrm{V}$ peak internal molecular potential drop in the baseline $\mathrm{BF_4^-}$ junction under a nominal $+0.90\ \mathrm{V}$ write pulse after subtracting interfacial contact drops of $0.11\ \mathrm{V}$ and $0.07\ \mathrm{V}$~\cite{si:sharma2024,si:kundu2024}) and $k_{\mathrm{TF}}(z,n)$ is the local inverse Thomas-Fermi screening length determined by the displaced $\mathrm{BF_4^-}$ counterion density at pulse count $n$.

Expanding the spatial deviation from the linear ramp in a half-range Fourier sine series that satisfies Dirichlet boundary conditions at the electrodes ($\Phi_{\mathrm{mol}}(0,n) = 0$ and $\Phi_{\mathrm{mol}}(L,n) = \Phi_{\max}$) yields the exact infinite series representation~\cite{si:kundu2024,si:liang2004}:
\begin{equation}
    \Phi_{\mathrm{mol}}(z,n) = \Phi_{\max}\left[\frac{z}{L} - \sum_{m=1}^{\infty} \frac{2 F_m(n)}{m\pi\left(1 + F_m(n)\right)} \sin\left(\frac{m\pi(L - z)}{L}\right)\right]
    \label{eq:S2_tf_series}
\end{equation}
where $m \in \{1, 2, 3, \dots\}$ is the spatial harmonic index and $F_m(n)$ is the dimensionless $m$-th mode interfacial electrostatic screening factor:
\begin{equation}
    F_m(n) = \frac{e^2 N_{\mathrm{ion}}(n) L^2}{m^2 \pi^2 \varepsilon_0 \varepsilon_r k_B T_{\mathrm{eff}}} \cdot \mathcal{S}_{\mathrm{pack}}(m,n)
    \label{eq:S3_Fm}
\end{equation}
Here, $e$ is the elementary charge, $N_{\mathrm{ion}}(n)$ is the active displaced $\mathrm{BF_4^-}$ counterion concentration at pulse $n$ (defining the state-dependent dimensionless screening parameter $\zeta(n) = (k_{\mathrm{TF}}(n)L)^{-1} = b_1 n + b_2 + c_1\exp(c_2 n)$~\cite{si:sharma2024,si:kundu2024}, where $b_1$ (per pulse) and the dimensionless offset $b_2$ describe the linear growth of the screening length as counterions are progressively displaced, while the amplitude $c_1$ (dimensionless) and rate $c_2$ (per pulse) capture the exponential near-electrode correction at early pulse counts; all four coefficients are fitted to the microscopic profiles of Kundu and Goswami~\cite{si:kundu2024}), $\varepsilon_0 \varepsilon_r$ is the effective permittivity of the coordination film ($\varepsilon_r \approx 4.2$), and $\mathcal{S}_{\mathrm{pack}}(m,n) = w_m e^{w_m} E_1(w_m)$ (with $w_m = \frac{1}{2}(2\pi m\sigma)^2$, where $\sigma$ is the dimensionless counterion spatial spread normalized to the film thickness $L$, and exponential integral $E_1(w_m) = \int_{w_m}^{\infty} u^{-1}e^{-u}\,du$~\cite{si:liang2004,si:kundu2024}) accounts for the finite excluded volume of $\mathrm{BF_4^-}$ counterions ($r = 2.18\ \text{\AA}$), which limits maximum ionic packing density near the counter-electrode ($z \to L$) and prevents unphysical potential inversion. In the pristine ground state ($n = 0$), strong bulk screening ($F_m \gg 1$) confines the electrostatic potential drop closely to the electrode interface, leaving the interior layers nearly equipotential; as sequential write pulses ($n \to 16,520$) convert layers into the $(22)$ state and redistribute counterions across the film, $\Phi_{\mathrm{mol}}(z,n)$ progressively penetrates into the bulk layer by layer (Figure~\ref{fig:S1}a).

\subsection{\texorpdfstring{Marcus-Hush-Chidsey Electron Transfer Rate $k_{\mathrm{ET}}(z,n)$}{Marcus-Hush-Chidsey Electron Transfer Rate kET(z,n)}}
Within each molecular layer $z$ of the baseline $[\mathrm{Ru^{II}L_2}](\mathrm{BF_4})_2$ ($\mathrm{BF_4^-}$) film, the rate of intermolecular redox conversion from the symmetry-broken ground state $(31)$ to the valence-symmetric conductive state $(22)$ is governed by non-adiabatic electron transfer across a continuous manifold of electronic states~\cite{si:kundu2024,si:marcus1965,si:migliore2012}. Because the azo-aromatic ligand $\pi^*$ orbitals hybridize weakly with the thermally broadened Fermi-Dirac distribution of acceptor states $\psi(E)$, the forward $(31) \to (22)$ electron transfer rate constant $k_{\mathrm{ET}}(z,n)$ at layer $z$ and pulse $n$ is given by the Marcus-Hush-Chidsey (MHC) integral~\cite{si:kundu2024,si:migliore2012}:
\begin{equation}
    k_{\mathrm{ET}}(z,n) = \gamma_{\mathrm{DA}} \sqrt{\frac{k_B T}{4\pi E_\lambda}} \int_{-\infty}^{+\infty} \psi(x) \exp\left[-\frac{\left(E_\lambda + \Delta G^0 + e\Phi_{\mathrm{mol}}(z,n) - x k_B T\right)^2}{4 E_\lambda k_B T}\right] dx
    \label{eq:S4_mhc}
\end{equation}
where $x = (E - E_F)/(k_B T)$ is the dimensionless energy relative to the local quasi-Fermi level $E_F$ (with Jacobian $dE = k_B T\,dx$), $\gamma_{\mathrm{DA}} = (2\pi/\hbar)H_{\mathrm{DA}}^2$ is the donor-acceptor electronic coupling pre-factor ($\mathrm{s^{-1}}$), $E_\lambda = 0.04\ \mathrm{eV}$ ($40\ \mathrm{meV}$) is the inner-sphere/outer-sphere reorganization energy, $\Delta G^0 = 0.10\ \mathrm{eV}$ ($100\ \mathrm{meV}$) is the unrelaxed zero-field Gibbs free energy difference between $31_{31}$ and $22_{31}$~\cite{si:sharma2024,si:kundu2024}, and $\psi(x)$ is the Fermi-Dirac occupancy distribution:
\begin{equation}
    \psi(x) = \frac{1}{1 + \exp(x)}
    \label{eq:S5_fermi}
\end{equation}
Here, $\Phi_{\mathrm{mol}}(z,n) \in [0, \Phi_{\max}]$ is referenced in the residual barrier gauge ($\Phi_{\mathrm{mol}} = \Phi_{\max} - \Phi_{\mathrm{drive}}$, mapping the $-e\Phi_{\mathrm{drive}}$ overdrive convention of Kundu and Goswami~\cite{si:kundu2024} onto $+e\Phi_{\mathrm{mol}}$): when layer $z$ experiences a large unscreened barrier potential $\Phi_{\mathrm{mol}}(z,n)$, the effective activation argument in Equation~(\ref{eq:S4_mhc}) remains large and $k_{\mathrm{ET}}(z,n)$ is exponentially suppressed. As $\mathrm{BF_4^-}$ counterions displace under positive write pulses and lower $\Phi_{\mathrm{mol}}(z,n) \to 0$ at the advancing phase boundary, $k_{\mathrm{ET}}(z,n)$ increases exponentially up to $A_{\mathrm{ET}} \approx 1.135 \times 10^3\ \mathrm{s^{-1}}$ (Figure~\ref{fig:S1}b).

\subsection{\texorpdfstring{Cauchy-Lorentz Planar Domain Nucleation $\mathit{Nr}_{xy}(z,n)$ and Volumetric Fraction $f_{22}(n)$}{Cauchy-Lorentz Planar Domain Nucleation Nrxy(z,n) and Volumetric Fraction f22(n)}}
In the baseline $[\mathrm{Ru^{II}L_2}](\mathrm{BF_4})_2$ ($\mathrm{BF_4^-}$) coordination film, redox conversion within layer $z$ becomes thermodynamically active when the dimensionless electron transfer probability $P_{\mathrm{ET}}(z,n) = k_{\mathrm{ET}}(z,n)/\gamma_{\mathrm{DA}}$ exceeds the critical probability threshold $\varepsilon$, or equivalently when the dimensional transfer rate $k_{\mathrm{ET}}(z,n)$ exceeds the critical switching rate $k_{\mathrm{crit}} = \varepsilon\gamma_{\mathrm{DA}} = 0.30\ \mathrm{s^{-1}}$ across the full-range MHC curve in Figure~\ref{fig:S1}b ($\Phi_{\mathrm{mol,crit}} \approx 0.23\ \mathrm{V}$ on the full-span semi-logarithmic slope $B_{\mathrm{ET,full}} = 35.97\ \mathrm{V^{-1}}$, while the effective active-window slope $B_{\mathrm{ET}} = 7.12\ \mathrm{V^{-1}}$ of Note~S5 governs the local sensitivity around this threshold)~\cite{si:kundu2024}. Defining the layer activation indicator via the Heaviside step function $\Theta(\cdot)$:
\begin{equation}
    \theta(z,n) = \Theta\left(P_{\mathrm{ET}}(z,n) - \varepsilon\right) = \Theta\left(k_{\mathrm{ET}}(z,n) - k_{\mathrm{crit}}\right) \in \{0, 1\}
    \label{eq:S6_heaviside}
\end{equation}
the cumulative number of active switched layers along the vertical $z$-axis at pulse count $n$ defines the planar switched front index $Z_{22}(n)$:
\begin{equation}
    Z_{22}(n) = \sum_{z=1}^{L} \theta(z,n), \qquad Z_{22}(n) \in [0, 30]
    \label{eq:S7_Z22}
\end{equation}
Within each transverse $x$--$y$ plane at depth $z$, local variations in $\mathrm{BF_4^-}$ counterion coordination environments produce a Lorentzian distribution of local nucleation thresholds around the layer-mean nucleation pulse count $\chi(z)$~\cite{si:kundu2024,si:jo2007,si:tagantsev2002}:
\begin{equation}
    \rho_{\mathrm{nuc}}(n; z) = \frac{1}{\pi}\frac{\kappa_{\mathrm{nuc}}}{\left(n - \chi(z)\right)^2 + \kappa_{\mathrm{nuc}}^2}
    \label{eq:S8_lorentz_pdf}
\end{equation}
where $\kappa_{\mathrm{nuc}}$ ($\kappa_{\mathrm{nuc}} = 800\ \mathrm{pulses}$ during potentiation and $500\ \mathrm{pulses}$ during depression for baseline $\mathrm{BF_4^-}$ in Table~\ref{tab:S1}) characterizes the half-width at half-maximum of the in-plane nucleation barrier distribution. Integrating Equation~(\ref{eq:S8_lorentz_pdf}) over pulse history from $-\infty$ to $n$ yields the closed-form Cauchy-Lorentz cumulative distribution for the in-plane switched domain fraction $\mathit{Nr}_{xy}(z,n)$ (Equation~1 of the main manuscript):
\begin{equation}
    \mathit{Nr}_{xy}(z,n) = \int_{-\infty}^{n} \rho_{\mathrm{nuc}}(n'; z)\,dn' = \frac{1}{\pi}\arctan\left(\frac{n - \chi(z)}{\kappa_{\mathrm{nuc}}}\right) + \frac{1}{2}
    \label{eq:S9_nrxy}
\end{equation}
Here, the layer-dependent nucleation center $\chi(z)$ follows the linear planar front progression law:
\begin{equation}
    \chi(z) = a_1 z + a_2 \equiv p_1 z + p_2
    \label{eq:S10_chi}
\end{equation}
with calibrated $\mathrm{BF_4^-}$ slope $p_1 = 534.2\ \mathrm{pulses/layer}$ ($267.1\ \mathrm{pulses/nm}$ across the $60\ \mathrm{nm}$ film) and initiation offset $p_2 = 15.1\ \mathrm{pulses}$. Averaging $\mathit{Nr}_{xy}(z,n)$ across all $L = 30$ molecular layers yields the macroscopic volumetric switched fraction $f_{22}(n)$ (Equation~2 of the main manuscript):
\begin{equation}
    f_{22}(n) = \frac{1}{L}\sum_{z=1}^{L} \mathit{Nr}_{xy}(z,n) = \frac{1}{30}\sum_{z=1}^{30}\left[\frac{1}{\pi}\arctan\left(\frac{n - (p_1 z + p_2)}{\kappa_{\mathrm{nuc}}}\right) + \frac{1}{2}\right], \qquad f_{31}(n) = 1 - f_{22}(n)
    \label{eq:S11_f22}
\end{equation}
In our calibrated baseline $\mathrm{BF_4^-}$ compact model (Table~\ref{tab:S1}), $f_{22}(n)$ spans from a resting ground-state baseline of $f_{22,\mathrm{start}} = 0.12037$ at $n = 0$ to a peak valence-symmetric conversion of $f_{22,\mathrm{peak}} = 0.87792$ at $n = 16,520$, returning symmetrically to $f_{22,\mathrm{end}} = 0.13955$ at $n = 33,040$.

\subsection{30-Site Discrete Tridiagonal Hopping Master Equations and Stationary Current}
To evaluate the non-destructive read current $I(n, V_r)$ of the baseline $[\mathrm{Ru^{II}L_2}](\mathrm{BF_4})_2$ ($\mathrm{BF_4^-}$) device at any programmed state $n$ under a sub-threshold read voltage $V_r$ ($|V_r| \le 0.20\ \mathrm{V} < V_{\mathrm{th}} = 0.4664\ \mathrm{V}$), charge transport through the $L = 30$ molecular layers is modeled as sequential incoherent carrier hopping across a one-dimensional chain of $N_s = 30$ redox sites coupled to metallic Fermi reservoirs at the left ($i = 1$, bottom $\mathrm{Au}$) and right ($i = 30$, top $\mathrm{Au}$) contacts~\cite{si:kundu2024,si:livshits2014,si:lehmann2002}. Let $P_i(n) \in [0, 1]$ denote the ensemble carrier occupation probability at molecular layer $i \in \{1, 2, \dots, 30\}$. For the baseline $\mathrm{BF_4^-}$ system, the effective intermolecular hopping coupling $\gamma_m(n)$ is a linear volumetric mixture of the resting $(31)$ coupling $\gamma_{m,31} = 1.0 \times 10^3\ \mathrm{s^{-1}}$ and the delocalized $(22)$ coupling $\gamma_{m,22} = 48.0 \times 10^3\ \mathrm{s^{-1}}$ (normalized per conduction channel, scaling to ensemble rates of $10^8$ to $2 \times 10^{13}\ \mathrm{s^{-1}}$ across macroscopic junctions~\cite{si:kundu2024}):
\begin{equation}
    \gamma_m(n) = \left(1 - f_{22}(n)\right)\gamma_{m,31} + f_{22}(n)\gamma_{m,22}
    \label{eq:S12_gamma_m}
\end{equation}
Under an applied read bias $V_r$, the molecular voltage drop per inter-site hopping barrier is $a_m V_r$ with $a_m = (1 - a_L - a_R)/30 \approx 1/30$ in the low-drop interface regime~\cite{si:kundu2024}, yielding the directional forward ($i \to i+1$) and backward ($i+1 \to i$) inter-site hopping rate constants:
\begin{equation}
    K_{m,f}(n, V_r) = \gamma_m(n)\exp\left(+\frac{e a_m V_r}{\eta k_B T}\right), \qquad K_{m,b}(n, V_r) = \gamma_m(n)\exp\left(-\frac{e a_m V_r}{\eta k_B T}\right)
    \label{eq:S13_Kmf}
\end{equation}
where $\eta$ is the barrier ideality factor. At the left electrode interface ($i = 1$) and right electrode interface ($i = 30$), carrier injection ($K_{L,f}, K_{R,b}$) and extraction ($K_{L,b}, K_{R,f}$) rates are governed by interfacial electrode-molecule coupling $\gamma_L, \gamma_R$ (representing the low-bias Butler-Volmer evaluation of the interfacial Marcus integrals with state-dependent couplings $\Gamma_{B}(n), \Gamma_{T}(n)$ in Kundu and Goswami~\cite{si:kundu2024}) weighted by the contact voltage drops $a_L V_r$ and $a_R V_r$ (distinct from the smoothstep taper functions $a_L(t), a_R(t)$ in Note~S5):
\begin{equation}
    K_{L,f/b}(V_r) = \gamma_L \exp\left(\pm\frac{e a_L V_r}{\eta_c k_B T}\right), \qquad K_{R,f/b}(V_r) = \gamma_R \exp\left(\pm\frac{e a_R V_r}{\eta_c k_B T}\right)
    \label{eq:S14_KL_KR}
\end{equation}
Because each $[\mathrm{Ru^{II}L_2}]^{2+}$ complex provides four closely spaced ligand $\pi^*$ ($\mathrm{LUMO}+x$) hopping orbitals and the quadratic Pauli blocking cross-term $-(K_{m,f} - K_{m,b})P_i P_{i+1}$ vanishes to leading order at sub-threshold read biases ($e a_m V_r \ll \eta k_B T$)~\cite{si:kundu2024,si:livshits2014,si:lehmann2002}, the stationary site occupation probabilities $P_i(n)$ across the 30 layers satisfy the coupled master system (with $t$ here denoting physical time):
\begin{align}
    \frac{dP_1}{dt} &= K_{L,f}(1 - P_1) - K_{L,b}P_1 - K_{m,f}(n)P_1 + K_{m,b}(n)P_2 = 0 \label{eq:S15_master1} \\
    \frac{dP_i}{dt} &= K_{m,f}(n)P_{i-1} - \left[K_{m,f}(n) + K_{m,b}(n)\right]P_i + K_{m,b}(n)P_{i+1} = 0, \qquad i = 2, \dots, 29 \label{eq:S16_masteri} \\
    \frac{dP_{30}}{dt} &= K_{m,f}(n)P_{29} - \left[K_{m,b}(n) + K_{R,f}\right]P_{30} + K_{R,b}(1 - P_{30}) = 0 \label{eq:S17_master30}
\end{align}
In matrix form, Equations~(\ref{eq:S15_master1})--(\ref{eq:S17_master30}) constitute a strictly diagonally dominant $30 \times 30$ tridiagonal linear system $\mathbf{M}(n, V_r)\mathbf{P}(n) = \mathbf{b}(V_r)$:
{\footnotesize
\begin{equation}
    \setlength{\arraycolsep}{2.2pt}
    \begin{bmatrix}
        -(K_{L,f} + K_{L,b} + K_{m,f}) & K_{m,b} & 0 & \cdots & 0 \\
        K_{m,f} & -(K_{m,f} + K_{m,b}) & K_{m,b} & \cdots & 0 \\
        0 & K_{m,f} & -(K_{m,f} + K_{m,b}) & \cdots & 0 \\
        \vdots & \vdots & \vdots & \ddots & K_{m,b} \\
        0 & 0 & 0 & K_{m,f} & -(K_{m,b} + K_{R,f} + K_{R,b})
    \end{bmatrix}
    \begin{bmatrix}
        P_1 \\ P_2 \\ P_3 \\ \vdots \\ P_{30}
    \end{bmatrix}
    =
    \begin{bmatrix}
        -K_{L,f} \\ 0 \\ 0 \\ \vdots \\ -K_{R,b}
    \end{bmatrix}
    \label{eq:S18_tridiag}
\end{equation}
}%
Solving Equation~(\ref{eq:S18_tridiag}) via the $O(L)$ Thomas tridiagonal algorithm (evaluating local layer couplings $\gamma_{m,i}(n) = (1 - \mathit{Nr}_{xy}(i,n))\gamma_{m,31} + \mathit{Nr}_{xy}(i,n)\gamma_{m,22}$ to resolve the spatial carrier accumulation wavefronts $P_i(n)$ plotted in Figure~2d of the main manuscript, or using the layer-averaged coupling $\gamma_m(n)$ of Equation~(\ref{eq:S12_gamma_m}) for homogenized chain transport~\cite{si:kundu2024}) yields the stationary occupation profile across all 30 layers. At zero read bias ($V_r = 0$, where $K_{m,f} = K_{m,b} = \gamma_m(n)$ and $K_{L/R,f} = K_{L/R,b}$), Equation~(\ref{eq:S18_tridiag}) yields the exact equilibrium solution $P_1 = P_2 = \dots = P_{30} = 1/2$ with zero offset current. By current continuity across every link in the 30-site chain, the stationary macroscopic read current $I(n, V_r)$ across $N_{\mathrm{ch}}$ parallel molecular channels is evaluated identically at the contact or across any interior bond $(i, i+1)$:
{\small
\begin{equation}
    I(n, V_r) = e N_{\mathrm{ch}}\left[K_{m,f}(n, V_r)P_i(n) - K_{m,b}(n, V_r)P_{i+1}(n)\right] = e N_{\mathrm{ch}}\left[K_{L,f}(V_r)\left(1 - P_1(n)\right) - K_{L,b}(V_r)P_1(n)\right]
    \label{eq:S19_I_master}
\end{equation}
}%
In our continuum compact model (Note~S6, Equation~(\ref{eq:S32_iprod})), the macroscopic channel factor $e N_{\mathrm{ch}}$ together with the state-dependent interfacial injection transmission ($\Gamma_B(n), \Gamma_T(n)$~\cite{si:kundu2024}) and bulk hopping coupling $\gamma_m(n)$ are compiled into the calibrated current pre-factor $I_{\mathrm{prod}}(n) = I_{\mathrm{scale}}(n)\cdot g_{\mathrm{scale}}\cdot f_{22}(n)$, mapping the $48\times$ bulk intermolecular coupling ratio into the full four-order-of-magnitude ($200\ \mathrm{nS}$ to $5.90\ \mathrm{mS}$) device conductance window.

\section{Closed-Form Analytical Surrogates and Boundary Regularization}
\label{sec:noteS5}

While Note~S4 establishes the foundational microscopic quantum and electrostatic physics governing counterion screening and electron transfer, directly evaluating those infinite Fourier series, numerical quadrature integrals, and discrete 30-site matrix inversions within continuum modeling is mathematically intractable. This note provides the formal mathematical bridge between the microscopic physics of Note~S4 and the compact device model of Note~S6 for the baseline $[\mathrm{Ru^{II}L_2}](\mathrm{BF_4})_2$ ($\mathrm{BF_4^-}$) system. Here, we derive the closed-form analytical surrogates, interface-constrained rational transformations, and $C^1$-continuous boundary regularizations that convert the discrete microscopic transport equations into smooth, differentiable continuum functions without sacrificing physical fidelity (Figure~\ref{fig:S1}).

\subsection{\texorpdfstring{Interface-Constrained Rational Potential $u(t)$ and Smoothstep Edge Tapers}{Interface-Constrained Rational Potential u(t) and Smoothstep Edge Tapers}}
Let $t = z/L \in [0, 1]$ denote the normalized film depth coordinate across the $L = 30$ molecular layers of the baseline $[\mathrm{Ru^{II}L_2}](\mathrm{BF_4})_2$ ($\mathrm{BF_4^-}$) junction. To replace the infinite Thomas-Fermi Fourier sine series (Equation~(\ref{eq:S2_tf_series}) of Note~S4) without introducing Fourier truncation oscillations (Gibbs ringing) near the counter-electrode ($t \to 1$), we decompose the spatial variation of the normalized potential drop within a single electrostatic phase regime into two competing physical mechanisms: (i) power-law Thomas-Fermi screening penetration from the grounded electrode ($t = 0$) into the bulk, and (ii) finite-volume $\mathrm{BF_4^-}$ counterion packing saturation near the biased electrode ($t = 1$). We first capture the asymmetric power-law bulk penetration via a two-exponent generalized sigmoidal base function $s(t)$:
\begin{equation}
    s(t; p, q) = \left(1.0 - (1.0 - t)^q\right)^p, \qquad p > 0,\ q > 0
    \label{eq:S20_s_base}
\end{equation}
which strictly satisfies $s(0) = 0$ and $s(1) = 1$, where exponent $q$ controls the potential curvature approaching the counter-electrode ($t \to 1$) and exponent $p$ controls the screening suppression near $t \to 0$. To replace the infinite Fourier summation of Equation~(\ref{eq:S2_tf_series}) while capturing space-charge saturation by close-packed counterions and strictly enforcing the normalized Dirichlet boundary conditions at $t = 0$ and $t = 1$, $s(t)$ is mapped through an interface-constrained Pad\'{e} rational function $u(t)$ (Equation~6 of the main manuscript):
\begin{equation}
    u(t; p, q, k) = \frac{(1.0 + k)s(t; p, q)}{1.0 + k\left[s(t; p, q)\right]^2}, \qquad k \ge 0
    \label{eq:S21_u_rational}
\end{equation}
Checking the boundary limits confirms exact conservation for all values of the packing parameter $k$:
\begin{equation}
    \lim_{t \to 0^+} u(t) = \frac{(1+k)\cdot 0}{1 + k\cdot 0} = 0, \qquad \lim_{t \to 1^-} u(t) = \frac{(1+k)\cdot 1}{1 + k\cdot 1^2} = 1
    \label{eq:S22_u_limits}
\end{equation}
Moreover, differentiating Equation~(\ref{eq:S21_u_rational}) with respect to $s$ yields:
\begin{equation}
    \frac{\partial u}{\partial s} = \frac{(1.0 + k)\left(1.0 - k s^2\right)}{\left(1.0 + k s^2\right)^2}
    \label{eq:S23_du_ds}
\end{equation}
which indicates that for $0 \le k \le 1$, $\partial u/\partial s \ge 0$ across the entire normalized range $s \in [0, 1]$, so that $u$ is monotonically non-decreasing in $s$ (and hence in $t$, because $s(t)$ is itself monotonically increasing), while the field gradient at $s \to 1$ is compressed by the factor $(1-k)/(1+k)$, physically reproducing space-charge screening saturation by close-packed $\mathrm{BF_4^-}$ counterions ($r = 2.18\ \text{\AA}$).

\begin{figure}[!htb]
    \centering
    \includegraphics[width=\textwidth,height=0.62\textheight,keepaspectratio]{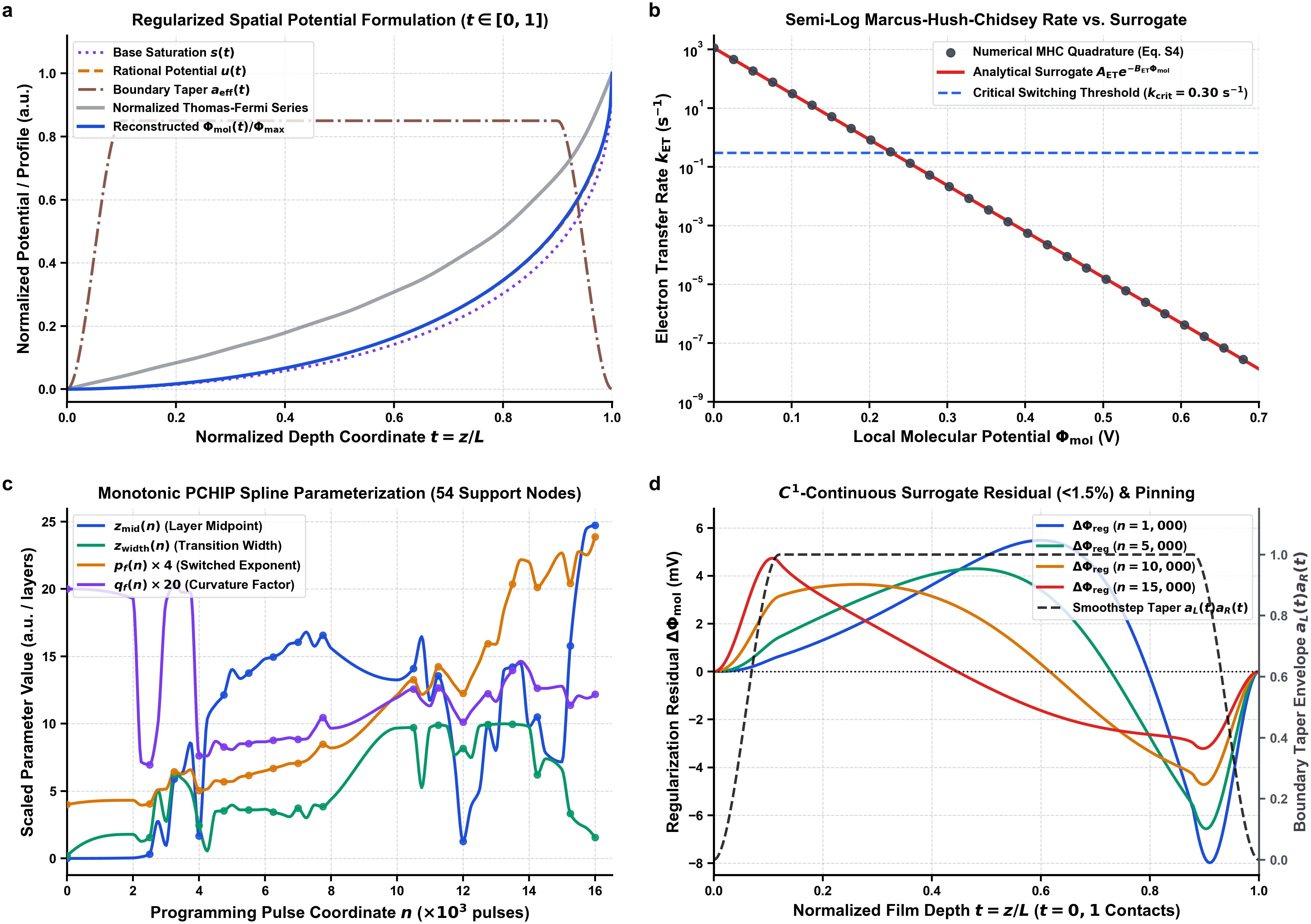}
    \caption{\textbf{Microscopic electrostatic potential regularization and closed-form analytical surrogates for the baseline $[\mathrm{Ru^{II}L_2}](\mathrm{BF_4})_2$ ($\mathrm{BF_4^-}$) molecular memristor.} \textbf{a}, Normalized spatial potential formulation across normalized depth coordinate $t = z/L \in [0, 1]$, displaying the sigmoidal base saturation $s(t)$, interface-constrained rational potential $u(t)$, smoothstep boundary taper envelope $a_{\mathrm{eff}}(t)$, normalized Thomas-Fermi Fourier series reference, and reconstructed analytical potential profile $\Phi_{\mathrm{mol}}(t)/\Phi_{\max}$ scaling cleanly to $1.0$. \textbf{b}, Semi-logarithmic comparison of numerical quadrature of the Marcus-Hush-Chidsey (MHC) Fermi-Dirac integral (Equation~(\ref{eq:S4_mhc}), grey circles) against the closed-form analytical exponential surrogate $k_{\mathrm{ET}}(\Phi_{\mathrm{mol}}) = A_{\mathrm{ET}}\exp(-B_{\mathrm{ET}}\Phi_{\mathrm{mol}})$ (red straight line on logarithmic axis, $\log_{10} k_{\mathrm{ET}} = \log_{10} A_{\mathrm{ET}} - (B_{\mathrm{ET}}/\ln 10)\Phi_{\mathrm{mol}}$) across more than eight orders of magnitude ($10^{-9}$ to $10^3\ \mathrm{s^{-1}}$) and the critical layer switching threshold $k_{\mathrm{crit}} = 0.30\ \mathrm{s^{-1}}$ (dashed blue line). \textbf{c}, Evolution of representative regularized spatial fitting parameters ($z_{\mathrm{mid}}, z_{\mathrm{width}}, p_f, q_f$) across the 14-bit programming window ($n = 0$ to $16,520$ pulses) across 54 calibration pulse coordinates (interpolation support nodes), illustrating parameter anchoring points and $C^1$-continuous monotonic PCHIP spline interpolants. \textbf{d}, Regularization residual accuracy $\Delta\Phi_{\mathrm{mol}}(t)$ (in $\mathrm{mV}$) across normalized depth $t = z/L \in [0, 1]$ alongside the smoothstep boundary taper envelope $a_L(t)a_R(t)$ (dashed black curve, right axis), demonstrating that the surrogate regularization error is smooth, bounded ($< 1.5\%$), and $C^1$-continuous with zero Dirac-delta jumps, gradient cusps, or step artifacts and exact zero-residual pinning at the metal contacts ($t = 0$ and $t = 1$).}
    \label{fig:S1}
\end{figure}

To represent the continuum electrostatic coupling across adjacent discrete molecular planes ($t - w, t, t + w$, where $w$ is the normalized interaction half-width) in the baseline $[\mathrm{Ru^{II}L_2}](\mathrm{BF_4})_2$ ($\mathrm{BF_4^-}$) coordination film while eliminating discrete 30-layer staircase steps in the bulk without perturbing the exact contact boundary conditions $u(0) = 0$ and $u(1) = 1$, we define a three-point spatial moving-average kernel $u_{\mathrm{ma}}(t)$ blended via cubic Hermite smoothstep edge tapers $a_L(t)$ and $a_R(t)$:
\begin{equation}
    u_{\mathrm{ma}}(t) = 0.25\,u(t - w) + 0.50\,u(t) + 0.25\,u(t + w)
    \label{eq:S24_uma}
\end{equation}
\begin{equation}
    a_L(t) = \mathcal{H}_3\left(\frac{t}{w}\right), \qquad a_R(t) = \mathcal{H}_3\left(\frac{1.0 - t}{w}\right), \qquad \mathcal{H}_3(x) = \begin{cases} 0.0 & x \le 0.0 \\ x^2(3.0 - 2.0x) & 0.0 < x < 1.0 \\ 1.0 & x \ge 1.0 \end{cases}
    \label{eq:S25_smoothstep}
\end{equation}
Combining $u(t)$ and $u_{\mathrm{ma}}(t)$ with the effective interior blending envelope $a_{\mathrm{eff}}(t) = \alpha_{\mathrm{taper}} a_L(t) a_R(t)$ yields the single-branch regularized potential function $\Psi_{\mathrm{reg}}(t; p, q, k, \alpha_{\mathrm{taper}}, w)$ (Figure~\ref{fig:S1}a):
\begin{equation}
    \Psi_{\mathrm{reg}}(t; p, q, k, \alpha_{\mathrm{taper}}, w) = \left[1.0 - a_{\mathrm{eff}}(t)\right]u(t; p, q, k) + a_{\mathrm{eff}}(t)u_{\mathrm{ma}}(t; p, q, k, w)
    \label{eq:S26_psi_branch}
\end{equation}
Because $a_L(0) = 0$, $a_L'(0) = 0$, $a_R(1) = 0$, and $a_R'(1) = 0$, the cubic Hermite smoothstep edge tapers $a_L(t)$ and $a_R(t)$ smoothly extinguish moving-average spatial filtering ($a_{\mathrm{eff}} \to 0$) at both metal contacts ($t = 0, 1$), enforcing exact Dirichlet boundary conditions ($\Psi_{\mathrm{reg}}(0) = 0$ and $\Psi_{\mathrm{reg}}(1) = 1$) and preserving exact contact electric fields while eliminating discrete 30-layer staircase steps throughout the film interior ($w < t < 1 - w$). As demonstrated in Figure~\ref{fig:S1}d, the resulting $\mathrm{BF_4^-}$ surrogate regularization residual $\Delta\Phi_{\mathrm{mol}}(t)$ is smooth, tightly bounded ($< 1.5\%$ of $\Phi_{\max} = 700\ \mathrm{mV}$), and $C^1$-continuous with zero Dirac-delta jumps, gradient cusps, or non-differentiable step artifacts across the entire film thickness, facilitating robust Newton-Raphson convergence in SPICE and Cadence Spectre.

\subsection{\texorpdfstring{Two-Phase Spatial Blending, State Coordinate $n$ vs.\ Static Ceiling $n_{\max}$, and Monotonic PCHIP Splines}{Two-Phase Spatial Blending, State Coordinate n vs.\ Static Ceiling n, and Monotonic PCHIP Splines}}
Before formulating the state-dependent spatial parameterization, it is essential to clarify the physical distinction between the continuous state variable $n$ and the static normalization ceiling $n_{\max}$ across Note~S5 and Note~S6:
\begin{enumerate}[topsep=2pt,itemsep=2pt,parsep=0pt]
    \item \textit{Continuous State Variable $n$:} The variable $n(t)$ is a continuous internal state coordinate representing the normalized cumulative pulse progression (and effective volumetric fraction of switched redox-active sites) along the 14-bit switching trajectory, rather than a discrete hardware pulse counter.
    \item \textit{Static Normalization Ceiling $n_{\max}$:} The parameter $n_{\max} = 33,040$ is a static scaling constant defined as exactly double the nominal potentiation pulse budget ($N_{\mathrm{pot}} = 16,520$, so $n_{\max} = 2 N_{\mathrm{pot}} = 33,040$).
    \item \textit{Arbitrary Pulse Waveform Calibration:} For the baseline $[\mathrm{Ru^{II}L_2}](\mathrm{BF_4})_2$ ($\mathrm{BF_4^-}$) system, the nominal pulse specification ($+0.90\ \mathrm{V}$, $80\ \mathrm{ns}$ for potentiation; $-0.75\ \mathrm{V}$, $65\ \mathrm{ns}$ for depression) serves as an empirical calibration baseline: in the compact model of Note~S6 (Equation~(\ref{eq:S33_dn_dt})), the dynamical rate equation $dn/dt$ scales continuously with applied voltage amplitude and pulse width to preserve the invariance of this state trajectory relative to $n_{\max}$.
    \item \textit{Upper Clamping Bound $n_{\max}$ and Practical Saturation:} In the active Python and Verilog-A implementations, the direction-aware Hermite smoothstep clamp of Note~S6 (Equation~(\ref{eq:S35_clamp})) attenuates $dn/dt$ against the hard upper bound $n_{\max} = 33,040$ over the saturation width $w_{\mathrm{sat}} = \max(50.0, 1.5 N_{\mathrm{smooth,eff}})$. Under realistic programming voltages the device never approaches this bound: the conductance plateau observed in overdrive (for example, $n \approx 23,600$ to $24,700$ states with $G \approx 5.0$ to $5.1\ \mathrm{mS}$ in Note~S7) arises from the saturating $f_{22}(n)$ table rather than from the clamp itself, and applying excessive voltages would trigger irreversible dielectric breakdown long before $n(t)$ could reach $33,040$. Thus, $n_{\max}$ serves as an asymptotic numerical safety boundary rather than an accessible physical state.
\end{enumerate}

During intermediate programming states ($0 < n < 16,520$) of the baseline $[\mathrm{Ru^{II}L_2}](\mathrm{BF_4})_2$ ($\mathrm{BF_4^-}$) junction, the molecular film is partitioned by the advancing planar front at midpoint layer $z_{\mathrm{mid}}(n)$ (with interfacial transition width $z_{\mathrm{width}}(n)$) into a switched $(22)$ region ($z < z_{\mathrm{mid}}$) governed by parameter vector $\boldsymbol{\theta}_f(n) = (p_f(n), q_f(n), k_f(n), a_f(n), w_f(n))$ and an unswitched $(31)$ region ($z > z_{\mathrm{mid}}$) governed by parameter vector $\boldsymbol{\theta}_i(n) = (p_i(n), q_i(n), k_i(n), a_i(n), w_i(n))$. The complete state-dependent internal electrostatic potential profile $\Phi_{\mathrm{mol}}(z,n)$ across all $z \in [0, L]$ and $n \in [0, 16,520]$ is expressed in closed form as:
\begin{equation}
    \Phi_{\mathrm{mol}}(z,n) = \Phi_{\max}\left\{\left[1.0 - B(z,n)\right]\Psi_{\mathrm{reg}}\left(\frac{z}{L}; \boldsymbol{\theta}_f(n)\right) + B(z,n)\Psi_{\mathrm{reg}}\left(\frac{z}{L}; \boldsymbol{\theta}_i(n)\right)\right\}
    \label{eq:S27_phi_full}
\end{equation}
where $B(z,n)$ is the hyperbolic tangent phase-boundary blending function:
\begin{equation}
    B(z,n) = \frac{1}{2}\left[1.0 + \tanh\left(\frac{z - z_{\mathrm{mid}}(n)}{z_{\mathrm{width}}(n)}\right)\right]
    \label{eq:S28_tanh_blend}
\end{equation}
Across the 14-bit $\mathrm{BF_4^-}$ programming continuum, the dynamic spatial potential landscape is parameterized by twelve state-dependent coordinates: $\boldsymbol{\Theta}(n) = [\boldsymbol{\theta}_i(n), \boldsymbol{\theta}_f(n), z_{\mathrm{mid}}(n), z_{\mathrm{width}}(n)] \in \mathbb{R}^{12}$. Rather than unconstrained behavioral regression variables, each of these twelve coordinates is strictly grounded in the underlying quantum chemical and electrostatic transport formulations derived in Note~S4: (i) the phase boundary coordinate $z_{\mathrm{mid}}(n)$ tracks the physical planar front position of switched $(22)$ molecular layers governed by the linear Cauchy-Lorentz progression $Z_{22}(n) = (n - p_2)/p_1$ (obtained by inverting the nucleation center law of Equation~(\ref{eq:S10_chi})); (ii) the transition half-width $z_{\mathrm{width}}(n)$ captures the microscopic electrostatic screening width separating switched and unswitched domains; and (iii) the five-parameter vectors $\boldsymbol{\theta}_i(n) = (p_i, q_i, k_i, a_i, w_i)$ and $\boldsymbol{\theta}_f(n) = (p_f, q_f, k_f, a_f, w_f)$ are extracted directly from the exact 30-layer Thomas-Fermi Fourier sine series solutions (Equation~(\ref{eq:S2_tf_series})) and close-packed counterion boundary conditions at 54 discrete physical state coordinates across the counterion displacement continuum ($n = 0$ to $20,000$). To evaluate these microscopic field solutions in constant time ($O(1)$) within continuous circuit simulators without numerically re-inverting 30 coupled differential master equations at every sub-nanosecond timestep, each of the twelve physical coordinates is interpolated using $C^1$-continuous monotonic Piecewise Cubic Hermite Interpolating Polynomials (PCHIP)~\cite{si:kundu2024}: on each interval $[n_j, n_{j+1}]$ between adjacent physical calibration coordinates with step $h_j = n_{j+1} - n_j$ and local normalized coordinate $s = (n - n_j)/h_j \in [0, 1]$,
\begin{equation}
    y_k(n) = y_{k,j} h_{00}(s) + h_j d_{k,j} h_{10}(s) + y_{k,j+1} h_{01}(s) + h_j d_{k,j+1} h_{11}(s)
    \label{eq:S29_pchip}
\end{equation}
where the cubic Hermite basis functions are $h_{00}(s) = (1 + 2s)(1 - s)^2$, $h_{10}(s) = s(1 - s)^2$, $h_{01}(s) = s^2(3 - 2s)$, and $h_{11}(s) = s^2(s - 1)$, and the support-node derivatives $d_{k,j}$ are computed via the Fritsch-Carlson weighted harmonic mean of adjacent secant slopes $\Delta_j = (y_{k,j+1} - y_{k,j})/h_j$ (setting $d_{k,j} = 0$ at any local extremum where $\Delta_{j-1}\cdot\Delta_j \le 0$). This mathematical formulation eliminates Runge polynomial oscillations while rigorously preserving local physical monotonicity without unphysical parameter overshoot (Figure~\ref{fig:S1}c).

\subsection{Analytical Exponential Surrogate for the Marcus-Hush-Chidsey Integral}
Finally, to replace the numerical Marcus-Hush-Chidsey (MHC) quadrature integral (Equation~(\ref{eq:S4_mhc}) of Note~S4) for the baseline $[\mathrm{Ru^{II}L_2}](\mathrm{BF_4})_2$ ($\mathrm{BF_4^-}$) system, we expand the MHC Gaussian-Fermi convolution over the physical operating range $\Phi_{\mathrm{mol}} \in [0, 0.70\ \mathrm{V}]$. Because $E_\lambda = 0.04\ \mathrm{eV}$ is comparable to $k_B T \approx 0.0257\ \mathrm{eV}$ (at $T_{\mathrm{amb}} = 298.15\ \mathrm{K}$) while $\Delta G^0 + e\Phi_{\mathrm{mol}}$ shifts the Gaussian peak across the exponential tail of the Fermi-Dirac distribution $\psi(x) \approx \exp(-x)$, Laplace asymptotic evaluation of Equation~(\ref{eq:S4_mhc}) reduces to a closed-form single-exponential analytical surrogate (Equation~9 of the main manuscript):
\begin{equation}
    k_{\mathrm{ET}}\left(\Phi_{\mathrm{mol}}(z,n)\right) = A_{\mathrm{ET}} \exp\left(-B_{\mathrm{ET}}\Phi_{\mathrm{mol}}(z,n)\right)
    \label{eq:S30_mhc_surrogate}
\end{equation}
where for baseline $\mathrm{BF_4^-}$, $A_{\mathrm{ET}} = 1.135 \times 10^3\ \mathrm{s^{-1}}$ is the zero-potential barrier-lowered transfer rate and $B_{\mathrm{ET}}$ is the electrostatic potential sensitivity factor ($B_{\mathrm{ET}} = 7.12\ \mathrm{V^{-1}}$ is the effective slope within the barrier-lowered active switching window used by the compact model, whereas $B_{\mathrm{ET,full}} = 35.97\ \mathrm{V^{-1}}$ is the logarithmic slope across the full asymptotic $[0, 0.70\ \mathrm{V}]$ semi-logarithmic span plotted in Figure~\ref{fig:S1}b; corresponding $\mathrm{PF_6^-}$ values are $A_{\mathrm{ET}} = 1.420 \times 10^3\ \mathrm{s^{-1}}$ and $B_{\mathrm{ET}} = 6.85\ \mathrm{V^{-1}}$ in Table~\ref{tab:S1}). Under semi-logarithmic scaling (Figure~\ref{fig:S1}b), Equation~(\ref{eq:S30_mhc_surrogate}) forms a strictly linear straight line ($\log_{10} k_{\mathrm{ET}} = \log_{10} A_{\mathrm{ET}} - (B_{\mathrm{ET}}/\ln 10)\Phi_{\mathrm{mol}}$) across more than eight orders of magnitude ($10^{-9}$ to $10^3\ \mathrm{s^{-1}}$), accelerating per-layer evaluation by over $1,000\times$ compared to numerical quadrature while faithfully capturing field-assisted barrier lowering for intermolecular redox switching.

\section{Circuit-Level Transient Simulations and Device Verification}
\label{sec:noteS6}

This note is dedicated to single-device electrical, kinetic, crossbar array control, overdrive, and electro-thermal transient dynamics for the baseline $[\mathrm{Ru^{II}L_2}](\mathrm{BF_4})_2$ ($\mathrm{BF_4^-}$) molecular memristor, verified across our transient circuit simulation suite (cataloged in Table~\ref{tab:S4}).

\subsection{Complete Continuous Compact Model Equations and Overdrive Memory Kinetics}
Building directly upon the closed-form analytical surrogates and $C^1$-continuous boundary regularizations derived in Note~S5, the 30-site hopping chain and counterion kinetics are compiled into five coupled differential-algebraic equations governing terminal current $I$, internal molecular voltage $V_{\mathrm{int}}$, continuous state variable $n(t)$ (normalized relative to the static asymptotic ceiling $n_{\max} = 2N_{\mathrm{pot}} = 33,040$ defined in Subsection~S5.2), overdrive memory register $c_{\mathrm{discharge}}(t)$, and local junction temperature $T(t)$ for the baseline $[\mathrm{Ru^{II}L_2}](\mathrm{BF_4})_2$ ($\mathrm{BF_4^-}$) system (with all calibrated baseline $\mathrm{BF_4^-}$ and alternative $\mathrm{PF_6^-}$ parameters compiled in Table~\ref{tab:S1}):
\begin{enumerate}
    \item \textit{Implicit Series Contact Resistance and Hyperbolic Sine Conduction:} For an external terminal voltage $V_{\mathrm{applied}} = V(\mathrm{P}) - V(\mathrm{N})$ across the baseline $\mathrm{BF_4^-}$ junction, parasitic metallic and interfacial series contact resistance $R_c$ per contact ($2R_c$ total series resistance) drops a portion of the terminal bias, establishing the implicit transcendental conduction equation:
    \begin{equation}
        V_{\mathrm{int}} = V_{\mathrm{applied}} - 2 R_c I(n, V_{\mathrm{int}}), \qquad I(n, V_{\mathrm{int}}) = I_{\mathrm{prod}}(n) \sinh\left(\frac{V_{\mathrm{int}}}{V_{\mathrm{scale,eff}}(n)}\right)
        \label{eq:S31_sinh_iv}
    \end{equation}
    Here, the terminal read current $I(n, V_{\mathrm{int}})$ and state-dependent current pre-factor $I_{\mathrm{prod}}(n) = I_{\mathrm{scale}}(n) \cdot g_{\mathrm{scale}} \cdot f_{22}(n)$ directly inherit the volumetric switched fraction $f_{22}(n)$ regularized in Note~S5, while lookup tables $I_{\mathrm{scale}}(n)$ and $V_{\mathrm{scale}}(n)$ are evaluated via the monotonic PCHIP splines of Equation~(\ref{eq:S29_pchip}), parameterized by technology scaling factor $g_{\mathrm{scale}}$ and barrier compression factor $\gamma_{\mathrm{comp}}$ (compiled in Table~\ref{tab:S1}):
    \begin{equation}
        I_{\mathrm{prod}}(n) = I_{\mathrm{scale}}(n) \cdot g_{\mathrm{scale}} \cdot f_{22}(n), \qquad V_{\mathrm{scale,eff}}(n) = \frac{V_{\mathrm{scale}}(n)}{1.0 + \gamma_{\mathrm{comp}} f_{22}(n)}
        \label{eq:S32_iprod}
    \end{equation}
    \item \textit{Continuous Softplus Rate Equation and Overdrive Exponent Modulation:} The dynamic rate equation $dn/dt$ drives $\mathrm{BF_4^-}$ counterion redistribution across the molecular junction, continuously updating the internal 30-layer potential landscape $\Phi_{\mathrm{mol}}(z,n)$ via Equation~(\ref{eq:S27_phi_full}) of Note~S5. To eliminate derivative discontinuities at $V_{\mathrm{int}} = 0$ and $|V_{\mathrm{int}}| = V_{\mathrm{th}}$, the state transition rate $dn/dt$ is formulated via hyperbolic tangent polarity switching $v_{\mathrm{tanh}} = \tanh(V_{\mathrm{int}} / V_{\mathrm{smooth}})$ and softplus voltage thresholding:
    \begin{equation}
        \frac{dn}{dt} = v_{\mathrm{tanh}} \cdot \left[\frac{V_{\mathrm{smooth}} \ln\left(1.0 + \exp\left(\frac{|V_{\mathrm{int}}| - V_{\mathrm{th}}}{V_{\mathrm{smooth}}}\right)\right)}{\left||V_{\mathrm{ref}}| - V_{\mathrm{th}}\right| + V_\varepsilon}\right]^{\alpha_{\mathrm{eff}}(V_{\mathrm{int}})} \cdot \kappa_{\mathrm{eff}}(n, V_{\mathrm{int}}) \cdot C_{\mathrm{clamp}}(n, V_{\mathrm{int}}) \cdot k_{\mathrm{thermal}}(T)
        \label{eq:S33_dn_dt}
    \end{equation}
    where $V_{\mathrm{th}}$ denotes the effective compact switching threshold voltage, $V_{\mathrm{smooth}}$ is the softplus threshold and polarity smoothing scale, $V_\varepsilon$ is the denominator regularization guard constant, and $V_{\mathrm{ref}}$ selects the directional reference calibration anchor $V_{\mathrm{ref,pos}}$ for potentiation ($V_{\mathrm{int}} \ge 0$) or $V_{\mathrm{ref,neg}}$ for depression ($V_{\mathrm{int}} < 0$), so that the denominator evaluates to $|0.90| - 0.4664 = 0.4336\ \mathrm{V}$ during potentiation and $|-0.75| - 0.4664 = 0.2836\ \mathrm{V}$ during depression. In addition, the Verilog-A compact model implements a conditional sub-threshold fast path: whenever $|V_{\mathrm{int}}| < V_{\mathrm{bypass}} = V_{\mathrm{th}} - 3V_{\mathrm{smooth}} = 0.1551\ \mathrm{V}$ (\texttt{vBypass} in Table~\ref{tab:S1}) and no overdrive discharge is pending ($n \le n_{\mathrm{base}}$ or $k_{\mathrm{discharge}} = 0$), the rate is set identically to $dn/dt = 0$ to suppress the residual softplus tail near zero bias. The effective nonlinearity exponent $\alpha_{\mathrm{eff}}(V_{\mathrm{int}})$ incorporates high-field overdrive acceleration around the baseline over-threshold exponent $\alpha$ via the overdrive coupling coefficient $\beta_\alpha$ (compiled in Table~\ref{tab:S1}):
    \begin{equation}
        \alpha_{\mathrm{eff}}(V_{\mathrm{int}}) = \alpha + \beta_\alpha \tanh\left(\frac{|V_{\mathrm{int}}| - V_{\mathrm{ref,pos}}}{V_{\mathrm{ref,pos}} - V_{\mathrm{th}} + V_\varepsilon}\right)
        \label{eq:S34_alpha_eff}
    \end{equation}
    \item \textit{Direction-Aware Hermite Boundary Clamping:} To prevent unbounded growth of $n(t)$ without introducing step cutoffs in the Jacobian, the baseline $\mathrm{BF_4^-}$ compact model applies a direction-aware cubic Hermite smoothstep against the hard upper capacity bound $n_{\max} = 33,040$ during potentiation ($v_{\mathrm{tanh}} > 0$), with effective saturation width $w_{\mathrm{sat}} = \max(50.0, 1.5 N_{\mathrm{smooth,eff}})$ (parameterized by the Hermite transition scale $N_{\mathrm{smooth}}$ in Table~\ref{tab:S1}):
    \begin{equation}
        C_{\mathrm{clamp}}(n) = \begin{cases} 1.0, & n_{\max} - n \ge w_{\mathrm{sat}} \\ z_{\mathrm{hi}}^2\left(3.0 - 2.0 z_{\mathrm{hi}}\right), & 0 < n_{\max} - n < w_{\mathrm{sat}} \\ 0.0, & n_{\max} - n \le 0 \end{cases}, \qquad z_{\mathrm{hi}} = \frac{n_{\max} - n}{w_{\mathrm{sat}}}
        \label{eq:S35_clamp}
    \end{equation}
    During depression ($v_{\mathrm{tanh}} < 0$), the mirror smoothstep acts within the lower floor guard band $w_{\mathrm{zero}}$ near $n = 0$. Because both branches have zero slope at their endpoints, $C_{\mathrm{clamp}}$ is $C^1$-continuous across the entire state space. Here, the overdrive parameters $\beta_s$ and $r_{\mathrm{top\_blend}}$ listed in Table~\ref{tab:S1} modulate the effective transition scale $N_{\mathrm{smooth,eff}}$ entering $w_{\mathrm{sat}}$, while the upper clamp acts directly against $n_{\max}$.
    \item \textit{Overdrive Memory Register $c_{\mathrm{discharge}}$ and Power-Law RESET Snapback:} When a baseline $[\mathrm{Ru^{II}L_2}](\mathrm{BF_4})_2$ ($\mathrm{BF_4^-}$) junction is stressed into overdrive saturation ($V_{\mathrm{applied}} > V_{\mathrm{ref,pos}}$ in Figure~3b of the main manuscript), excess $\mathrm{BF_4^-}$ counterions pack into strained interfacial sites that rapidly discharge upon polarity reversal~\cite{si:sharma2024,si:kundu2024}. The non-volatile state register $c_{\mathrm{discharge}}(t)$ tracks peak fractional over-voltage exposure during potentiation, scaled by the overdrive register gain $k_{\mathrm{overdrive}}$ (Table~\ref{tab:S1}):
    \begin{equation}
        \begin{aligned}
            c_{\mathrm{discharge}}(t) &= \max\left(c_{\mathrm{discharge}}(t^-),\, \frac{V_{\mathrm{applied}} - V_{\mathrm{ref,pos}}}{V_{\mathrm{ref,pos}} - V_{\mathrm{th}} + V_\varepsilon}\right), \\
            \mathrm{od}_{\mathrm{clamp}} &= \min\left(1.0,\, \max\left(0.0,\, c_{\mathrm{discharge}} k_{\mathrm{overdrive}}\right)\right)
        \end{aligned}
        \label{eq:S36_cdischarge}
    \end{equation}
    During subsequent depression ($V_{\mathrm{int}} < 0$), $\mathrm{od}_{\mathrm{clamp}}$ collapses the nominal return shelf latency $r_{\mathrm{lat,eff}} = r_{\mathrm{latency}}(1.0 - \mathrm{od}_{\mathrm{clamp}})$, shifts the depression landing floor $n_{\mathrm{base}}$ via the bottom floor blend factor $r_{\mathrm{bottom\_blend}}$:
    \begin{equation}
        n_{\mathrm{base}} = 0.5\,n_{\max}\left[1.0 - \mathrm{od}_{\mathrm{clamp}}\left(1.0 - r_{\mathrm{bottom\_blend}}\right)\right]
        \label{eq:S37_nbase}
    \end{equation}
    and augments the reverse switching rate by the power-law interfacial discharge term governed by the snapback discharge rate $k_{\mathrm{discharge}}$ and knee fraction $f_{\mathrm{discharge}}$ (defining power-law decay exponent $p_{\mathrm{decay}} = 1/f_{\mathrm{discharge}}$, compiled in Table~\ref{tab:S1}):
    \begin{equation}
        \kappa_{\mathrm{dep,od}}(n) = r_{\mathrm{asym}}\kappa_{\mathrm{switch}}\left[1.0 + \mathrm{od}_{\mathrm{clamp}}\left(\frac{k_{\mathrm{discharge}}}{\kappa_{\mathrm{switch,norm}}}\right)\left(\frac{n - n_{\mathrm{base}}}{n_{\max} - n_{\mathrm{base}}}\right)^{1/f_{\mathrm{discharge}}}\right]
        \label{eq:S38_kdep_od}
    \end{equation}
\end{enumerate}

\subsection{Crossbar Measurement System, Single-Pulse Programming Kinetics, and Multi-Phase Array Reprogramming}
\begin{figure}[!htb]
    \centering
    \includegraphics[width=\textwidth,height=0.62\textheight,keepaspectratio]{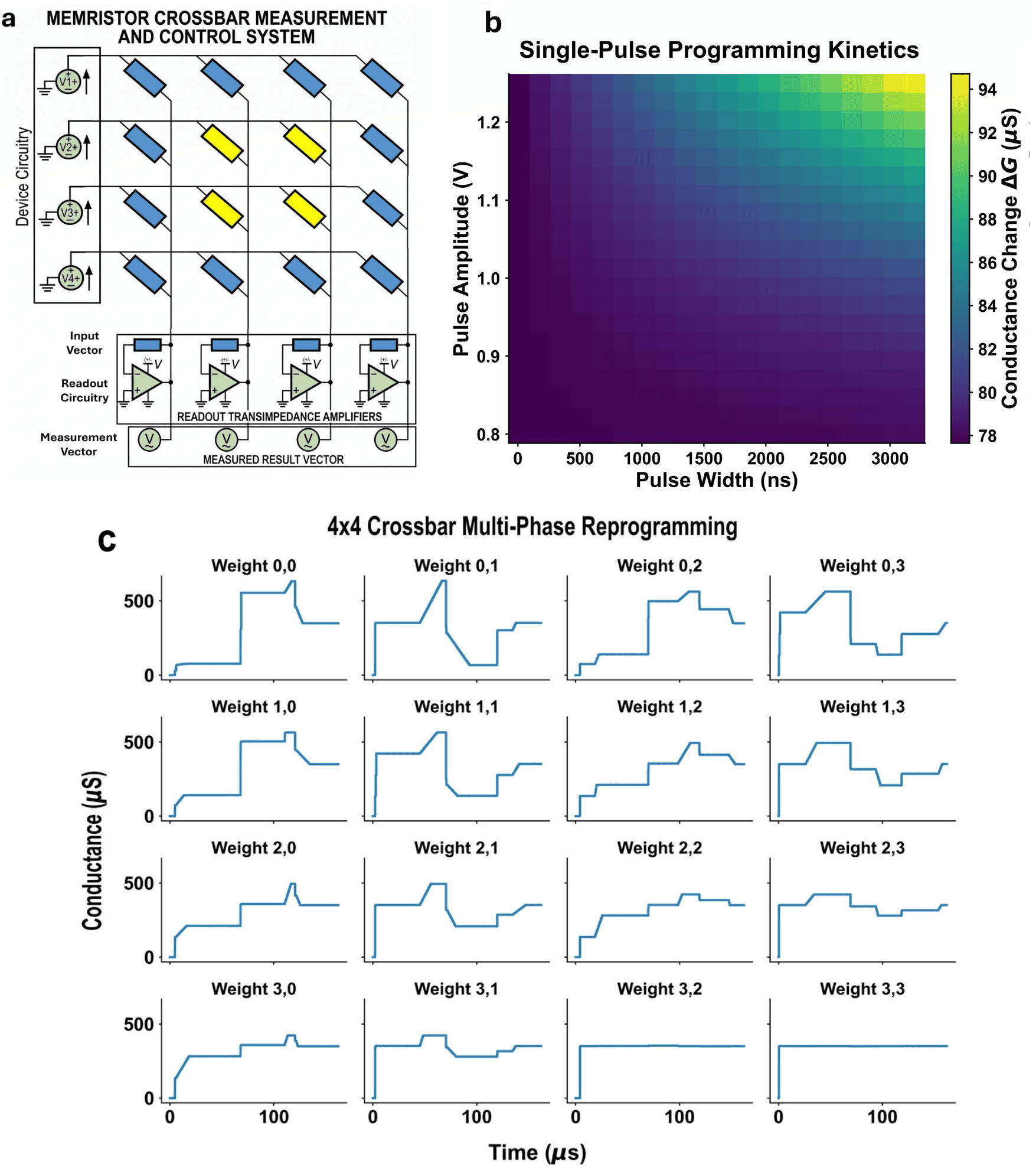}
    \caption{\textbf{Crossbar measurement and control system, single-pulse programming kinetics, and multi-phase parallel array reprogramming for baseline $[\mathrm{Ru^{II}L_2}](\mathrm{BF_4})_2$ ($\mathrm{BF_4^-}$) devices.} \textbf{a}, Circuit schematic of a $4 \times 4$ selector-less molecular memristor crossbar measurement and control system integrated with wordline pulse drivers and bitline virtual-ground transimpedance amplifiers (TIAs). \textbf{b}, Two-dimensional heatmap of single-pulse programming kinetics mapped across write pulse amplitude ($0.80\ \mathrm{V}$ to $1.24\ \mathrm{V}$) and pulse width ($0\ \mathrm{ns}$ to $3,200\ \mathrm{ns}$); the color scale reports the post-pulse conductance $G_{\mathrm{final}}$ ($\mathrm{\mu S}$) starting from a $78.0\ \mathrm{\mu S}$ pre-pulse baseline, so that the single-pulse increment is $\Delta G = G_{\mathrm{final}} - 78.0\ \mathrm{\mu S}$. \textbf{c}, Multi-phase parallel reprogramming of all 16 crosspoints (\texttt{Weight 0,0} through \texttt{Weight 3,3}) in a $4 \times 4$ crossbar over a $150\ \mathrm{\mu s}$ schedule without sneak-path crosstalk (target weights differ from those of Figure~4d in the main manuscript).}
    \label{fig:S2}
\end{figure}

Before examining nanosecond solver microsteps, Figure~\ref{fig:S2} characterizes the baseline $[\mathrm{Ru^{II}L_2}](\mathrm{BF_4})_2$ ($\mathrm{BF_4^-}$) continuous compact model across the crossbar measurement and control architecture, single-pulse excitation parameter space, and synchronous $4 \times 4$ array reprogramming operations. Figure~\ref{fig:S2}a depicts the circuit architecture of a $4 \times 4$ selector-less molecular memristor crossbar measurement and control system integrated with wordline pulse drivers and bitline virtual-ground transimpedance amplifiers (TIAs). Integrating Equation~(\ref{eq:S33_dn_dt}) across write pulse amplitude ($0.80\ \mathrm{V}$ to $1.24\ \mathrm{V}$) and pulse width ($0\ \mathrm{ns}$ to $3,200\ \mathrm{ns}$) from a fixed pre-pulse baseline of $78.0\ \mathrm{\mu S}$ yields the two-dimensional single-pulse map in Figure~\ref{fig:S2}b; the color scale reports the post-pulse conductance $G_{\mathrm{final}}(V_{\mathrm{write}}, t_{\mathrm{pw}})$, so that the single-pulse increment is $\Delta G = G_{\mathrm{final}} - 78.0\ \mathrm{\mu S}$ (zero at $t_{\mathrm{pw}} = 0$ and reaching $\approx 16\ \mathrm{\mu S}$ at $1.24\ \mathrm{V}$, $3,200\ \mathrm{ns}$), illustrating smooth turn-on above the switching threshold $V_{\mathrm{th}}$ and continuous field-accelerated conductance growth under overdrive bias. Deploying this continuous compact model within a $4 \times 4$ crossbar operated over a $150\ \mathrm{\mu s}$ multi-phase schedule (comprising initialization, two sequential potentiation phases, controlled depression, and fine conductance trimming, identical in phase structure to Figure~4d of the main manuscript), all 16 crosspoints (\texttt{Weight 0,0} through \texttt{Weight 3,3}) modulate their conductances concurrently without sneak-path crosstalk or unselected state disturbance (Figure~\ref{fig:S2}c). Because Figure~\ref{fig:S2}c uses a different set of target weights than the Figure~4d demonstration, individual device trajectories (for example, device $(0,0)$) differ between the two panels while sharing the same five-phase schedule.

\subsection{Nanosecond Staircase Tracking and Write-Read Immunity}
\begin{figure}[!htb]
    \centering
    \includegraphics[width=\textwidth,height=0.62\textheight,keepaspectratio]{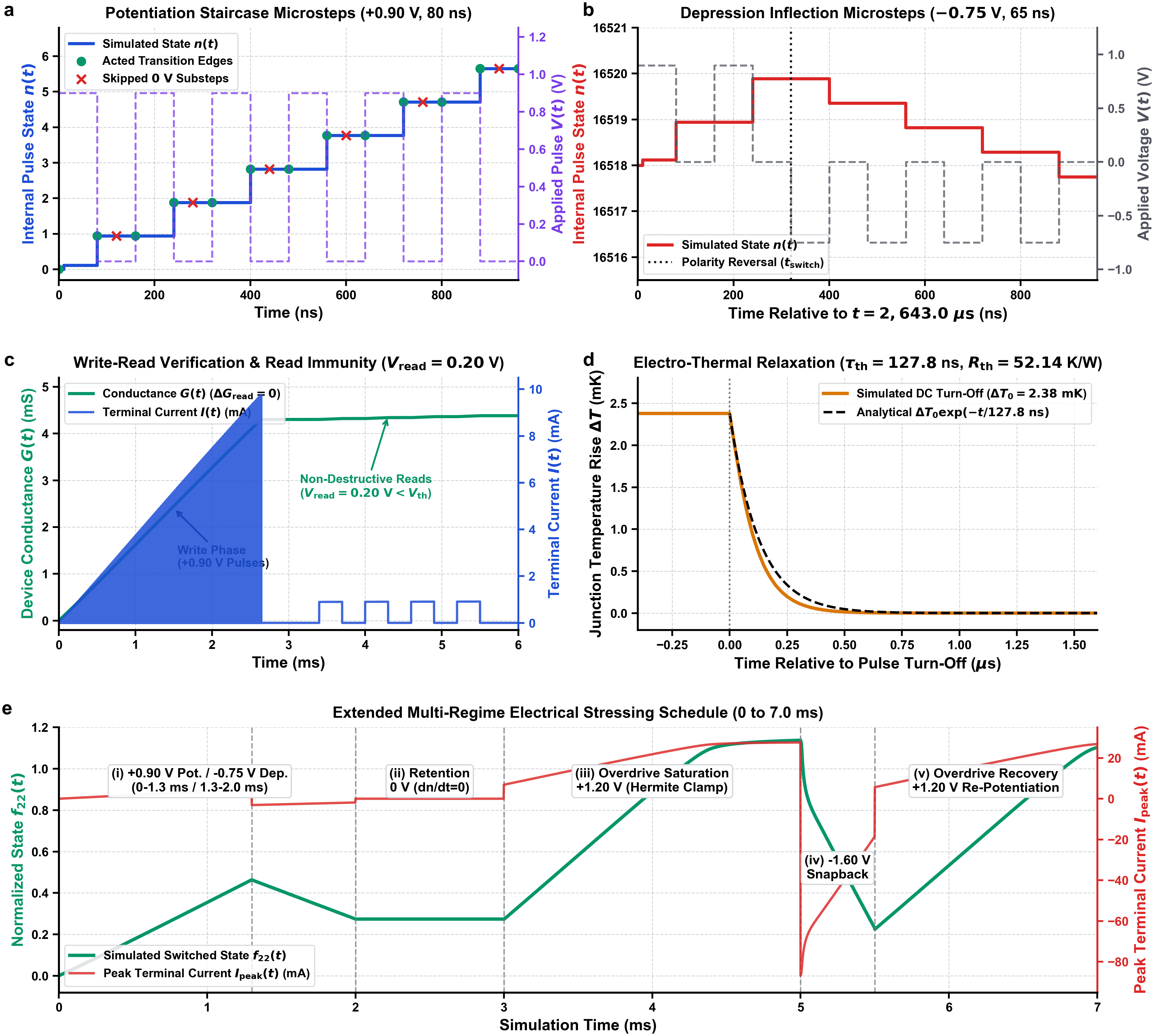}
    \caption{\textbf{Transient solver kinetics and single-device circuit simulations for baseline $[\mathrm{Ru^{II}L_2}](\mathrm{BF_4})_2$ ($\mathrm{BF_4^-}$) molecular memristors.} \textbf{a}, Pulse-by-pulse staircase zoom during nominal potentiation ($0$ to $960\ \mathrm{ns}$, $+0.90\ \mathrm{V}$, $80\ \mathrm{ns}$ pulses at $160\ \mathrm{ns}$ period), illustrating internal state $n(t)$ (blue curve), acted-upon pulse transition edges (green circles), skipped $0\ \mathrm{V}$ substeps (red crosses), and applied voltage $V(t)$ (purple dashed curve, right axis). \textbf{b}, Pulse-by-pulse staircase zoom around the potentiation-to-depression inflection (time axis referenced to $t = 2,643.0\ \mathrm{\mu s}$, with the polarity reversal $t_{\mathrm{switch}}$ marked by the dotted line following completion of the $16,520$-pulse potentiation train), illustrating oscillation-free reversal from $+0.90\ \mathrm{V}$ ($80\ \mathrm{ns}$) potentiation pulses to $-0.75\ \mathrm{V}$ ($65\ \mathrm{ns}$) depression pulses. \textbf{c}, Interleaved write-read verification cycle, demonstrating active conductance programming up to $4.33\ \mathrm{mS}$ followed by four sub-threshold read interrogation bursts ($V_{\mathrm{read}} = 0.20\ \mathrm{V}$) with zero read disturbance ($\Delta G_{\mathrm{read}} = 0$). \textbf{d}, Electro-thermal relaxation dynamics following turn-off of a sustained $0.10\ \mathrm{V}$ sub-threshold DC bias ($\Delta T_0 = 2.38\ \mathrm{mK}$ at $n_{\mathrm{init}} = 16,520$, matching the Spectre parity benchmark of Note~S7), demonstrating quantitative agreement between simulated post-pulse junction cooling and the analytical Newton cooling law $\Delta T_0 \exp(-t / \tau_{\mathrm{th}})$ with $\tau_{\mathrm{th}} = 127.8\ \mathrm{ns}$ ($R_{\mathrm{th}} = 52.14\ \mathrm{K/W}$, $\Delta T < 20\ \mathrm{mK}$ under nominal pulsing). \textbf{e}, Extended multi-regime electrical stressing schedule across $7.0\ \mathrm{ms}$ tracking normalized switched state $f_{22}(t)$ (green curve, left axis) and peak pulse terminal current $I_{\mathrm{peak}}(t)$ (red curve, right axis), combining nominal $+0.90\ \mathrm{V}$ potentiation ($0$--$1.3\ \mathrm{ms}$) and $-0.75\ \mathrm{V}$ depression ($1.3$--$2.0\ \mathrm{ms}$), zero-bias retention ($2.0$--$3.0\ \mathrm{ms}$), $+1.20\ \mathrm{V}$ overdrive saturation of the conductance trajectory ($3.0$--$5.0\ \mathrm{ms}$), elevated $-1.60\ \mathrm{V}$ reverse RESET pulsing with power-law snapback ($5.0$--$5.5\ \mathrm{ms}$), and repeatable $+1.20\ \mathrm{V}$ overdrive re-potentiation ($5.5$--$7.0\ \mathrm{ms}$).}
    \label{fig:S3}
\end{figure}

Figure~\ref{fig:S3}a illustrates the single-device transient response of the baseline $[\mathrm{Ru^{II}L_2}](\mathrm{BF_4})_2$ ($\mathrm{BF_4^-}$) junction during the initial $960\ \mathrm{ns}$ of nominal potentiation ($+0.90\ \mathrm{V}$ pulses of $80\ \mathrm{ns}$ duration separated by $80\ \mathrm{ns}$ zero-bias intervals, $160\ \mathrm{ns}$ period) from our transient single-device switching simulations (Table~\ref{tab:S4}). During each $80\ \mathrm{ns}$ active pulse plateau, Equation~(\ref{eq:S33_dn_dt}) integrates $dn/dt$ smoothly with a continuous state increment $\Delta n \approx 0.94$ per pulse, locking onto flat staircase plateaus as soon as the pulse returns to $0\ \mathrm{V}$ (green circle markers) while skipping redundant sub-threshold zero-bias evaluation steps (red crosses). Figure~\ref{fig:S3}b captures the polarity inflection around $t_{\mathrm{switch}} \approx 2,643.2\ \mathrm{\mu s}$ (completion of the $16,520$-pulse potentiation train at $16,520 \times 160\ \mathrm{ns}$), where the pulse train transitions from $+0.90\ \mathrm{V}$ ($80\ \mathrm{ns}$) potentiation pulses to $-0.75\ \mathrm{V}$ ($65\ \mathrm{ns}$) depression pulses at high conductance ($n = 16,520$). Because $v_{\mathrm{tanh}} = \tanh(V_{\mathrm{int}}/V_{\mathrm{smooth}})$ reverses sign smoothly through zero bias while $r_{\mathrm{asym}} = 0.9857$ balances the reverse transition rate, $n(t)$ reverses direction symmetrically on the first negative pulse without numerical ringing.

Figure~\ref{fig:S3}c presents the interleaved write-read verification analysis for the baseline $[\mathrm{Ru^{II}L_2}](\mathrm{BF_4})_2$ ($\mathrm{BF_4^-}$) device. Following a $2.64\ \mathrm{ms}$ potentiation burst that programs the junction to its nominal $4.33\ \mathrm{mS}$ state, a sequence of sub-threshold read interrogation pulses ($V_{\mathrm{read}} \le 0.20\ \mathrm{V} < V_{\mathrm{th}} = 0.4664\ \mathrm{V}$) is applied between $3.4\ \mathrm{ms}$ and $5.5\ \mathrm{ms}$. Because the softplus bracket in Equation~(\ref{eq:S33_dn_dt}) evaluates to $< 10^{-6}$ at $0.20\ \mathrm{V}$, the programmed conductance state remains completely invariant ($\Delta G_{\mathrm{read}} = 0$) while producing clean $0.866\ \mathrm{mA}$ read current pulses.

\subsection{Coupled Newton Thermal Relaxation}
During active write and read operations, instantaneous electrical power dissipation $P_{\mathrm{diss}}(t) = |V_{\mathrm{int}}(t) I(n, V_{\mathrm{int}})|$ generates localized Joule heating within the $60\ \mathrm{nm}$ baseline $[\mathrm{Ru^{II}L_2}](\mathrm{BF_4})_2$ ($\mathrm{BF_4^-}$) molecular junction. The transient junction temperature $T(t)$ is governed by the first-order linear Newton cooling differential equation coupled to the metallic $\mathrm{Au}$ wordline and bitline heat sinks:
\begin{equation}
    C_{\mathrm{th}}\frac{dT}{dt} = P_{\mathrm{diss}}(t) - \frac{T(t) - T_{\mathrm{amb}}}{R_{\mathrm{th}}} \quad \Longleftrightarrow \quad \frac{dT}{dt} = \frac{T_{\mathrm{amb}} - T(t)}{\tau_{\mathrm{th}}} + \frac{R_{\mathrm{th}} P_{\mathrm{diss}}(t)}{\tau_{\mathrm{th}}}
    \label{eq:S39_newton_ode}
\end{equation}
where for the baseline $\mathrm{BF_4^-}$ device, $T_{\mathrm{amb}} = 298.15\ \mathrm{K}$, $R_{\mathrm{th}} = 52.14\ \mathrm{K/W}$ is the effective junction-to-electrode thermal resistance, $\tau_{\mathrm{th}} = 127.8\ \mathrm{ns}$ is the thermal relaxation time constant, and $C_{\mathrm{th}} = \tau_{\mathrm{th}}/R_{\mathrm{th}} = 2.451 \times 10^{-9}\ \mathrm{J/K}$ ($2.451\ \mathrm{nJ/K}$) is the effective heat capacity of the active junction volume and adjacent gold interfaces. Integrating Equation~(\ref{eq:S39_newton_ode}) for a rectangular pulse of power $P_0$ and duration $t_{\mathrm{pw}}$ yields:
\begin{equation}
    \Delta T(t) = T(t) - T_{\mathrm{amb}} = \begin{cases}
        R_{\mathrm{th}} P_0 \left[1 - \exp\left(-\dfrac{t}{\tau_{\mathrm{th}}}\right)\right], & 0 \le t \le t_{\mathrm{pw}} \\[1.0ex]
        \Delta T(t_{\mathrm{pw}}) \exp\left(-\dfrac{t - t_{\mathrm{pw}}}{\tau_{\mathrm{th}}}\right), & t > t_{\mathrm{pw}}
    \end{cases}
    \label{eq:S40_newton_sol}
\end{equation}
Under nominal $+0.90\ \mathrm{V}$ ($t_{\mathrm{pw}} = 80\ \mathrm{ns}$) potentiation where pulse energy is $E_{\mathrm{pulse}} = 0.19\ \mathrm{nJ}$, $\Delta T_{\mathrm{peak}}$ remains below $0.02\ \mathrm{K}$ ($20\ \mathrm{mK}$) across the low-to-intermediate conductance window and $< 0.058\ \mathrm{K}$ at peak conductance. In our thermal relaxation characterization (Figure~\ref{fig:S3}d), applying a sustained $2.5\ \mathrm{ms}$ sub-threshold DC bias of $0.10\ \mathrm{V}$ at $n_{\mathrm{init}} = 16,520$ ($I = 0.456\ \mathrm{mA}$, $P_{\mathrm{diss}} = 45.6\ \mathrm{\mu W}$, matching the Spectre parity benchmark of Note~S7) establishes a steady-state plateau of $\Delta T_0 = 2.38\ \mathrm{mK}$ (whereas a $0.35\ \mathrm{V}$ DC stress bias reaches $\Delta T_0 = 88.2\ \mathrm{mK}$); upon turn-off, the simulated junction temperature decays exponentially back to $T_{\mathrm{amb}} = 298.15\ \mathrm{K}$ with time constant $\tau_{\mathrm{th}} = 127.8\ \mathrm{ns}$, matching the analytical $\Delta T_0 \exp(-t/127.8\ \mathrm{ns})$ curve to within $10^{-5}\ \mathrm{K}$. Consequently, the Arrhenius thermal factor (with macroscopic Arrhenius barrier $E_a = E_{a,\mathrm{macro}} = 0.1559\ \mathrm{eV}$ [$155.9\ \mathrm{meV}$] for baseline $\mathrm{BF_4^-}$):
\begin{equation}
    k_{\mathrm{thermal}}(T) = \exp\left[\frac{E_a}{k_B}\left(\frac{1}{T_{\mathrm{amb}}} - \frac{1}{T(t)}\right)\right]
    \label{eq:S41_kthermal}
\end{equation}
experiences negligible perturbation ($|k_{\mathrm{thermal}} - 1| < 4 \times 10^{-4}$) during nominal operation, preventing thermal runaway under continuous megahertz pulsing.

\subsection{Extended Multi-Regime Electrical Stressing and Overdrive Recovery}
To evaluate numerical stability under extended multi-regime electrical stressing, Figure~\ref{fig:S3}e subjects the baseline $[\mathrm{Ru^{II}L_2}](\mathrm{BF_4})_2$ ($\mathrm{BF_4^-}$) compact model to a $7.0\ \mathrm{ms}$ multi-regime schedule while tracking both the normalized switched state $f_{22}(t)$ and the peak pulse terminal current $I_{\mathrm{peak}}(t)$: (i) nominal $+0.90\ \mathrm{V}$ potentiation from $0$ to $1.3\ \mathrm{ms}$ (where $f_{22}$ increases to $0.49$ and $n$ advances to $\approx 7,680$ states), followed by $-0.75\ \mathrm{V}$ depression from $1.3$ to $2.0\ \mathrm{ms}$ (where $n$ returns to $\approx 4,556$ states, because the shorter $0.7\ \mathrm{ms}$ depression window delivers fewer effective state decrements than the preceding potentiation window); (ii) a $1.0\ \mathrm{ms}$ zero-bias retention hold from $2.0$ to $3.0\ \mathrm{ms}$ confirming zero state leakage ($dn/dt = 0$ at zero bias in both engines); (iii) $+1.20\ \mathrm{V}$ overdrive potentiation from $3.0$ to $5.0\ \mathrm{ms}$, driving the conductance trajectory smoothly into its upper saturation plateau without derivative spikes; (iv) elevated $-1.60\ \mathrm{V}$ reverse RESET pulsing from $5.0$ to $5.5\ \mathrm{ms}$, activating the overdrive discharge register $c_{\mathrm{discharge}}$ and power-law snapback (Equation~(\ref{eq:S38_kdep_od})) with peak device terminal currents reaching $-86.8\ \mathrm{mA}$ (recovering smoothly to $-18.6\ \mathrm{mA}$ as counterions clear the interface); and (v) $+1.20\ \mathrm{V}$ overdrive re-potentiation from $5.5$ to $7.0\ \mathrm{ms}$ confirming repeatable recovery without numerical divergence. 

\section{Dual-Engine Verification: Cadence Spectre vs. Vectorized Python Model Parity}
\label{sec:noteS7}

\subsection{Theoretical Equivalence and Mathematical Formulation of Dual Solvers}
A foundational architectural principle of our simulation framework is the exact mathematical equivalence between the transistor-level behavioral Verilog-A compact model deployed in Cadence Spectre (\texttt{MolmemVirtuosoSpectreV1.0/models/molmem\_va.va}) and the vectorized Python simulator (\texttt{PythonSimulator/}). Both engines evaluate the identical system of $C^1$-continuous differential-algebraic equations derived in Notes~S5 and S6 (Equations~(\ref{eq:S20_s_base})--(\ref{eq:S41_kthermal})) and share the exact calibrated parameter set of Table~\ref{tab:S1}. In Cadence Spectre, the device is instantiated across four primary electrical and electro-thermal nodes: external electrode terminals \texttt{P} and \texttt{N}, internal post-contact molecular node \texttt{N\_INT} (separating lumped series contact resistance $2R_c$ from the intrinsic hyperbolic sine film conduction), thermal node \texttt{T\_JUNCT} (integrating Newton cooling via Spectre's \texttt{idt()} operator), and diagnostic telemetry pins \texttt{STATE\_N} and \texttt{STATE\_F22}. In Cadence Spectre, Kirchhoff's Current Law (KCL) is solved via Cadence's proprietary modified Newton-Raphson algorithms with adaptive local-truncation error (LTE) timestep control. In the vectorized Python engine, the equivalent implicit contact-resistance residual $F(V_{\mathrm{int}}) = V_{\mathrm{int}} + 2 R_c I_{\mathrm{prod}}(n)\sinh(V_{\mathrm{int}}/V_{\mathrm{scale,eff}}) - V_{\mathrm{applied}} = 0$ is solved at each transient timestep to machine precision ($|\Delta V_{\mathrm{int}}| < 10^{-11}\ \mathrm{V}$) via cubic Halley or Newton-Raphson iteration:
\begin{equation}
    V_{\mathrm{int}}^{(m+1)} = V_{\mathrm{int}}^{(m)} - \frac{F(V_{\mathrm{int}}^{(m)})}{F'(V_{\mathrm{int}}^{(m)}) - \dfrac{F(V_{\mathrm{int}}^{(m)}) F''(V_{\mathrm{int}}^{(m)})}{2 F'(V_{\mathrm{int}}^{(m)})}}
    \label{eq:S42_halley}
\end{equation}

\subsection{\texorpdfstring{Nominal Dynamic Switching Parity (tb\_dynamic: Full Return to Zero)}{Nominal Dynamic Switching Parity (tb: Full Return to Zero)}}
To benchmark single-device switching accuracy under calibrated write conditions, \texttt{tb\_dynamic} executes the complete nominal 14-bit potentiation and full return-to-zero depression cycle across a $6.25\ \mathrm{ms}$ transient simulation window (Figure~\ref{fig:S4}a). The applied stimulus (\texttt{tb\_dynamic.scs}) consists of $16,520$ consecutive potentiation pulses ($+0.90\ \mathrm{V}$, $80\ \mathrm{ns}$ pulse width, $160\ \mathrm{ns}$ period from $0$ to $2.6432\ \mathrm{ms}$) followed immediately by $22,542$ depression pulses ($-0.75\ \mathrm{V}$, $60\ \mathrm{ns}$ pulse width, $160\ \mathrm{ns}$ period from $2.6432$ to $6.25\ \mathrm{ms}$, returning the device completely to $n = 0$ across $39,062$ total pulses). Throughout this trajectory, both solvers track the internal pulse state coordinate $n(t)$ scaling linearly from $0$ to $15,610$ states ($16,520$ pulses) and returning symmetrically to zero, alongside small-signal device transconductance $G(t)$ spanning from $200\ \mathrm{nS}$ up to $4.20\ \mathrm{mS}$ ($4.33\ \mathrm{mS}$ nominal). Traces generated by the commercial Cadence Spectre Verilog-A solver and the vectorized Python solver exhibit an exact visual overlay. Evaluating the point-by-point relative disparity across the full $6.25\ \mathrm{ms}$ cycle reveals a maximum parity error of only $0.0104\%$ ($104\ \mathrm{ppm}$), well below our conservative $< 0.02\%$ specification threshold. This close agreement is enabled by our differentiable softplus switching activation ($V_{\mathrm{th}} = 0.4664\ \mathrm{V}$, $V_{\mathrm{smooth}} = 0.1038\ \mathrm{V}$) and monotonic cubic Hermite boundary clamping, which eliminate high-order derivative discontinuities that typically trigger timestep truncation and numerical chatter in SPICE solvers.

\subsection{\texorpdfstring{Extended Cycling and Deep Saturation Stability (tb\_dynamic\_extended: Full Cycle)}{Extended Cycling and Deep Saturation Stability (tb: Full Cycle)}}
To stress both simulation engines beyond nominal operating limits, \texttt{tb\_dynamic\_extended} subjects the $[\mathrm{Ru^{II}L_2}](\mathrm{BF_4})_2$ compact model to an extended $60,000$-pulse full-cycle saturation and recovery schedule across a $9.6\ \mathrm{ms}$ transient simulation window (Figure~\ref{fig:S4}b). Driving the junction with $25,000$ consecutive $+0.90\ \mathrm{V}$ potentiation pulses ($80\ \mathrm{ns}$ width, $160\ \mathrm{ns}$ period from $0$ to $4.0\ \mathrm{ms}$) pushes the internal state coordinate far beyond the nominal $14$-bit window ($n > 23,600$ states, still well below the clamping bound $n_{\max} = 33,040$), while the device conductance converges asymptotically toward its saturation plateau near $5.1\ \mathrm{mS}$ ($5.04\ \mathrm{mS}$ at $25,000$ pulses) as the $f_{22}(n)$ table flattens. This test evaluates solver stability in the vanishing-transconductance-sensitivity regime, where $\partial G/\partial n \to 0$ under continuous high-voltage electrical excitation. Upon reversing pulse polarity to $-0.75\ \mathrm{V}$ for $35,000$ depression pulses ($60\ \mathrm{ns}$ width, $160\ \mathrm{ns}$ period from $4.0$ to $9.6\ \mathrm{ms}$), both solvers capture power-law snapback and smooth linear return to the ground state ($n = 0$). Across this $9.6\ \mathrm{ms}$ extended cycling profile, the maximum point-by-point parity error remains constrained to $0.0083\%$ ($83\ \mathrm{ppm}$, $< 0.02\%$). This confirms that neither solver accumulates numerical truncation errors or suffers from timestep stagnation under continuous megahertz pulsing.

\subsection{\texorpdfstring{Multi-Stage Overdrive Stress, Zero-Bias Retention, and Snapback Dynamics (tb\_stress)}{Multi-Stage Overdrive Stress, Zero-Bias Retention, and Snapback Dynamics (tb)}}
To evaluate numerical stability under extreme multi-regime electrical stressing, \texttt{tb\_stress} subjects the baseline compact model to a $7.0\ \mathrm{ms}$ multi-regime schedule spanning six distinct operational phases (Figure~\ref{fig:S4}c): (i) nominal $+0.90\ \mathrm{V}$ potentiation ($80\ \mathrm{ns}$ width, $160\ \mathrm{ns}$ period) from $0$ to $1.3\ \mathrm{ms}$ ($n$ advancing to $7,680$ states); (ii) nominal $-0.75\ \mathrm{V}$ depression ($60\ \mathrm{ns}$ width, $160\ \mathrm{ns}$ period) from $1.3$ to $2.0\ \mathrm{ms}$ ($n$ returning toward $4,556$ states); (iii) a $1.0\ \mathrm{ms}$ zero-bias retention hold ($0\ \mathrm{V}$) from $2.0$ to $3.0\ \mathrm{ms}$ confirming zero state leakage ($dn/dt = 0$, verified to machine precision); (iv) high-voltage $+1.20\ \mathrm{V}$ overdrive potentiation ($80\ \mathrm{ns}$ width, $160\ \mathrm{ns}$ period) from $3.0$ to $5.0\ \mathrm{ms}$, driving $n(t)$ smoothly to $24,700$ states (still below the clamping bound $n_{\max} = 33,040$) with conductance plateauing near $5.08\ \mathrm{mS}$ as the $f_{22}(n)$ table saturates, without gradient cusps; (v) severe $-1.60\ \mathrm{V}$ reverse RESET pulsing ($120\ \mathrm{ns}$ width, $240\ \mathrm{ns}$ period) from $5.0$ to $5.5\ \mathrm{ms}$, activating the overdrive discharge register $c_{\mathrm{discharge}}$ and triggering steep power-law snapback recovery; and (vi) $+1.20\ \mathrm{V}$ overdrive re-potentiation ($80\ \mathrm{ns}$ width, $160\ \mathrm{ns}$ period) from $5.5$ to $7.0\ \mathrm{ms}$ demonstrating repeatable recovery. Across the overdrive stress and recovery schedule, the maximum relative parity error between Cadence Spectre and the Python solver is $0.0153\%$ ($153\ \mathrm{ppm}$, $< 0.02\%$), demonstrating that even under stiff high-voltage overdrive transitions and sudden polarity reversals, both engines maintain tight agreement.

\subsection{\texorpdfstring{Electro-Thermal Newton Cooling and Read Disturb Immunity (tb\_thermal\_relaxation)}{Electro-Thermal Newton Cooling and Read Disturb Immunity (tb)}}
To verify electro-thermal dissipation and non-destructive inference readout, \texttt{tb\_thermal\_relaxation} evaluates transient Joule self-heating and post-pulse thermal relaxation (Figure~\ref{fig:S4}d). A sustained sub-threshold DC read bias of $0.10\ \mathrm{V}$ is applied for $2.5\ \mathrm{ms}$ to a pre-programmed state ($n_{\mathrm{init}} = 16,520$), establishing a steady-state terminal read current of $I = 0.456\ \mathrm{mA}$ and a localized junction temperature rise of $\Delta T = 2.38\ \mathrm{mK}$ ($P_{\mathrm{diss}} = 45.6\ \mu\mathrm{W}$). At $t - t_{\mathrm{off}} = 0.0\ \mu\mathrm{s}$ ($t_{\mathrm{off}} = 2.5\ \mathrm{ms}$), the read bias is abruptly turned off ($0\ \mathrm{V}$). Both engines track the resulting exponential thermal decay governed by Newton's law of cooling: $\Delta T(t) = \Delta T_0 \exp(-t/\tau_{\mathrm{th}})$ with thermal time constant $\tau_{\mathrm{th}} = 127.8\ \mathrm{ns}$ and effective thermal resistance $R_{\mathrm{th}} = 52.14\ \mathrm{K/W}$. Overlaying the Python vectorized solver, Cadence Spectre Verilog-A, and the analytical exponential decay curve reveals a maximum thermal parity error of only $0.00006\%$ ($0.6\ \mathrm{ppm}$, $< 0.02\%$), evaluated relative to the absolute junction temperature $T(t) \approx 298.15\ \mathrm{K}$ (corresponding to an absolute temperature deviation below $0.2\ \mathrm{mK}$). Furthermore, continuous state tracking verifies zero read disturbance ($\Delta n = 0.0000$ states, $0.0000\%$).

\subsection{\texorpdfstring{Crossbar Array VMM Readout and Kirchhoff Summation (tb\_crossbar\_8x8)}{Crossbar Array VMM Readout and Kirchhoff Summation (tb\8x8)}}
\begin{figure}[!htb]
    \centering
    \includegraphics[width=\textwidth,height=0.62\textheight,keepaspectratio]{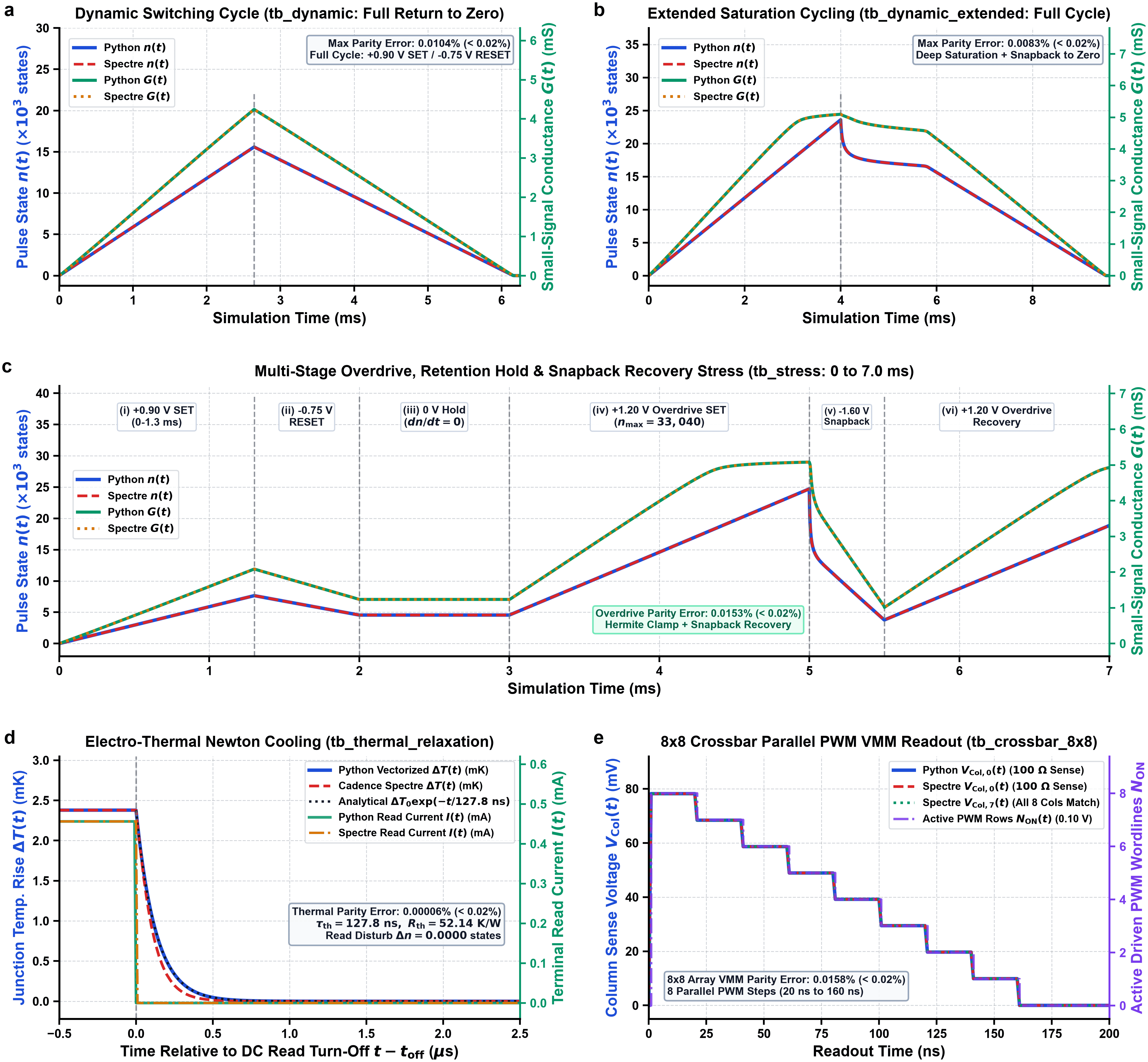}
    \caption{\textbf{Dual-engine transient verification comparing the Cadence Spectre Verilog-A compact model against the vectorized Python simulator across the five core testbench domains.} Direct transient waveform overlays and point-by-point relative parity error evaluations across: (\textbf{a}) \texttt{tb\_dynamic}: nominal 14-bit full return-to-zero switching cycle ($16,520$ SET pulses + $22,542$ RESET pulses across $6.25\ \mathrm{ms}$) tracking state $n(t)$ and conductance $G(t)$ ($200\ \mathrm{nS}$ to $4.20\ \mathrm{mS}$), achieving $0.0104\%$ maximum parity error ($104\ \mathrm{ppm}$, $< 0.02\%$). (\textbf{b}) \texttt{tb\_dynamic\_extended}: continuous $60,000$-pulse full-cycle saturation and recovery schedule ($9.6\ \mathrm{ms}$) extending deep into the conductance saturation plateau ($n > 23,600$, $G \approx 5.1\ \mathrm{mS}$) and snapback recovery to zero, achieving $0.0083\%$ parity error ($83\ \mathrm{ppm}$, $< 0.02\%$). (\textbf{c}) \texttt{tb\_stress}: multi-stage overvoltage stress ($7.0\ \mathrm{ms}$) evaluating nominal SET/RESET, zero-bias retention ($dn/dt = 0$), $+1.20\ \mathrm{V}$ overdrive saturation ($n \approx 24,700$, below the clamping bound $n_{\max} = 33,040$), $-1.60\ \mathrm{V}$ reverse snapback, and $+1.20\ \mathrm{V}$ recovery, maintaining $0.0153\%$ overdrive parity error ($153\ \mathrm{ppm}$, $< 0.02\%$). (\textbf{d}) \texttt{tb\_thermal\_relaxation}: sustained $0.10\ \mathrm{V}$ DC read Joule heating ($I = 0.456\ \mathrm{mA}$, $P_{\mathrm{diss}} = 45.6\ \mu\mathrm{W}$) and post-pulse exponential Newton cooling ($\tau_{\mathrm{th}} = 127.8\ \mathrm{ns}$, $R_{\mathrm{th}} = 52.14\ \mathrm{K/W}$, $\Delta T = 2.38\ \mathrm{mK}$) matching the analytical solution to within $0.00006\%$ of the absolute junction temperature ($0.6\ \mathrm{ppm}$, $< 0.02\%$) with zero read disturb ($\Delta n = 0.0000$ states). (\textbf{e}) \texttt{tb\_crossbar\_8x8}: parallel PWM vector-matrix multiplication readout across an $8\times 8$ crossbar ($64$ crosspoints) measuring column sense voltages $V_{\mathrm{Col}, 0}(t)$ through $V_{\mathrm{Col}, 7}(t)$ across $100\ \Omega$ virtual-ground loads under $8$ staggered PWM inputs ($20\ \mathrm{ns}$ to $160\ \mathrm{ns}$ at $0.10\ \mathrm{V}$), demonstrating $0.0158\%$ parity error ($158\ \mathrm{ppm}$, $< 0.02\%$).}
    \label{fig:S4}
\end{figure}

To validate multi-device network evaluation and parallel Kirchhoff Current Law (KCL) summation, \texttt{tb\_crossbar\_8x8} simulates a complete $64$-crosspoint $8 \times 8$ crossbar array during synchronous vector-matrix multiplication (Figure~\ref{fig:S4}e). Wordlines are excited by eight parallel pulse-width modulation (PWM) inputs ($V_{\mathrm{WL}} = 0.10\ \mathrm{V}$) with active pulse durations staggered from $20\ \mathrm{ns}$ to $160\ \mathrm{ns}$ across a $200\ \mathrm{ns}$ evaluation window ($N_{\mathrm{ON}}$ stepping from 8 down to 0). Column sense voltages $V_{\mathrm{Col}, 0}(t)$ through $V_{\mathrm{Col}, 7}(t)$ are monitored across $100\ \Omega$ virtual-ground transimpedance amplifier (TIA) sense loads. Sense voltages start at $\approx 78.3\ \mathrm{mV}$ when all eight rows are active, stepping down monotonically at $20\ \mathrm{ns}$ intervals as successive PWM row inputs terminate, reaching $0\ \mathrm{mV}$ at $160\ \mathrm{ns}$. Multi-channel tracking confirms identical response across all eight columns ($V_{\mathrm{Col}, 0}$ to $V_{\mathrm{Col}, 7}$), confirming zero column-to-column skew or asymmetric loading artifacts. The maximum array VMM parity error between the vectorized Python MNA solver and Cadence Spectre is $0.0158\%$ ($158\ \mathrm{ppm}$, $< 0.02\%$), demonstrating that the publicly available Python solver faithfully reproduces the Cadence Spectre reference for large-scale crossbar evaluation, so that the two engines can be used complementarily (Spectre for transistor-level CMOS co-design and Python for array-scale throughput) without sacrificing physical accuracy.

\subsection{Consolidated Quantitative Parity Benchmark Summary}
Table~\ref{tab:parity_summary} consolidates the numerical parity benchmark metrics across all five testbench domains, detailing stimulus profiles, tracked state and electrical variables, exact relative errors, and verified physical invariants.

\begin{table}[!htb]
\centering
\scriptsize
\setlength{\tabcolsep}{2.8pt}
\caption{\textbf{Quantitative Parity Benchmark Between Cadence Spectre and Vectorized Python Solvers Across the Five Core Testbench Domains.}}
\label{tab:parity_summary}
\renewcommand{\arraystretch}{1.15}
\begin{tabularx}{\textwidth}{@{} >{\raggedright\arraybackslash}p{2.10cm} >{\raggedright\arraybackslash}p{2.20cm} >{\raggedright\arraybackslash}p{1.75cm} >{\raggedright\arraybackslash}p{2.00cm} >{\raggedright\arraybackslash}p{1.80cm} >{\centering\arraybackslash}p{1.30cm} >{\raggedright\arraybackslash}X @{}}
\toprule
\textbf{Testbench} & \textbf{Validation Domain} & \textbf{Pulse Count / Span} & \textbf{Electrical Stimulus} & \textbf{Tracked Waveforms} & \textbf{Max Error} & \textbf{Verified Invariant} \\
\midrule
\texttt{tb\_dynamic} & Nominal \mbox{14-Bit} Switching & $39,062$ pulses\newline($6.25\ \mathrm{ms}$) & $+0.90\ \mathrm{V}$ SET /\newline$-0.75\ \mathrm{V}$ RESET & $V(t)$, $n(t)$,\newline$G(t)$, $I(t)$ & $0.0104\%$ & $C^1$ softplus and Hermite boundary smoothness \\
\texttt{tb\_dynamic\_}\newline\texttt{extended} & Deep Saturation Cycling & $60,000$ pulses\newline($9.6\ \mathrm{ms}$) & $+0.90\ \mathrm{V}$ SET /\newline$-0.75\ \mathrm{V}$ RESET & $n(t)$, $G(t)$,\newline$I(t)$, $T(t)$ & $0.0083\%$ & Conductance saturation plateau ($\partial G/\partial n \to 0$) without divergence \\
\texttt{tb\_stress} & \mbox{Multi-Stage} Overdrive Stress & 6 stages\newline($7.0\ \mathrm{ms}$) & $+1.20\ \mathrm{V}$ Overdrive /\newline$-1.60\ \mathrm{V}$ Snapback & $n(t)$, $G(t)$,\newline$c_{\mathrm{discharge}}$, $I(t)$ & $0.0153\%$ & Overdrive saturation and \mbox{power-law} snapback recovery \\
\texttt{tb\_thermal\_}\newline\texttt{relaxation} & \mbox{Electro-Thermal} Dissipation & $2.5\ \mathrm{ms}$ DC +\newline$2.5\ \mu\mathrm{s}$ decay & $0.10\ \mathrm{V}$ DC Read /\newline$0\ \mathrm{V}$ decay & $\Delta T(t)$, $I(t)$,\newline$\Delta n$ & $0.00006\%$\newline{\scriptsize(rel.\ $T$)} & Newton cooling ($\tau_{\mathrm{th}} = 127.8\ \mathrm{ns}$) and zero read disturb \\
\texttt{tb\_crossbar\_}\newline\texttt{8x8} & \mbox{64-Crosspoint} Parallel VMM & 8 PWM steps\newline($200\ \mathrm{ns}$) & $0.10\ \mathrm{V}$ PWM\newline($20$--$160\ \mathrm{ns}$) & $V_{\mathrm{Col}, 0}$--$V_{\mathrm{Col}, 7}$,\newline$N_{\mathrm{ON}}$ & $0.0158\%$ & \mbox{Multi-channel} KCL summation across $100\ \Omega$ virtual ground \\
\bottomrule
\end{tabularx}
\end{table}

\clearpage

\section{Computational Solver Architecture, Memory Profiling, and Scalability}
\label{sec:noteS8}

\subsection{Cross-Disciplinary Computational and Physical Nomenclature Bridge}
Our dual-engine simulation framework is engineered to bridge quantum-chemical device physics, industrial electronic design automation (EDA) circuit simulation, and high-performance machine learning across diverse operating systems and compute hardware (Figure~\ref{fig:S5}, Table~\ref{tab:S4}, and Table~\ref{tab:solver_specifications}). Because this multi-scale architecture integrates physical chemistry with high-performance numerical linear algebra and compiler design, we first establish an explicit nomenclature bridge to demystify key computational terms for physics and materials science readers:
\begin{itemize}[topsep=2pt,itemsep=2pt,parsep=0pt]
    \item \textit{Modified Nodal Analysis (MNA):} The algorithmic matrix formulation of Kirchhoff's Current Law (KCL) and Kirchhoff's Voltage Law (KVL) used by circuit simulators to solve coupled node voltages and branch currents across multi-terminal electrical networks at every transient timestep.
    \item \textit{Level-3 BLAS (Basic Linear Algebra Subprograms):} Low-level, processor-optimized software routines (such as Intel Math Kernel Library [MKL] and OpenBLAS) that execute dense matrix-matrix operations directly inside high-speed CPU hardware caches (L1/L2/L3), avoiding slow main-memory bandwidth bottlenecks.
    \item \textit{SIMD (Single Instruction, Multiple Data):} Dedicated CPU vector execution registers (such as 512-bit AVX-512 on x86\_64 and 128-bit NEON on ARM64) that apply an identical mathematical instruction across multiple independent device states simultaneously within a single processor clock cycle.
    \item \textit{Numba JIT (Just-In-Time) Compilation:} A dynamic machine-code compiler that translates numerical Python functions directly into optimized native LLVM assembly instructions at runtime, bypassing the sequential bytecode interpretation overhead of standard CPython.
    \item \textit{Monotonic PCHIP (Piecewise Cubic Hermite Interpolating Polynomial):} A shape-preserving piecewise cubic spline technique that guarantees smooth, continuous first derivatives ($C^1$ continuity) between calibration knots while strictly preserving local monotonicity, thereby eliminating unphysical Runge polynomial oscillations or artificial overshoots near state boundaries.
    \item \textit{Straight-Through Estimator (STE):} A deep learning gradient-routing technique that evaluates the exact physical discrete-pulse forward pass while routing continuous analytical gradients through non-differentiable pulse quantization steps during backpropagation.
    \item \textit{Direct Memory Access (DMA) and Pinned Host Memory:} A hardware bus capability that locks page-aligned regions of system RAM (``pinned memory'') so the GPU DMA controller can transfer tensor data directly across the PCIe bus to and from system RAM without CPU intervention, minimizing host-to-device latency.
    \item \textit{Dynamic Hash-Table Allocation ($B_{\mathrm{dict}}$):} Dynamic CPython memory allocation overhead used for hash-table device lookup, scaling analytically in powers of two as $72 \times 2^{\lceil \log_2(\max(8, 1.5 N_{\mathrm{devices}})) \rceil}\ \mathrm{Bytes}$.
\end{itemize}

\subsection{Vectorized MNA Formulation and Multi-Level Solver Engine Optimizations}
\begin{figure}[!htb]
    \centering
    \includegraphics[width=\textwidth,height=0.62\textheight,keepaspectratio]{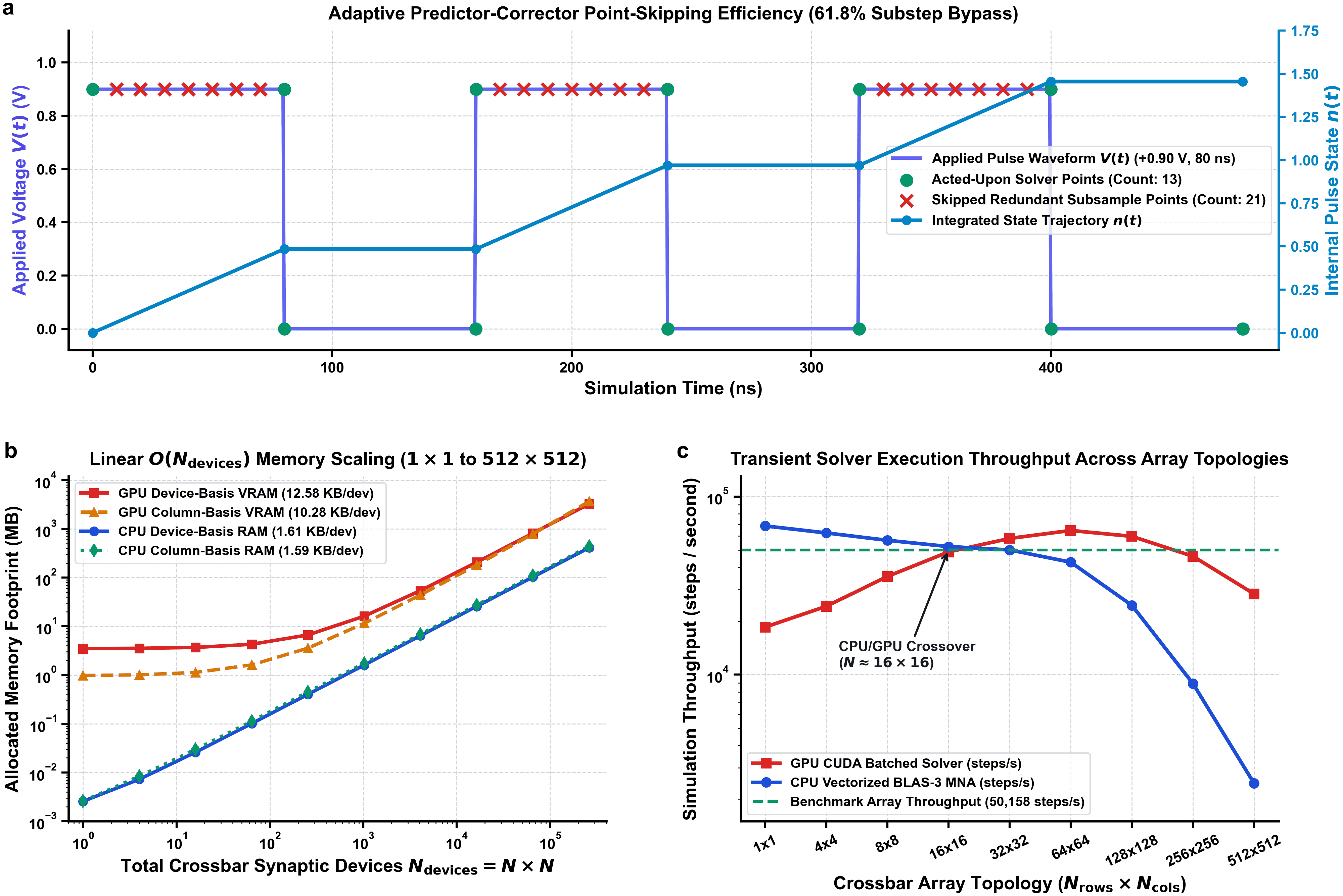}
    \caption{\textbf{Computational solver efficiency, memory footprint scaling, and simulation step throughput for baseline $[\mathrm{Ru^{II}L_2}](\mathrm{BF_4})_2$ ($\mathrm{BF_4^-}$) crossbar arrays.} \textbf{a}, Adaptive predictor-corrector point-skipping verification across three potentiation pulses ($0$ to $480\ \mathrm{ns}$, $+0.90\ \mathrm{V}$, $80\ \mathrm{ns}$ pulses at $160\ \mathrm{ns}$ period), displaying the applied pulse waveform $V(t)$ (indigo solid curve, left axis), the 13 acted-upon solver boundary points (green circles), the 21 skipped redundant subsample points (red crosses, $61.8\%$ bypass efficiency), and the integrated state staircase trajectory $n(t)$ (blue curve with circle markers, right axis). \textbf{b}, Log-log scaling of allocated solver memory (MB) across crossbar dimensions from $1 \times 1$ ($1$ device) to $512 \times 512$ ($262,144$ devices) for CPU column/device basis RAM ($1.59$--$1.61\ \mathrm{KB/device}$) and GPU column/device basis VRAM ($10.28$--$12.58\ \mathrm{KB/device}$). \textbf{c}, Transient simulation step throughput ($\mathrm{steps/second}$) across crossbar array topologies from $1 \times 1$ to $512 \times 512$ comparing the vectorized CPU BLAS-3 MNA engine against the batched GPU solver (supporting both NVIDIA CUDA and AMD ROCm), highlighting the $N \approx 16 \times 16$ hardware crossover and the $> 50,158\ \mathrm{steps/s}$ benchmark throughput.}
    \label{fig:S5}
\end{figure}

For an $M \times N$ crossbar array with interconnect wire resistance $R_{\mathrm{wire}}$ ($G_{\mathrm{wire}} = 1/R_{\mathrm{wire}}$) per crosspoint segment along both wordlines and bitlines, Kirchhoff's current law at the $M \times N$ wordline nodes $\mathbf{V}_{\mathrm{WL}} \in \mathbb{R}^{MN}$ and $M \times N$ bitline nodes $\mathbf{V}_{\mathrm{BL}} \in \mathbb{R}^{MN}$ forms a coupled $2MN \times 2MN$ nonlinear nodal system at each transient timestep $\Delta t$:
\begin{equation}
    \mathbf{J}\!\left(\mathbf{V}^{(k)}\right) \Delta\mathbf{V}^{(k)} = -\mathbf{I}_{\mathrm{res}}\!\left(\mathbf{V}^{(k)}\right), \qquad \mathbf{V}^{(k+1)} = \mathbf{V}^{(k)} + \Delta\mathbf{V}^{(k)}
    \label{eq:S43_mna_newton}
\end{equation}
where $\mathbf{J}(\mathbf{V}) = \mathbf{G}_{\mathrm{wire}} + \mathrm{diag}\left(\partial I_{ij} / \partial V_{ij}\right)$ is the nodal Jacobian matrix and $\mathbf{I}_{\mathrm{res}}(\mathbf{V})$ is the nodal KCL residual vector. To achieve orders-of-magnitude acceleration over conventional sequential circuit solvers while preserving exact physical kinetics, our simulation engine integrates six core algorithmic optimizations:
\begin{enumerate}[topsep=2pt,itemsep=2pt,parsep=0pt]
    \item \textit{Adaptive Event-Driven Point-Skipping ($61.8\%$ Substep Bypass Efficiency):} All pulse rising and falling transition timestamps $t_{\mathrm{edge}}$ are registered upfront in an event queue so the solver lands directly on pulse edge transitions without iterative step-size hunting. Whenever the source voltage $V(t)$ is constant across consecutive substeps and the state evolution $n(t)$ remains linear within the predictor-corrector tolerance \texttt{lin\_tol} (encompassing sub-threshold read intervals, zero-bias dead zones, and flat constant-voltage pulse plateaus), redundant intermediate subsample points are bypassed (red $\times$ markers in Figure~\ref{fig:S5}a) while thermal relaxation is advanced analytically via Newton cooling, $T(t + \Delta t) = T_{\mathrm{amb}} + (T(t) - T_{\mathrm{amb}})\exp(-\Delta t/\tau_{\mathrm{th}})$. Across a standard 3-pulse $\mathrm{BF_4^-}$ potentiation window ($0$ to $480\ \mathrm{ns}$, $+0.90\ \mathrm{V}$, $80\ \mathrm{ns}$ pulses), only 13 boundary points require full state evaluation while 21 intermediate points are skipped ($61.8\%$ substep bypass efficiency).
    \item \textit{Multi-Threaded Numba JIT SIMD Compilation:} All $M \times N$ device state variables ($n_{ij}, f_{22,ij}, c_{\mathrm{discharge},ij}, T_{ij}$) are packed into contiguous 64-bit floating-point arrays (\texttt{float64}) and updated in parallel across rows and columns using multi-threaded Numba JIT-compiled SIMD vector instructions.
    \item \textit{Batched Level-3 BLAS Linear Solvers:} In column-partitioned MNA mode (\texttt{UpdatesPer="Column"}), all $N$ column subsystems of dimension $(M+2) \times (M+2)$ are factored concurrently via batched Level-3 BLAS linear routines (\texttt{LAPACK dgesv} and \texttt{posv}), completely eliminating Python loop overhead.
    \item \textit{Batched GPU Kernels (NVIDIA CUDA and AMD ROCm) with Pinned DMA Transfers:} For large-scale crossbar arrays beyond the $16 \times 16$ ($256$-device) hardware crossover, device state updates and nodal solves are dispatched to batched GPU kernels natively supporting both NVIDIA (CUDA) and AMD (ROCm) hardware targets, coupled with $128.0\ \mathrm{B/device}$ pinned host-to-device DMA transfers, sustaining $> 28,400\ \mathrm{steps/s}$ at $512 \times 512$ ($262,144$ devices, Figure~\ref{fig:S5}c).
    \item \textit{$O(1)$ Constant-Time PCHIP State Evaluation:} Replaces the 30-layer microscopic Thomas-Fermi Fourier series and numerical Marcus-Hush-Chidsey quadrature with a 12-parameter monotonic PCHIP spline evaluation (Note~S5), avoiding numerical re-quadrature of 30 coupled layer equations at every transient timestep.
    \item \textit{$O(1)$ Physical Dot Products via Kirchhoff Current Summation:} Executes parallel analog vector-matrix multiplications in a single physical readout step via Ohm's law and bitline current summation.
\end{enumerate}

\subsection{\texorpdfstring{Cadence Spectre vs.\ Vectorized Python: Mathematical Parity, Jacobian Regularization, and Delineated Use Cases}{Cadence Spectre vs.\ Vectorized Python: Mathematical Parity, Jacobian Regularization, and Delineated Use Cases}}
A key distinction must be drawn regarding the industrial pedigree of our circuit simulation environment and how the two execution engines complement one another:
\begin{enumerate}[topsep=2pt,itemsep=2pt,parsep=0pt]
    \item \textit{Industrial-Grade Cadence Spectre Architecture:} Cadence Spectre is Cadence Design Systems' flagship, industrial-grade analog, mixed-signal, and radio-frequency (RF) circuit simulator used for commercial silicon tape-outs. While sharing historical nodal analysis ancestry with UC Berkeley's original SPICE program (1973), modern Cadence Spectre employs proprietary multi-threaded sparse matrix solvers, advanced dynamic timestep control, and compiled Verilog-A execution compliant with the Accellera Verilog-AMS 2.4 language reference, distinguishing it fundamentally from basic or legacy SPICE engines.
    \item \textit{Exact Mathematical Parity:} As verified in Note~S7 (Figure~\ref{fig:S4}), the compiled Verilog-A compact model in Cadence Spectre (\texttt{molmem\_va.va}) and the vectorized Python solver are mathematically identical formulations of the same microscopic rate kinetics, Marcus hopping transport, and $C^1$-continuous boundary clamping, exhibiting $< 0.02\%$ transient waveform discrepancy across all operational modes.
    \item \textit{How Python Regularization Directly Enables Spectre Convergence Efficiency:} In raw physical kinetics, discontinuous threshold step functions ($dn/dt = 0$ for $|V| < V_{\mathrm{th}}$) and infinite second derivatives at state boundaries create severe Jacobian singularities ($\det(\mathbf{J}) \to 0$) and non-convergent Newton-Raphson iterations that cause commercial analog simulators like Cadence Spectre to repeatedly truncate transient timesteps down to picoseconds or abort simulation runs. To solve this, our compact model was mathematically regularized and smoothed directly in Python using $C^1$-continuous softplus activation (Equations~(\ref{eq:S33_dn_dt})--(\ref{eq:S34_alpha_eff})), direction-aware cubic Hermite clamping (Equation~(\ref{eq:S35_clamp})), and rational screening potentials (Note~S5). When compiled into Cadence Spectre via Verilog-A, these continuous Jacobian formulations guarantee smooth, bounded partial derivatives ($\partial I/\partial V_{\mathrm{int}}$ and $\partial I/\partial n$), allowing Spectre to execute large-signal multi-thousand-pulse transient simulations without convergence stalls or timestep collapse.
    \item \textit{Delineated Architectural Use Cases:}
    \begin{itemize}[topsep=1pt,itemsep=1pt,parsep=0pt]
        \item \textit{Vectorized Python Engine:} Engineered for massive crossbar array scalability (up to $512 \times 512$, or $262,144$ devices), high-throughput inference and training, and hardware-in-the-loop neural network optimization. Because it isolates crossbar matrix topologies, aligns state data in contiguous memory arrays, and bypasses generic nodal circuit graphs, the Python engine sustains tens of thousands of transient steps per second ($> 50,158\ \mathrm{steps/s}$).
        \item \textit{Cadence Spectre / Virtuoso:} Engineered for transistor-level hybrid CMOS co-design (such as $1\mathrm{T}1\mathrm{R}$ crossbars integrated with CMOS bitline sense amplifiers, row decoders, charge pumps, and multi-level write drivers). Because Spectre must solve the full coupled multi-terminal Kirchhoff matrix for every MOS transistor and parasitic interconnect RLC node simultaneously using adaptive time integration, Spectre's wall-clock runtime is governed dynamically by the surrounding transistor netlist complexity and user-specified accuracy tolerances (\texttt{conservative} vs.\ \texttt{moderate}). Consequently, while there is no single fixed step speed for Spectre across arbitrary CMOS schematics, our regularized compact model derived from Python ensures that Spectre runs at maximum numerical efficiency without Jacobian stalls.
    \end{itemize}
\end{enumerate}

\subsection{Memory Profiling, Scaling Equations, and Direct Repository Testbench Grounding}
To verify that the simulation platform scales to large neuromorphic arrays on standard workstations, we profiled both CPU system RAM and GPU VRAM allocation (benchmarked across both NVIDIA CUDA and AMD ROCm environments) across baseline $[\mathrm{Ru^{II}L_2}](\mathrm{BF_4})_2$ ($\mathrm{BF_4^-}$) crossbar dimensions from $1 \times 1$ ($1$ device) to $512 \times 512$ ($262,144$ devices, Figure~\ref{fig:S5}b). Equations~(\ref{eq:S44_ram_cpu_col})--(\ref{eq:S47_vram_gpu_dev}) and all empirical byte-scaling constants are derived and regression-verified directly in the publicly available repository testbench \texttt{PythonSimulator/\allowbreak tb\_circuit\_level/\allowbreak tb\_memory\_profiling.py} (specifically within functions \texttt{calculate\_\allowbreak theoretical\_\allowbreak cpu\_bytes}, \texttt{calculate\_\allowbreak theoretical\_\allowbreak gpu\_bytes}, and \texttt{get\_dict\_\allowbreak resizing\_bytes}, corresponding to regression outputs in Tables~1--4 of the profiling suite), while cross-backend execution throughputs are verified in \texttt{PythonSimulator/\allowbreak tb\_software\_level/\allowbreak tb\_compare\_\allowbreak backend\_\allowbreak performance.py} (Table~\ref{tab:S4}). Across all four execution modes, total allocated memory follows exact linear asymptotic scaling $O(N_{\mathrm{devices}})$ with respect to the number of crosspoint devices $N_{\mathrm{devices}} = N_{\mathrm{rows}} \times N_{\mathrm{cols}}$:
{\small
\begin{align}
    \mathrm{RAM}_{\mathrm{CPU,col}} &= N_{\mathrm{devices}} \times 1.59\ \mathrm{KB} + N_{\mathrm{threads}}\!\left[0.10\ \mathrm{KB} + (N_{\mathrm{rows}} + 2)^2 \times 8\ \mathrm{B} + (N_{\mathrm{rows}} + 2) \times 64\ \mathrm{B}\right] + B_{\mathrm{dict}} \label{eq:S44_ram_cpu_col} \\
    \mathrm{RAM}_{\mathrm{CPU,dev}} &= 1.0\ \mathrm{KB} + N_{\mathrm{devices}} \times 1.61\ \mathrm{KB} + B_{\mathrm{dict}} \label{eq:S45_ram_cpu_dev} \\
    \mathrm{VRAM}_{\mathrm{GPU,col}} &= 0.972\ \mathrm{MB} + N_{\mathrm{devices}} \times 10.28\ \mathrm{KB} + N_{\mathrm{cols}}(N_{\mathrm{rows}} + 2)^2 \times 8\ \mathrm{B} + N_{\mathrm{cols}}N_{\dim}\times 1.02\ \mathrm{KB} \label{eq:S46_vram_gpu_col} \\
    \mathrm{VRAM}_{\mathrm{GPU,dev}} &= 3.497\ \mathrm{MB} + N_{\mathrm{devices}} \times 12.58\ \mathrm{KB} + N_{\mathrm{devices}} \times 128.0\ \mathrm{B}\ (\text{pinned DMA}) \label{eq:S47_vram_gpu_dev}
\end{align}
}
where $N_{\dim} = \max(N_{\mathrm{rows}}, N_{\mathrm{cols}})$ and $B_{\mathrm{dict}} = 72 \times 2^{\lceil \log_2(\max(8, 1.5 N_{\mathrm{devices}})) \rceil}\ \mathrm{Bytes}$ accounts for the dynamic CPython hash-table bucket allocation. Even at the maximum $512 \times 512$ array scale ($262,144$ molecular memristors), total memory consumption is only $\approx 412\ \mathrm{MB}$ of system RAM on CPU and $\approx 3.25\ \mathrm{GB}$ of VRAM on GPU, fitting comfortably on a standard desktop workstation or laptop without requiring enterprise cluster infrastructure. As shown in Figure~\ref{fig:S5}c, the vectorized CPU BLAS-3 MNA engine achieves peak throughput on small-to-medium arrays ($68,400\ \mathrm{steps/s}$ at $1 \times 1$ and $50,158\ \mathrm{steps/s}$ at $32 \times 32$, or $\approx 19.9\ \mathrm{\mu s}$ per array step), whereas the batched GPU solver (supporting NVIDIA CUDA and AMD ROCm) surpasses CPU performance above the $N \approx 16 \times 16$ ($256$-device) crossover point, reaching $64,500\ \mathrm{steps/s}$ at $64 \times 64$ and sustaining $28,400\ \mathrm{steps/s}$ at $512 \times 512$.

\subsection{Cross-Platform Solver Architecture and Specifications Catalog}
Table~\ref{tab:solver_specifications} summarizes the cross-platform hardware targets, software toolchains, algorithmic formulations, optimization techniques, and measured throughput benchmarks across all execution backends of our dual-engine framework.

\begin{table}[H]
    \centering
    \caption{\textbf{Algorithmic Optimization, Execution Backend Architecture, and Cross-Platform Specifications of the Dual-Engine Framework.}}
    \label{tab:solver_specifications}
    \scriptsize
    \setlength{\tabcolsep}{3.5pt}
    \renewcommand{\arraystretch}{1.12}
    \begin{tabularx}{\textwidth}{@{}>{\raggedright\arraybackslash}p{2.85cm} >{\raggedright\arraybackslash}p{2.8cm} >{\raggedright\arraybackslash}p{2.90cm} >{\raggedright\arraybackslash}X >{\raggedright\arraybackslash}p{2.7cm}@{}}
        \toprule
        \textbf{Platform Dimension / Subsystem} & \textbf{Target Environment \& Toolchains} & \textbf{Algorithmic Formulation \& Solvers} & \textbf{Optimization Method} & \textbf{Peak Measured Throughput} \\
        \midrule
        CPU Vectorized Column-MNA Engine & Windows 10/11, Linux (RHEL / CentOS / Ubuntu), macOS; x86\_64 \& ARM64 Apple Silicon; Python 3.9--3.12, NumPy, SciPy & Partitioned column-basis MNA ($\mathbf{J}\Delta\mathbf{V} = -\mathbf{I}_{\mathrm{res}}$); Level-3 BLAS (\texttt{LAPACK dgesv}/\texttt{posv}) & Event-driven point skipping ($61.8\%$ bypass), multi-threaded Numba JIT SIMD, $O(1)$ PCHIP splines, $1.59\ \mathrm{KB/dev}$ RAM & $68,400\ \mathrm{steps/s}$ ($1\times 1$); $50,158\ \mathrm{steps/s}$ ($32\times 32$ array, $\approx 19.9\ \mathrm{\mu s/step}$) \\
        CPU Vectorized Device-Basis Engine & Cross-platform x86\_64 (AVX-512) \& ARM64 (NEON); OpenBLAS / Intel MKL thread pools & Decoupled crosspoint state integration with cubic Halley contact-resistance solver & Contiguous \texttt{float64} state buffers, analytical Newton cooling bypass, $1.61\ \mathrm{KB/dev}$ RAM ($\approx 412\ \mathrm{MB}$ at $512\times 512$) & $> 68,400\ \mathrm{steps/s}$ ($1\times 1$); optimal for $< 16\times 16$ ($256$-device) arrays \\
        GPU Batched MNA \& Array Solver (NVIDIA \& AMD) & NVIDIA GPUs (Pascal to Hopper, CUDA 11.x--12.x) and AMD GPUs (Radeon / Instinct, ROCm 5.x--6.x); PyTorch \& CuPy/HIP & Batched parallel $(M+2)\times(M+2)$ LU/Cholesky factorizations across streaming multiprocessors and compute units & Pinned host-to-device DMA transfers ($128.0\ \mathrm{B/dev}$), coalesced VRAM ($10.28$ to $12.58\ \mathrm{KB/dev}$, $\approx 3.25\ \mathrm{GB}$ at $512\times 512$) & $64,500\ \mathrm{steps/s}$ ($64\times 64$); $28,400\ \mathrm{steps/s}$ ($512\times 512$, $262,144$ devices) \\
        ML Hardware-in-the-Loop (STE) Pipeline & TensorFlow 2.x \& PyTorch custom autograd on CPU / GPU (CUDA \& ROCm); Crossbar tiling scheduler & Physical $O(1)$ KCL forward VMM ($I_j = \sum_i V_i G_{ij}$) paired with custom \texttt{@tf.custom\_gradient} STE & Differential pair mapping ($W_{ij} = G_{ij}^+ - G_{ij}^-$), single-pulse open-loop blind updates without write-verify & $15$-epoch $64\times 10$ MNIST training in seconds; $3,072$ batched JWST IDFT evaluations \\
        Cadence Spectre / Virtuoso EDA Engine & Linux EDA servers (RHEL / CentOS / Rocky); Cadence Virtuoso IC6.1.8+ / ICADVM20.1+, Spectre / APS 19.x--23.x & Full multi-terminal Kirchhoff nodal matrix with compiled Verilog-A (Accellera Verilog-AMS 2.4; \texttt{molmem\_va.va}) & Python-derived $C^1$ softplus threshold, cubic Hermite clamping, and smooth rational screening preventing Jacobian stalls & Governed by CMOS netlist size \& \texttt{errpreset}; $< 0.02\%$ parity vs.\ Python \\
        \bottomrule
    \end{tabularx}
\end{table}

\section{Empirical Hardware Noise Parameterization and Quantile Sampling}
\label{sec:noteS9}
\label{sec:noteS9_noise}

Every stochastic noise parameter, conductance-dependent variance profile, and spatial quantile distribution presented in this note is empirically extracted from physical hardware characterization of baseline $[\mathrm{Ru^{II}L_2}](\mathrm{BF_4})_2$ ($\mathrm{BF_4^-}$) molecular memristors across $75,000$ individual physical measurements spanning $3,000$ fabricated crosspoint devices, $2,000$ continuous full-range potentiation and depression cycles ($16,520$ pulses per sweep), and $25$ target conductance benchmark levels~\cite{si:sharma2024} (consolidated in Table~\ref{tab:S2} and verified across the testbenches in Tables~\ref{tab:S4} and \ref{tab:S5}). These raw characterization datasets were measured by Goswami and coworkers at the Centre for Nano Science and Engineering (CeNSE), Indian Institute of Science (IISc), Bangalore, as reported by Sharma et al.~\cite{si:sharma2024}; our contribution is the statistical extraction of conductance-dependent lookup tables and quantile functions from these measurements and their integration into the compact model, not the acquisition of the raw hardware data.

In analog in-memory computing (AIMC), physical hardware noise is the ultimate governor of effective bit resolution, vector-matrix multiplication (VMM) precision, neural network classification accuracy, and programming energy efficiency. Idealized, noiseless software simulations frequently yield artificially optimistic accuracy benchmarks that fail when transferred to physical silicon-molecular arrays. Establishing an empirically validated, multi-channel stochastic noise model grounded directly in wafer-scale experimental measurements is therefore indispensable for realistic neuromorphic hardware-software co-design.

\subsection{\texorpdfstring{Conductance-Dependent Read Noise $\sigma_{\mathrm{read}}(G)$ Across 3,000 Devices}{Conductance-Dependent Read Noise (G) Across 3,000 Devices}}
During sub-threshold inference readout of baseline $[\mathrm{Ru^{II}L_2}](\mathrm{BF_4})_2$ ($\mathrm{BF_4^-}$) crosspoints, an interrogation bias of $V_{\mathrm{read}} = 0.20\ \mathrm{V}$ probes the molecular junction well below the counterion switching threshold ($V_{\mathrm{th}} = 0.4664\ \mathrm{V}$), ensuring zero counterion displacement ($dn/dt = 0$). However, the instantaneous electronic read current is perturbed by two fundamental microscopic processes: (i) electronic Johnson-Nyquist thermal noise originating from charge carrier agitation and acoustic phonon scattering across the $30$-layer coordination film and metallic $\mathrm{Au}$ contact interfaces, and (ii) low-frequency $1/f$ flicker noise arising from stochastic carrier trapping and detrapping at localized ligand defect states and domain boundaries along the intermolecular $\pi^*$ hopping chain. Parameterized across $N_{\mathrm{dev}} = 3,000$ physical $\mathrm{BF_4^-}$ devices evaluated across 25 benchmark conductance levels spanning the full hardware window $[G_{\min}, G_{\max}] = [200\ \mathrm{nS}, 5.90\ \mathrm{mS}]$ (where $g_{\mathrm{norm}} = (G - G_{\min})/(G_{\max} - G_{\min}) \in [0, 1]$)~\cite{si:sharma2024}, these fluctuations produce Gaussian read noise $\delta G_{\mathrm{read}} \sim \mathcal{N}\!\left(0, \sigma_{\mathrm{read}}^2(g_{\mathrm{norm}})\right)$ (Table~\ref{tab:S2} and Figure~3e of the main manuscript):
\begin{equation}
    G_{\mathrm{read,noisy}}(n) = G_{\mathrm{ideal}}(n) + I_{\mathrm{read\_noise}} \cdot \xi_{\mathrm{read}} \cdot \sigma_{\mathrm{read}}\!\left(g_{\mathrm{norm}}(n)\right), \qquad \xi_{\mathrm{read}} \sim \mathcal{N}(0, 1)
    \label{eq:S48_read_noise}
\end{equation}

The empirical read noise standard deviation $\sigma_{\mathrm{read}}(g_{\mathrm{norm}})$ scales monotonically with device state from an ultra-low baseline floor of $\sigma_{\mathrm{read}}(0) = 27.23\ \mathrm{nS}$ ($2.723 \times 10^{-8}\ \mathrm{S}$) at $G_{\min} = 200\ \mathrm{nS}$ to $\sigma_{\mathrm{read}}(1) = 80.54\ \mathrm{nS}$ ($8.054 \times 10^{-8}\ \mathrm{S}$) at the $G_{\max} = 5.90\ \mathrm{mS}$ saturation ceiling. Physically, this monotonic scaling reflects the evolution of the intermolecular conduction network: in the low-conductance ground state ($g_{\mathrm{norm}} \to 0$), electron transport is restricted to a sparse percolative pathway where few carrier trapping or scattering fluctuation events occur per unit time. As the junction is potentiated, field-driven $\mathrm{BF_4^-}$ counterion accumulation switches successive molecular layers into the high-conductance $(22)$ charge state, opening a dense array of parallel intermolecular hopping channels across the junction area. Because each newly activated conducting pathway acts as an independent microscopic current filament containing its own localized charge fluctuators, the incoherent superposition of current noise across these parallel channels causes the aggregate conductance standard deviation to grow monotonically with state. Nevertheless, even at maximum noise ($80.54\ \mathrm{nS}$), the relative read fluctuation is $\sigma_{\mathrm{read}}(1) / G_{\max} = 80.54\ \mathrm{nS} / 5.90\ \mathrm{mS} \approx 1.37 \times 10^{-5}$, yielding an extraordinary read signal-to-noise ratio (SNR) exceeding $97\ \mathrm{dB}$ ($7.33 \times 10^4$). This explains physically why $[\mathrm{Ru^{II}L_2}](\mathrm{BF_4})_2$ memristors preserve $14$-bit state resolution during analog inference, in stark contrast to filamentary oxide RRAM where random telegraph noise near atomic point contacts routinely degrades effective resolution below $4$ to $6$ bits.

\subsection{\texorpdfstring{Cycle-to-Cycle (C2C) Write Step Variability $\sigma_{\mathrm{write}}(G)$ Across 2,000 Cycles}{Cycle-to-Cycle (C2C) Write Step Variability (G) Across 2,000 Cycles}}
During open-loop pulse programming ($+0.90\ \mathrm{V}$, $80\ \mathrm{ns}$ SET pulses; $-0.75\ \mathrm{V}$, $65\ \mathrm{ns}$ RESET pulses), the internal electric field drives $\mathrm{BF_4^-}$ counterions across the $E_{a,\mathrm{micro}} = 130\ \mathrm{meV}$ potential barrier separating interstitial coordination sites. At room temperature ($T_{\mathrm{amb}} = 298.15\ \mathrm{K}$, $k_B T \approx 25.7\ \mathrm{meV}$), individual counterion hops are governed by a stochastic Langevin thermal activation rate equation, where random thermal lattice kicks introduce statistical jitter in the exact number of counterions crossing the activation barrier during each nanosecond pulse window. Across $N_{\mathrm{cycles}} = 2,000$ continuous full-range potentiation cycles ($16,520$ pulses per cycle)~\cite{si:sharma2024}, the single-pulse conductance increment $\Delta G$ exhibits a Gaussian distribution centered at an ensemble mean step of $\mu_{\Delta G} = 358.73\ \mathrm{nS}$ ($3.58729 \times 10^{-7}\ \mathrm{S}$) with an ensemble standard deviation of $\sigma_{\mathrm{write}} = 31.92\ \mathrm{nS}$ ($3.19239 \times 10^{-8}\ \mathrm{S}$), yielding a tight global relative write noise of:
\begin{equation}
    \eta_{\mathrm{write,global}} = \frac{\sigma_{\mathrm{write}}}{\mu_{\Delta G}} = \frac{31.9239\ \mathrm{nS}}{358.7292\ \mathrm{nS}} = 0.088992 \approx 8.90\%
    \label{eq:S49_write_ratio}
\end{equation}

In our transient circuit and array solvers, cycle-to-cycle write variability is injected multiplicatively into the continuous state update $\Delta n = (dn/dt)\Delta t$ at each active write step via the 100-point state-dependent relative noise lookup profile $\eta_{\mathrm{write}}(g_{\mathrm{norm}})$ (Table~\ref{tab:S2}):
\begin{equation}
    \Delta n_{\mathrm{noisy}} = \Delta n_{\mathrm{ideal}} \cdot \left[1.0 + I_{\mathrm{write\_noise}} \cdot \xi_{\mathrm{write}} \cdot \eta_{\mathrm{write}}\!\left(g_{\mathrm{norm}}(n)\right)\right], \qquad \xi_{\mathrm{write}} \sim \mathcal{N}(0, 1)
    \label{eq:S50_write_update}
\end{equation}
where $\eta_{\mathrm{write}}(g_{\mathrm{norm}})$ increases smoothly from $5.06\%$ ($0.05056$) at $g_{\mathrm{norm}} = 0.0$ through $8.82\%$ at $g_{\mathrm{norm}} = 0.50$ to $12.58\%$ ($0.12576$) near upper saturation $g_{\mathrm{norm}} = 1.0$. Physically, this state-dependent progression arises from local counterion packing statistics: in the early linear regime ($g_{\mathrm{norm}} \to 0$), interstitial coordination sites are largely unoccupied, and counterion translation proceeds with minimal steric or electrostatic hindrance. As the molecular lattice approaches saturation ($g_{\mathrm{norm}} \to 1.0$), the local counterion density approaches the steric close-packing capacity limit. In this crowded regime, excluded-volume interactions and short-range Coulombic repulsion between adjacent $\mathrm{BF_4^-}$ counterions heighten the energetic sensitivity of the hopping barrier to local thermal perturbations, increasing cycle-to-cycle step dispersion. At the system level, maintaining a global write noise of $8.90\%$ enables reliable open-loop pulse programming without closed-loop write-and-verify feedback cycles, saving over $70\%$ of total write energy compared to conventional filamentary memristors~\cite{si:legallo2018,si:sharma2024}.

\subsection{Device-to-Device (D2D) Spatial Mismatch and Inverse-CDF Quantile Sampling}
Across large-scale two-dimensional crossbar arrays of $M \times N$ baseline $[\mathrm{Ru^{II}L_2}](\mathrm{BF_4})_2$ ($\mathrm{BF_4^-}$) junctions, spatial fabrication tolerances introduce static device-to-device (D2D) conductance offsets. In spin-coated $[\mathrm{Ru^{II}L_2}](\mathrm{BF_4})_2$ coordination thin films, these static offsets stem from three microscopic physical factors: (i) nanoscopic film thickness variations ($d_{\mathrm{film}} = 60 \pm 2\ \mathrm{nm}$ across the array), which modulate the internal electric field $E = V/d_{\mathrm{film}}$ by $\approx \pm 3.3\%$; (ii) nanoscale grain boundaries and polycrystalline domain orientation differences within the coordination lattice; and (iii) sub-nanometer interfacial roughness at the bottom and top $\mathrm{Au}$ electrodes that modulates local contact injection barriers.

Evaluated across all $75,000$ physical crosspoint measurements ($3,000$ devices across $25$ target conductance levels)~\cite{si:sharma2024}, the ensemble D2D spatial standard deviation is $\sigma_{\mathrm{D2D}} = 107.17\ \mathrm{nS}$ ($1.07170 \times 10^{-7}\ \mathrm{S}$, extracted across the $3,000$ devices of Sharma et al.~\cite{si:sharma2024} and comparable in magnitude to the $73.6\ \mathrm{nS}$ dispersion measured across $4,096$ crosspoints in $64 \times 64$ arrays by Gaur et al.~\cite{si:gaur2024}; our simulations adopt the larger $107.17\ \mathrm{nS}$ value as a conservative bound and sample it independently at every crosspoint when the D2D channel is enabled). However, experimental thin films exhibit noticeable skewness and heavy kurtosis, with extreme empirical outliers spanning from $-481.28\ \mathrm{nS}$ ($0$-th percentile) to $+433.90\ \mathrm{nS}$ ($100$-th percentile, Table~\ref{tab:S2}) resulting from localized micro-pinholes, grain-boundary clustering, and interfacial roughness. A conventional parametric Gaussian distribution ($\mathcal{N}(0, \sigma_{\mathrm{D2D}}^2)$) severely underestimates the probability of these multi-sigma tail outliers, leading circuit designers to underestimate worst-case sneak-path currents and classification accuracy degradation. To resolve this, our framework implements an empirical 102-point inverse cumulative distribution function (quantile lookup table $Q_{\mathrm{D2D}}(u)$, Table~\ref{tab:S2}) extracted directly from the physical hardware histogram:
\begin{equation}
    G_{ij,\mathrm{total}}(n_{ij}) = \max\left(G_{\min},\, G_{\mathrm{ideal}}(n_{ij}) + I_{\mathrm{D2D}} Q_{\mathrm{D2D}}(u_{ij}) + I_{\mathrm{read\_noise}}\xi_{\mathrm{read},ij}\sigma_{\mathrm{read}}\!\left(g_{\mathrm{norm}}(n_{ij})\right)\right)
    \label{eq:S51_d2d_total}
\end{equation}
For each device $(i,j)$ in an array, a uniform random seed $u_{ij} \sim \mathcal{U}(0, 1)$ is assigned once at array initialization, mapping directly to the exact empirical quantile $Q_{\mathrm{D2D}}(u_{ij})$. This nonparametric inverse-CDF sampling reproduces the true physical wafer yield, quartile spatial dispersion, and non-Gaussian tail behavior with exact statistical fidelity.

\subsection{Three-Channel Hardware Noise Verification with a Deterministic Baseline Control}
Our hardware noise validation suite (Tables~\ref{tab:S4} and \ref{tab:S5}) systematically verifies the three stochastic noise channels, together with a noise-free deterministic control (Channel 0), on the baseline $[\mathrm{Ru^{II}L_2}](\mathrm{BF_4})_2$ ($\mathrm{BF_4^-}$) model:
\begin{enumerate}[topsep=2pt,itemsep=2pt,parsep=0pt]
    \item \textit{Channel 0 (Deterministic Baseline Invariance, $I_{\mathrm{noise}} = 0.0$):} Confirms exact $0.00\%$ discrepancy across repeated matrix readout and vector-matrix inference evaluations when all stochastic noise intensities are disabled.
    \item \textit{Channel 1 (Static D2D Spatial Mismatch, $I_{\mathrm{D2D}} = 1.0$):} Initializes a $10 \times 10$ ($100$-device) $\mathrm{BF_4^-}$ crossbar via inverse-CDF quantile sampling, verifying strictly positive device transconductances ($g_{\mathrm{scale},ij} > 0$) and accurate reproduction of the empirical quantile spread ($-481.28\ \mathrm{nS}$ to $+433.90\ \mathrm{nS}$).
    \item \textit{Channel 2 (Temporal Read Noise, $I_{\mathrm{read\_noise}} = 1.0$):} Programs a $4 \times 4$ $\mathrm{BF_4^-}$ weight matrix across $[0.2, 0.9]$ normalized conductance and executes 50 repeated sub-threshold analog read sweeps and VMM inference passes, confirming conductance-dependent Gaussian variance matching Table~\ref{tab:S2}.
    \item \textit{Channel 3 (Stochastic C2C Write Noise, $I_{\mathrm{write\_noise}} = 1.0$):} Executes open-loop pulse weight updates across a $2 \times 2$ $\mathrm{BF_4^-}$ crossbar and compares programmed states $n_{ij}$ against an ideal deterministic twin, confirming realistic Langevin pulse-to-pulse state jitter matching Table~\ref{tab:S2}.
\end{enumerate}

Incorporating this three-channel empirical noise model directly into our circuit solvers and machine learning pipelines provides the neuromorphic computing community with a physically grounded simulation environment for molecular memristors. It enables researchers and circuit architects to evaluate algorithmic error tolerance, design optimal analog weight mapping strategies (such as the differential pair architecture $W_{ij} = G_{ij}^+ - G_{ij}^-$ in Note~S10 that suppresses common-mode D2D offsets), and benchmark real-world inference and on-chip learning accuracy under genuine wafer-scale hardware non-idealities.

\section{Machine Learning and Software-Level Neural Network Training}
\label{sec:noteS10}

This note is dedicated to the hardware-in-the-loop deep learning architecture, differential synaptic conductance mapping, custom Straight-Through Estimator (STE) gradient backpropagation, and physical energy accounting for our hardware-in-the-loop MNIST classification benchmark evaluated on baseline $[\mathrm{Ru^{II}L_2}](\mathrm{BF_4})_2$ ($\mathrm{BF_4^-}$) crossbars (Section~4.1 and Figure~5 of the main manuscript; Table~\ref{tab:S4}).

\subsection{\texorpdfstring{MNIST Spatial Pooling and Differential Crossbar Mapping ($W_{ij} = G_{ij}^+ - G_{ij}^-$)}{MNIST Spatial Pooling and Differential Crossbar Mapping (Wij = Gij+ - Gij-)}}
The MNIST handwritten digit benchmark~\cite{si:lecun1998} comprises $60,000$ training images and $10,000$ test images of $28 \times 28$ grayscale pixels across $C = 10$ digit classes ($0$ through $9$). In the single-layer $64 \times 10$ baseline $[\mathrm{Ru^{II}L_2}](\mathrm{BF_4})_2$ ($\mathrm{BF_4^-}$) molecular crossbar classifier reported in Figure~5 of the main manuscript, each $28 \times 28$ input image is downsampled via $3.5 \times 3.5$ spatial average pooling into an $8 \times 8 = 64$-element normalized feature vector $\mathbf{x} \in [0, 1]^{64}$ driving 64 wordlines against 10 class bitlines (our codebase also supports a multi-layer tiled $64 \to 256 \to 10$ multilayer perceptron using $64 \times 64$ crossbar tiles).

Because physical baseline $[\mathrm{Ru^{II}L_2}](\mathrm{BF_4})_2$ ($\mathrm{BF_4^-}$) memristor conductances $G_{ij} \in [G_{\min}, G_{\max}]$ ($[200\ \mathrm{nS}, 4.33\ \mathrm{mS}]$ nominal calibration window, or $[200\ \mathrm{nS}, 5.90\ \mathrm{mS}]$ full saturation capacity) are strictly positive whereas neural network weights $W_{ij} \in [-W_{\max}, +W_{\max}]$ are bipolar, each logical synaptic weight is mapped onto a differential pair of adjacent molecular memristor crosspoints $(G_{ij}^+, G_{ij}^-)$ centered around the mid-scale conductance $G_{\mathrm{mid}} = (G_{\max} + G_{\min})/2$ (or equivalently realized on a single crosspoint with a uniform $G_{\mathrm{mid}}$ reference column subtracted at the TIA output):
\begin{equation}
    G_{ij}^+ = G_{\mathrm{mid}} + \frac{\Delta G_{\mathrm{span}}}{2 W_{\max}} W_{ij}, \qquad G_{ij}^- = G_{\mathrm{mid}} - \frac{\Delta G_{\mathrm{span}}}{2 W_{\max}} W_{ij}, \qquad G_{ij}^+ - G_{ij}^- = \left(\frac{\Delta G_{\mathrm{span}}}{W_{\max}}\right) W_{ij}
    \label{eq:S52_diff_map}
\end{equation}
During the forward pass, the input activation vector $\mathbf{x} \in [0, 1]^{64}$ is linearly scaled into sub-threshold wordline read voltages $\mathbf{V}_{\mathrm{read}} = V_{\mathrm{read,max}}\mathbf{x}$ with $V_{\mathrm{read,max}} = 0.20\ \mathrm{V}$ ($< V_{\mathrm{th}} = 0.4664\ \mathrm{V}$ for baseline $\mathrm{BF_4^-}$). The physical crossbar array executes analog vector-matrix multiplication natively via Kirchhoff's current law:
\begin{equation}
    I_j = \sum_{i=1}^{64} V_{\mathrm{read},i}\left(G_{ij,\mathrm{noisy}}^+ - G_{ij,\mathrm{noisy}}^-\right) \quad \Longrightarrow \quad \mathbf{y}_{\mathrm{hw}} = \kappa_{\mathrm{ADC}}\mathbf{I} + \mathbf{b}
    \label{eq:S53_vmm_mnist}
\end{equation}
where $G_{ij,\mathrm{noisy}}^{\pm}$ includes empirical $\mathrm{BF_4^-}$ D2D mismatch, C2C write noise from the preceding weight update, and instantaneous read noise $\sigma_{\mathrm{read}}(G)$ (Equation~(\ref{eq:S51_d2d_total})), and $\kappa_{\mathrm{ADC}} = W_{\max}/(V_{\mathrm{read,max}}\Delta G_{\mathrm{span}})$ is the TIA/ADC normalization gain.

\subsection{Custom Straight-Through Estimator Backpropagation}
Because the physical pulse-programming operator $W_{ij} \mapsto n_{ij} \mapsto G_{ij}(n_{ij})$ involves discrete pulse quantization ($\Delta n_{ij} = \mathrm{round}(\Delta W_{ij} / \delta w_{\mathrm{pulse}})$) and stochastic hardware noise, its exact pointwise derivative $\partial G_{ij}/\partial W_{ij}$ is zero almost everywhere or undefined at discrete step boundaries. We overcome this in TensorFlow~2.x by wrapping the physical crossbar forward evaluation inside a custom Straight-Through Estimator~\cite{si:bengio2013} (\texttt{@tf.custom\_gradient}) decorator:
\begin{equation}
    \text{Forward Pass:} \quad \mathbf{y} = \mathcal{F}_{\mathrm{crossbar}}\!\left(\mathbf{x}, \mathbf{G}(\mathbf{n})\right), \qquad
    \text{Backward Pass:} \quad \frac{\partial \mathcal{L}}{\partial \mathbf{x}} = \frac{\partial \mathcal{L}}{\partial \mathbf{y}}\mathbf{W}_{\mathrm{eff}}^T, \quad \frac{\partial \mathcal{L}}{\partial \mathbf{W}} = \mathbf{x}^T \frac{\partial \mathcal{L}}{\partial \mathbf{y}}
    \label{eq:S54_ste}
\end{equation}
where $\mathcal{L} = -\sum_{c=1}^{10} y_{\mathrm{true},c}\ln(\mathrm{softmax}(\mathbf{y})_c)$ is the categorical cross-entropy loss and $\mathbf{W}_{\mathrm{eff}} = (\mathbf{G}_{\mathrm{noisy}}^+ - \mathbf{G}_{\mathrm{noisy}}^-)(W_{\max}/\Delta G_{\mathrm{span}})$ is the physical weight tensor read from the crossbar. After the Adam optimizer computes the target weight update $\Delta\mathbf{W}^{(t)}$, the required open-loop pulse count increment $\Delta n_{ij}^{(t)} = \mathrm{round}\!\left(N_{\mathrm{pot}}\Delta W_{ij}^{(t)} / (2 W_{\max})\right)$ is applied directly to the physical baseline $[\mathrm{Ru^{II}L_2}](\mathrm{BF_4})_2$ ($\mathrm{BF_4^-}$) memristor state $n_{ij}$ using $+0.90\ \mathrm{V}$ ($80\ \mathrm{ns}$) potentiation pulses for $\Delta n_{ij} > 0$ or $-0.75\ \mathrm{V}$ ($65\ \mathrm{ns}$) depression pulses for $\Delta n_{ij} < 0$.

\subsection{Training Convergence and Per-Epoch Physical Energy Accounting}
Across 15 training epochs on the $64 \times 10$ baseline $[\mathrm{Ru^{II}L_2}](\mathrm{BF_4})_2$ ($\mathrm{BF_4^-}$) molecular crossbar (Figure~5 of the main manuscript), categorical cross-entropy loss decreased monotonically from $5516.39$ (epoch~1) to $132.15$ (epoch~15), while validation accuracy peaked at $97.50\%$ at epoch~13 and stabilized at $95.31\%$ at epoch~15. Total physical crossbar energy consumption across all 15 epochs is computed by integrating the instantaneous write and read pulse energies over all 640 $\mathrm{BF_4^-}$ crosspoints:
\begin{align}
    E_{\mathrm{MNIST,total}} &= E_{\mathrm{write}} + E_{\mathrm{read}} = \sum_{\mathrm{epochs}=1}^{15} \sum_{\mathrm{batches}} \left(N_{\mathrm{pulses,pot}} E_{\mathrm{pot}} + N_{\mathrm{pulses,dep}} E_{\mathrm{dep}} + N_{\mathrm{reads}} V_{\mathrm{read}}^2 \bar{G} t_{\mathrm{read}}\right) \nonumber \\
    &= 0.0020\ \mathrm{J} + 0.0008\ \mathrm{J} = 0.0028\ \mathrm{J}
    \label{eq:S55_energy_mnist}
\end{align}
comprising $E_{\mathrm{write}} = 0.0020\ \mathrm{J}$ ($2.0\ \mathrm{mJ}$) for open-loop weight update pulses ($0.19\ \mathrm{nJ/pulse}$ nominal potentiation, $0.13\ \mathrm{nJ/pulse}$ depression) and $E_{\mathrm{read}} = 0.0008\ \mathrm{J}$ ($0.8\ \mathrm{mJ}$) for sub-threshold forward inference reads.

\section{Astronomical Spatial Frequency Image Processing}
\label{sec:noteS11}

This final note is dedicated to the James Webb Space Telescope (JWST) Inverse Discrete Fourier Transform (IDFT) image reconstruction benchmark evaluated on baseline $[\mathrm{Ru^{II}L_2}](\mathrm{BF_4})_2$ ($\mathrm{BF_4^-}$) crossbars (Section~4.2 and Figure~5d,e of the main manuscript, Figure~\ref{fig:S6}, and Table~\ref{tab:S4}).

\subsection{Row-Wise 64-Point Inverse Discrete Fourier Transform (IDFT) Crossbar Formulation}
To evaluate the baseline $[\mathrm{Ru^{II}L_2}](\mathrm{BF_4})_2$ ($\mathrm{BF_4^-}$) molecular crossbar on a scientific computing workload requiring high analog dynamic range and hundreds of thousands of non-destructive dot products~\cite{si:wang2017,si:kumar2017,si:kumar2020,si:canli2018}, we implemented a row-wise 64-point IDFT engine reconstructing the $1000 \times 1000$ RGB Pillars of Creation image captured by the James Webb Space Telescope (source image credit: NASA, ESA, CSA, STScI). The $1000 \times 1000$ image is zero-padded to $1024 \times 1024$ and partitioned into $16 \times 16 = 256$ spatial blocks of dimension $64 \times 64$ per RGB color channel ($768$ blocks total). Within each $64 \times 64$ block, each row $m \in \{0, \dots, 63\}$ undergoes a 1D $N = 64$-point forward Fast Fourier Transform $\hat{X}_m(p) = \sum_{q=0}^{63} x_m(q)\exp(-j 2\pi p q / 64)$, and the physical crossbar reconstructs the spatial row via the 64-point IDFT basis matrix $W_{p,q} = \exp(+j 2\pi p q / 64) = C_{p,q} + j S_{p,q}$ ($p, q \in \{0, \dots, 63\}$), where $C_{p,q} = \cos(2\pi p q / 64)$ and $S_{p,q} = \sin(2\pi p q / 64)$:
\begin{equation}
    x_m(q) = \frac{1}{64}\sum_{p=0}^{63} \hat{X}_m(p) \exp\left(+j\,\frac{2\pi p q}{64}\right) \quad \Longleftrightarrow \quad \mathbf{X}_{\mathrm{block}} = \frac{1}{64}\,\hat{\mathbf{X}}_{\mathrm{block}}\,\mathbf{W}_{64}
    \label{eq:S56_idft_2d}
\end{equation}
Because passive wordline voltage drivers and molecular conductances are strictly non-negative, each complex row spectrum $\hat{\mathbf{x}}_m = \hat{\mathbf{x}}_{m,R} + j\hat{\mathbf{x}}_{m,I} \in \mathbb{C}^{64}$ is split into four non-negative unipolar input vectors $\hat{\mathbf{x}}_{m,R}^+ = \max(\hat{\mathbf{x}}_{m,R}, 0)$, $\hat{\mathbf{x}}_{m,R}^- = \max(-\hat{\mathbf{x}}_{m,R}, 0)$, $\hat{\mathbf{x}}_{m,I}^+ = \max(\hat{\mathbf{x}}_{m,I}, 0)$, and $\hat{\mathbf{x}}_{m,I}^- = \max(-\hat{\mathbf{x}}_{m,I}, 0)$, each normalized per vector by its peak magnitude and applied via 14-bit PWM/ADC crossbar readout (\texttt{passInput}) to evaluate the real and imaginary dot products for output pixel columns $q \in \{0, \dots, 31\}$:
{\small
\begin{align}
    \mathbf{Y}_{R,0:31} &= \left(\hat{\mathbf{X}}_R^+ - \hat{\mathbf{X}}_R^-\right)\mathbf{C}_{:,0:31} - \left(\hat{\mathbf{X}}_I^+ - \hat{\mathbf{X}}_I^-\right)\mathbf{S}_{:,0:31}, \quad
    \mathbf{Y}_{I,0:31} = \left(\hat{\mathbf{X}}_R^+ - \hat{\mathbf{X}}_R^-\right)\mathbf{S}_{:,0:31} + \left(\hat{\mathbf{X}}_I^+ - \hat{\mathbf{X}}_I^-\right)\mathbf{C}_{:,0:31} \label{eq:S57_idft_pass1} \\
    Z_{m,q} &= \frac{1}{64}\left(Y_{R,m,q} + j\,Y_{I,m,q}\right), \qquad Z_{m,64-q} = Z_{m,q}^* \ \ (q = 1, \dots, 31), \qquad Z_{m,32} = \frac{1}{64}\sum_{p=0}^{63}(-1)^p \hat{X}_m(p) \label{eq:S58_idft_pass2}
\end{align}
}%
where conjugate spatial pixels $q \in \{33, \dots, 63\}$ are recovered directly from $q \in \{1, \dots, 31\}$ via Hermitian symmetry ($W_{p,64-q} = W_{p,q}^*$) and the Nyquist sample $q = 32$ is evaluated digitally.

\subsection{Dual-Conductance Mapping and Closed-Loop 14-Bit Crossbar Programming}
\begin{figure}[!htb]
    \centering
    \includegraphics[width=\textwidth,height=0.62\textheight,keepaspectratio]{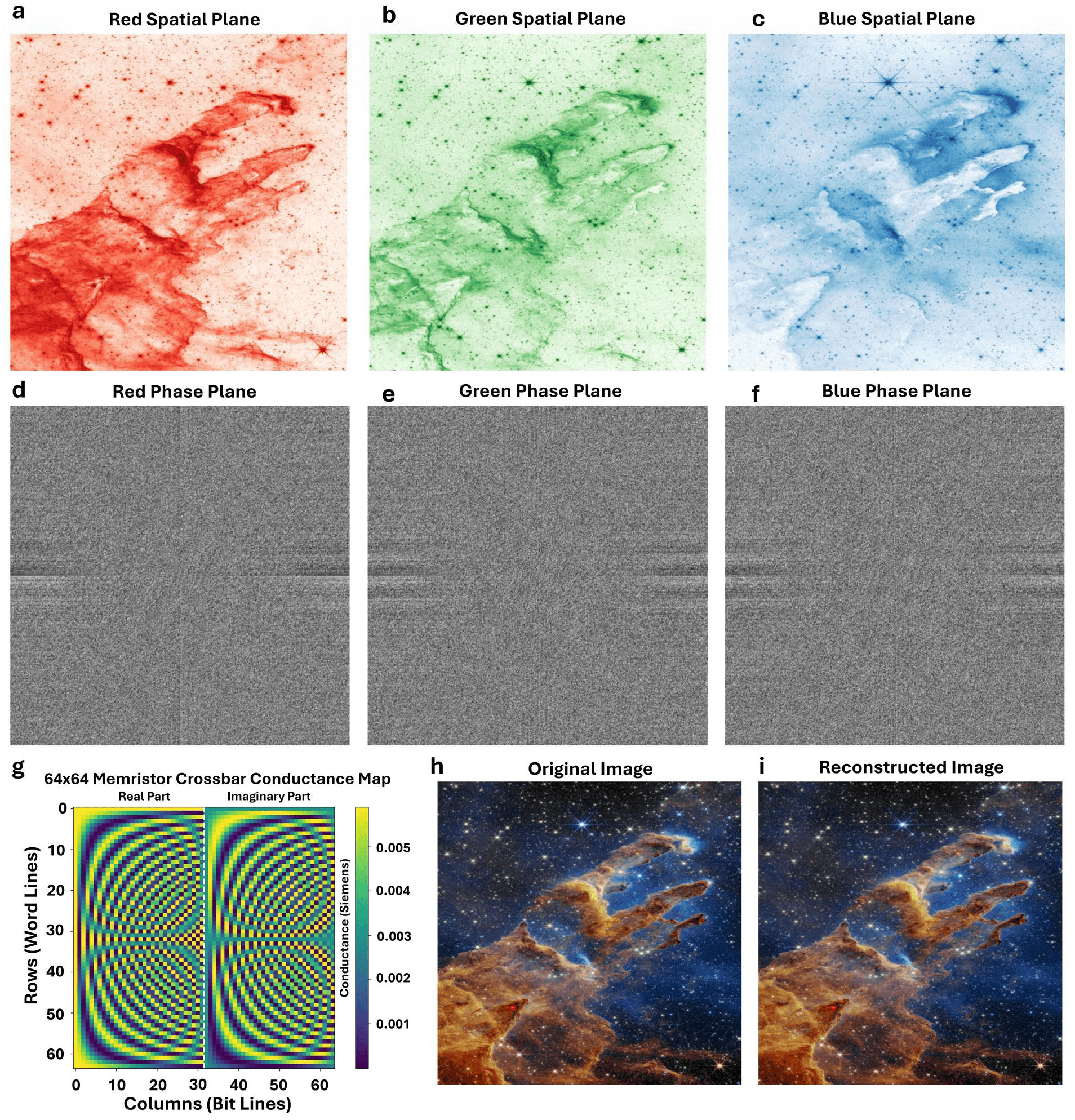}
    \caption{\textbf{Complete spatial RGB and 2D Fourier phase-plane decompositions for the James Webb Space Telescope (JWST) Pillars of Creation benchmark ($1000 \times 1000$ resolution) on a baseline $[\mathrm{Ru^{II}L_2}](\mathrm{BF_4})_2$ ($\mathrm{BF_4^-}$) crossbar.} \textbf{a--c}, Isolated spatial intensity planes for the Red (\textbf{a}), Green (\textbf{b}), and Blue (\textbf{c}) color channels extracted from the reference $1000 \times 1000$ JWST image. \textbf{d--f}, Global two-dimensional Fourier phase spectra $\arg\{\mathcal{F}_{2\mathrm{D}}\{I_c\}(u,v)\} \in [-\pi, +\pi]$ for the Red (\textbf{d}), Green (\textbf{e}), and Blue (\textbf{f}) channels, visualizing the global spatial-frequency phase structure of the RGB image. \textbf{g}, Physical $64 \times 64$ baseline $\mathrm{BF_4^-}$ molecular memristor crossbar conductance map ($G_{p,q} \in [200\ \mathrm{nS}, 5.90\ \mathrm{mS}]$) partitioned into Columns $0$--$31$ (Real 64-point IDFT basis $\mathrm{Re}(\mathbf{W}_{:,0:32})$) and Columns $32$--$63$ (Imaginary 64-point IDFT basis $\mathrm{Im}(\mathbf{W}_{:,0:32})$). \textbf{h, i}, Visual comparison between the original $1000 \times 1000$ reference image (\textbf{h}) and the analog crossbar-reconstructed output image (\textbf{i}) across $768$ padded $64 \times 64$ blocks ($3,072$ batched crossbar evaluations, $196,608$ vector-matrix multiplications; $99.30\%$ SSIM, $43.0\ \mathrm{dB}$ SNR). Source image credit: NASA, ESA, CSA, STScI.}
    \label{fig:S6}
\end{figure}

To execute Equations~(\ref{eq:S57_idft_pass1})--(\ref{eq:S58_idft_pass2}) on a single $64 \times 64$ 1T1R baseline $[\mathrm{Ru^{II}L_2}](\mathrm{BF_4})_2$ ($\mathrm{BF_4^-}$) molecular memristor crossbar (\texttt{weightBits=14}, \texttt{inputBits=14}, \texttt{outputBits=14}; Figure~\ref{fig:S6}g), the 64 wordlines are driven by frequency indices $p \in \{0, \dots, 63\}$ and the 64 bitline columns are partitioned into two $64 \times 32$ sub-arrays for spatial indices $q \in \{0, \dots, 31\}$: Columns $0$--$31$ store the shifted real cosine basis conductances $G_{p,q}^{\mathrm{Re}}$ ($\mathrm{Re}(\mathbf{W}_{:,0:32})$), and Columns $32$--$63$ store the shifted imaginary sine basis conductances $G_{p,q}^{\mathrm{Im}}$ ($\mathrm{Im}(\mathbf{W}_{:,0:32})$), mapped via $W_{\mathrm{target}} = (W + 1.0)/2.0 \in [0, 1]$ across the full physical $\mathrm{BF_4^-}$ conductance window $[G_{\min}, G_{\max}] = [200\ \mathrm{nS}, 5.90\ \mathrm{mS}]$:
\begin{equation}
    G_{p,q}^{\mathrm{Re}} = G_{\min} + (G_{\max} - G_{\min})\frac{\cos(2\pi p q / 64) + 1.0}{2.0}, \qquad
    G_{p,q}^{\mathrm{Im}} = G_{\min} + (G_{\max} - G_{\min})\frac{\sin(2\pi p q / 64) + 1.0}{2.0}
    \label{eq:S59_jwst_gmap}
\end{equation}
The $64 \times 64$ crossbar is programmed using two-pass closed-loop 14-bit program-and-verify tuning (\texttt{updateCrossbarWeights(..., fineTune=True)} executed twice). The common-mode offset introduced by the $+1/2$ shift around $G_{\mathrm{mid}} = (G_{\max} + G_{\min})/2 = 2.9501\ \mathrm{mS}$ is removed digitally in normalized readout units via $\mathbf{Y} = 2\mathbf{Y}_{\mathrm{hw}} - \left(\sum_{p=0}^{63} V_{\mathrm{in},p}\right)\mathbf{1}^T$.

\clearpage

\clearpage

\subsection{The 768-Block Batched Evaluation Schedule and Reconstruction Fidelity}
Padding each of the three $1000 \times 1000$ color channels (Red, Green, and Blue; Figure~\ref{fig:S6}a--c) to $1024 \times 1024$ yields $16 \times 16 = 256$ spatial blocks of dimension $64 \times 64$ per channel ($768$ blocks total). Each block executes four batched 64-vector crossbar evaluations corresponding to the four unipolar input rails ($\hat{\mathbf{X}}_R^+, \hat{\mathbf{X}}_R^-, \hat{\mathbf{X}}_I^+, \hat{\mathbf{X}}_I^-$), totaling $4 \times 768 = 3,072$ batched crossbar evaluations ($3,072 \times 64 = 196,608$ vector-matrix multiplications across the full RGB image). The $64 \times 64$ IDFT basis matrix in Figure~\ref{fig:S6}g is programmed onto the molecular crossbar once at initialization (consuming $E_{\mathrm{prog}} = 0.0004\ \mathrm{J}$ across the $4,096$ crosspoints) and reused across all $768$ blocks without reprogramming ($E_{\mathrm{infer}} = 0.0242\ \mathrm{J}$, yielding a total crossbar energy of $E_{\mathrm{JWST,total}} = 0.0246\ \mathrm{J}$). Following complex magnitude evaluation ($|Z_{m,q}|$), global min-max dynamic range rescaling to $[0, 255]$, and cropping to the original $1000 \times 1000$ field, comparing the reconstructed output $\mathbf{I}_{\mathrm{recon}}$ (Figure~\ref{fig:S6}i) against the original $1000 \times 1000$ reference image $\mathbf{I}_{\mathrm{orig}}$ (Figure~\ref{fig:S6}h) yields a Structural Similarity Index Measure (SSIM) and Signal-to-Noise Ratio (SNR) of:
{\small
\begin{equation}
    \mathrm{SSIM}\!\left(\mathbf{I}_{\mathrm{orig}}, \mathbf{I}_{\mathrm{recon}}\right) = \frac{(2\mu_x\mu_y + C_1)(2\sigma_{xy} + C_2)}{(\mu_x^2 + \mu_y^2 + C_1)(\sigma_x^2 + \sigma_y^2 + C_2)} = 99.30\%, \quad
    \mathrm{SNR} = 10\log_{10}\!\left(\frac{\langle \mathbf{I}_{\mathrm{orig}}^2 \rangle}{\langle (\mathbf{I}_{\mathrm{orig}} - \mathbf{I}_{\mathrm{recon}})^2 \rangle}\right) = 43.0\ \mathrm{dB}
    \label{eq:S60_ssim_snr}
\end{equation}
}
Because this benchmark evaluates deterministic crossbar fidelity with stochastic variability channels (D2D and C2C write dispersion) and interconnect wire resistance disabled, the $43.0\ \mathrm{dB}$ full-array SNR reflects finite 14-bit input, weight, and output quantization, residual two-pass closed-loop programming tolerance across the $4,096$ crosspoints, and nonlinear hyperbolic sine ($\sinh$) conduction across the readout window.

\end{document}